%% file: SUS-16-033_temp.tex
\pdfoutput=1

\documentclass[11pt,twoside,a4paper,cmspaper,final,collab]{cms-tdr}
\def\svnVersion{423249P}\def\cmsCernNoTag{CERN-EP-2017-072}\def\cmsCernDate{\today}\def\cmsMessage{Published in Physical Review D as \href{http://dx.doi.org/10.1103/PhysRevD.96.032003}{\doi{10.1103/PhysRevD.96.032003.}}}

\begin{document}\cmsNoteHeader{SUS-16-033}

\hyphenation{had-ron-i-za-tion}
\hyphenation{cal-or-i-me-ter}
\hyphenation{de-vices}
\RCS$HeadURL: svn+ssh://svn.cern.ch/reps/tdr2/papers/SUS-16-033/trunk/SUS-16-033.tex $
\RCS$Id: SUS-16-033.tex 413333 2017-06-28 20:03:28Z bgary $

\newcolumntype{x}{D{,}{\,\pm\,}{4.2}}

\newlength\cmsFigWidth
\ifthenelse{\boolean{cms@external}}{\setlength\cmsFigWidth{0.49\textwidth}}{\setlength\cmsFigWidth{0.9\textwidth}}
\ifthenelse{\boolean{cms@external}}{\providecommand{\cmsLeft}{top\xspace}}{\providecommand{\cmsLeft}{left\xspace}}
\ifthenelse{\boolean{cms@external}}{\providecommand{\cmsRight}{bottom\xspace}}{\providecommand{\cmsRight}{right\xspace}}
\ifthenelse{\boolean{cms@external}}{\providecommand{\cmsAppendix}{}}{\providecommand{\cmsAppendix}{Appendix~}}

\newcommand{\jtrans}{\ensuremath{{\vec p}_{\text{T},\text{j}_{\text{1}}}}\xspace}
\newcommand{\MHT}{\ensuremath{H_{\text{T}}^{\text{miss}}}\xspace}
\newcommand{\htvecmiss}{\ensuremath{\vec{H}_{\text{T}}^{\text{miss}}}\xspace}
\newcommand{\njets}{\ensuremath{N_{\text{jet}}}\xspace}
\newcommand{\nbjets}{\ensuremath{N_{{\cPqb}\text{-jet}}}\xspace}
\newcommand{\rqcd}{\ensuremath{R^{\text{QCD}}}\xspace}
\newcommand{\khtnj}{\ensuremath{K_{ij}^{\text{data}}}\xspace}
\newcommand{\smhtsim}{\ensuremath{S_{ik}^{\text{sim}}}\xspace}
\newcommand{\mt}{\ensuremath{m_{\mathrm{T}}}\xspace}
\newcommand{\mlsp}{\ensuremath{m_{{\PSGcz}_1}}\xspace}
\newcommand{\mgluino}{\ensuremath{m_{\PSg}}\xspace}
\newcommand{\msquark}{\ensuremath{m_{\PSQ}}\xspace}
\newcommand{\mstop}{\ensuremath{m_{\PSQt}}\xspace}
\newcommand{\zll}{\ensuremath{\cPZ\to \ell^{+}\ell^{-}}\xspace}
\newcommand{\gjets}{{{\cPgg}+jets}\xspace}
\newcommand{\ngdata}{\ensuremath{N_{\gamma}^\text{obs}}\xspace}
\newcommand{\nznn}{\ensuremath{N_{\znn}^\text{pred}}\xspace}
\newcommand{\rznn}{\ensuremath{\mathcal{R}_{\znn/\gamma}}\xspace}
\newcommand{\rznnsim}{\ensuremath{\mathcal{R}_{\znn/\gamma}^{\text{sim}}}\xspace}
\newcommand{\znn}{\ensuremath{\cPZ \to \cPgn \cPagn}\xspace}
\newcommand{\mtop}{\ensuremath{m_{\text{top}}}\xspace}
\newcommand{\dmass}{\ensuremath{\Delta m}\xspace}
\newcommand{\neles}{\ensuremath{N_{\text{electron}}}\xspace}
\newcommand{\nmuons}{\ensuremath{N_{\text{muon}}}\xspace}
\newcommand{\nisomuons}{\ensuremath{N_{\text{isolated tracks}}^{\text{(muon)}}}\xspace}
\newcommand{\nisoeles}{\ensuremath{N_{\text{isolated tracks}}^{\text{(electron)}}}\xspace}
\newcommand{\nisohads}{\ensuremath{N_{\text{isolated tracks}}^{\text{(hadron)}}}\xspace}
\newcommand{\dphimht}{\ensuremath{\Delta\phi_{\MHT,\text{j}_{i}}}\xspace}
\newcommand{\dpmht}[1]{\ensuremath{\Delta\phi_{\MHT,\text{j}_{#1}}}\xspace}
\newcommand{\dphimhtb}{\ensuremath{\Delta\phi_{\MHT,\text{j}_{1(b)}}}\xspace}
\newcommand{\SFdmcg}{\ensuremath{\mathcal{C}_{\text{data/sim}}^\gamma}\xspace}
\newcommand{\SFdmcll}{\ensuremath{\mathcal{C}_{\text{data/sim}}^{\ell\ell}}\xspace}
\newcommand{\Fdir}{\ensuremath{\mathcal{F}^{\text{sim}}_{\text{dir}}}\xspace}
\newcommand{\DblR}{\ensuremath{\rho}\xspace}
\newcommand{\njetsisr}{\ensuremath{N_{\text{jet}}^{\text{ISR}}}\xspace}
\newcommand{\rands}{R\&S\xspace}
\newcommand{\dphi}{\ensuremath{\Delta \phi}\xspace}
\newcommand{\nvtx}{\ensuremath{N_{\text{vtx}}}\xspace}
\newcommand{\rscale}{\ensuremath{\mu_{\text{R}}}\xspace}
\newcommand{\fscale}{\ensuremath{\mu_{\text{F}}}\xspace}
\newcommand{\colspacea}{\hphantom{33333333}}
\newcommand{\colspaceb}{\hphantom{33333333}}
\newcommand{\colspacec}{\hphantom{33333333}}
\newcommand\zjets{{{\cPZ}+jets}\xspace}
\newcommand\wjets{{{\PW}+jets}\xspace}
\newcommand\zlljets{{\ensuremath{\cPZ (\to \ell^{+} \ell^{-} )}+jets}\xspace}
\newcommand\znnjets{{\ensuremath{\cPZ(\to\cPgn \cPagn )}+jets}\xspace}
\newcommand\imini{\ensuremath{I}\xspace}

\newcolumntype{R}{>{$}r<{$}}
\newcolumntype{L}{>{$}l<{$}}
\newcolumntype{M}{R@{$\;$}L}
\newcolumntype{S}{r@{$\,\pm\,$}r}

\cmsNoteHeader{SUS-15-002}
\title{Search for supersymmetry in multijet events with missing transverse momentum in proton-proton collisions at \texorpdfstring{13\TeV}{13 TeV}}

\date{\today}

\abstract{
A search for supersymmetry is presented based on multijet events with large missing transverse momentum produced in proton-proton collisions at a center-of-mass energy of $\sqrt{s}=13\TeV$. The data, corresponding to an integrated luminosity of 35.9\fbinv, were collected with the CMS detector at the CERN LHC in 2016. The analysis utilizes four-dimensional exclusive search regions defined in terms of the number of jets, the number of tagged bottom quark jets, the scalar sum of jet transverse momenta, and the magnitude of the vector sum of jet transverse momenta. No evidence for a significant excess of events is observed relative to the expectation from the standard model. Limits on the cross sections for the pair production of gluinos and squarks are derived in the context of simplified models. Assuming the lightest supersymmetric particle to be a weakly interacting neutralino, 95\% confidence level lower limits on the gluino mass as large as 1800 to 1960\GeV are derived, and on the squark mass as large as 960 to 1390\GeV, depending on the production and decay scenario.
}

\hypersetup{%
pdfauthor={CMS Collaboration},%
pdftitle={Search for supersymmetry in multijet events with missing transverse momentum in proton-proton collisions at 13 TeV},%
pdfsubject={CMS},%
pdfkeywords={CMS, physics, supersymmetry, multijets}}

\maketitle

\section{Introduction}
\label{sec:introduction}

The standard model (SM) of particle physics
describes many aspects of weak, electromagnetic,
and strong interactions.
However,
it requires fine tuning~\cite{Barbieri:1987fn}
to explain the observed value
of the Higgs boson mass~\cite{Aad:2015zhl},
and it does not provide an explanation for dark matter.
Supersymmetry (SUSY)~\cite{Ramond:1971gb,Golfand:1971iw,Neveu:1971rx,
Volkov:1972jx,Wess:1973kz,Wess:1974tw,Fayet:1974pd,Nilles:1983ge},
a widely studied extension of the SM,
potentially solves these problems through the introduction
of a new particle,
called a superpartner, for each SM particle,
with a spin that differs from that of its SM counterpart by a half unit.
Additional Higgs bosons and their superpartners are also introduced.
The superpartners of quarks and gluons are squarks {\sQua}
and gluinos {\PSg}, respectively,
while neutralinos \PSGcz and charginos \PSGcpm
are mixtures of the superpartners of
the Higgs and electroweak gauge bosons.
Provided that the masses of gluinos, top squarks, and bottom squarks
are no heavier than a few TeV,
SUSY can resolve the fine-tuning
problem~\cite{Barbieri:1987fn,Dimopoulos:1995mi,Barbieri:2009ev,Papucci:2011wy}.
Furthermore,
in $R$-parity~\cite{bib-rparity} conserving SUSY models,
the lightest SUSY particle (LSP)
is stable and might interact only weakly,
thus representing a dark matter candidate.

In this paper,
we present a search
for squarks and gluinos produced in proton-proton ($\Pp\Pp$)
collisions at $\sqrt{s}=13\TeV$.
Squark and gluino production have large potential cross sections
in $\Pp\Pp$ collisions,
thus motivating this search.
The study is performed in the multijet final state,
i.e., the visible elements consist solely of jets.
Other $\sqrt{s}=13\TeV$ inclusive multijet SUSY searches are presented in
Refs.~\cite{Aad:2016eki,Aaboud:2016zdn,Khachatryan:2016kdk,Khachatryan:2016xvy,
Khachatryan:2016epu,Khachatryan:2016dvc}.
We assume the conservation of $R$-parity,
meaning that the squarks and gluinos are produced in pairs.
The events are characterized by the presence
of jets and undetected, or ``missing,'' transverse momentum,
where the missing transverse momentum arises
from the weakly interacting and unobserved LSPs.
The data,
corresponding to an integrated luminosity of 35.9\fbinv,
were collected in 2016 with the CMS detector at the CERN LHC.
The analysis is performed in four-dimensional exclusive
regions in the number of jets \njets,
the number of tagged bottom quark jets \nbjets,
the scalar sum \HT of the transverse momenta \pt of jets,
and the magnitude \MHT of the vector \pt sum of jets.
The number of observed events in each region is compared
with the expected number of SM events to
search for excesses in the data.

The study is an extension of that presented
in Ref.~\cite{Khachatryan:2016kdk},
using improved analysis techniques and
around 16 times more data.
Relative to Ref.~\cite{Khachatryan:2016kdk},
the following principal modifications have been made.
Firstly,
the search intervals in \njets and \HT are given by
$\njets\geq2$ and $\HT>300\GeV$,
compared with $\njets\geq4$ and $\HT>500\GeV$
in Ref.~\cite{Khachatryan:2016kdk}.
Inclusion of events with $\njets=2$ and 3
increases the sensitivity to squark pair production.
The lower threshold in \HT provides better
sensitivity to scenarios with small mass differences between
the LSP and the squark or gluino.
Secondly,
the rebalance-and-smear technique~\cite{Collaboration:2011ida,Chatrchyan:2014lfa}
is introduced as a complementary means to evaluate the
quantum chromodynamics (QCD) background,
namely the background from SM events with multijet final states
produced exclusively through the strong interaction.
Thirdly, the search interval in \MHT is given by $\MHT>300\GeV$,
rather than the previous $\MHT>200\GeV$,
in order to reserve the QCD-dominated $250<\MHT<300\GeV$ region
for a QCD background control sample in data.
A final principal change is that finer segmentation
than in Ref.~\cite{Khachatryan:2016kdk}
is used to define exclusive intervals in \HT and \MHT,
to profit from the increased sensitivity afforded by the
larger data sample.

\begin{figure*}[tb]
\centering
\includegraphics[width=0.45\linewidth]{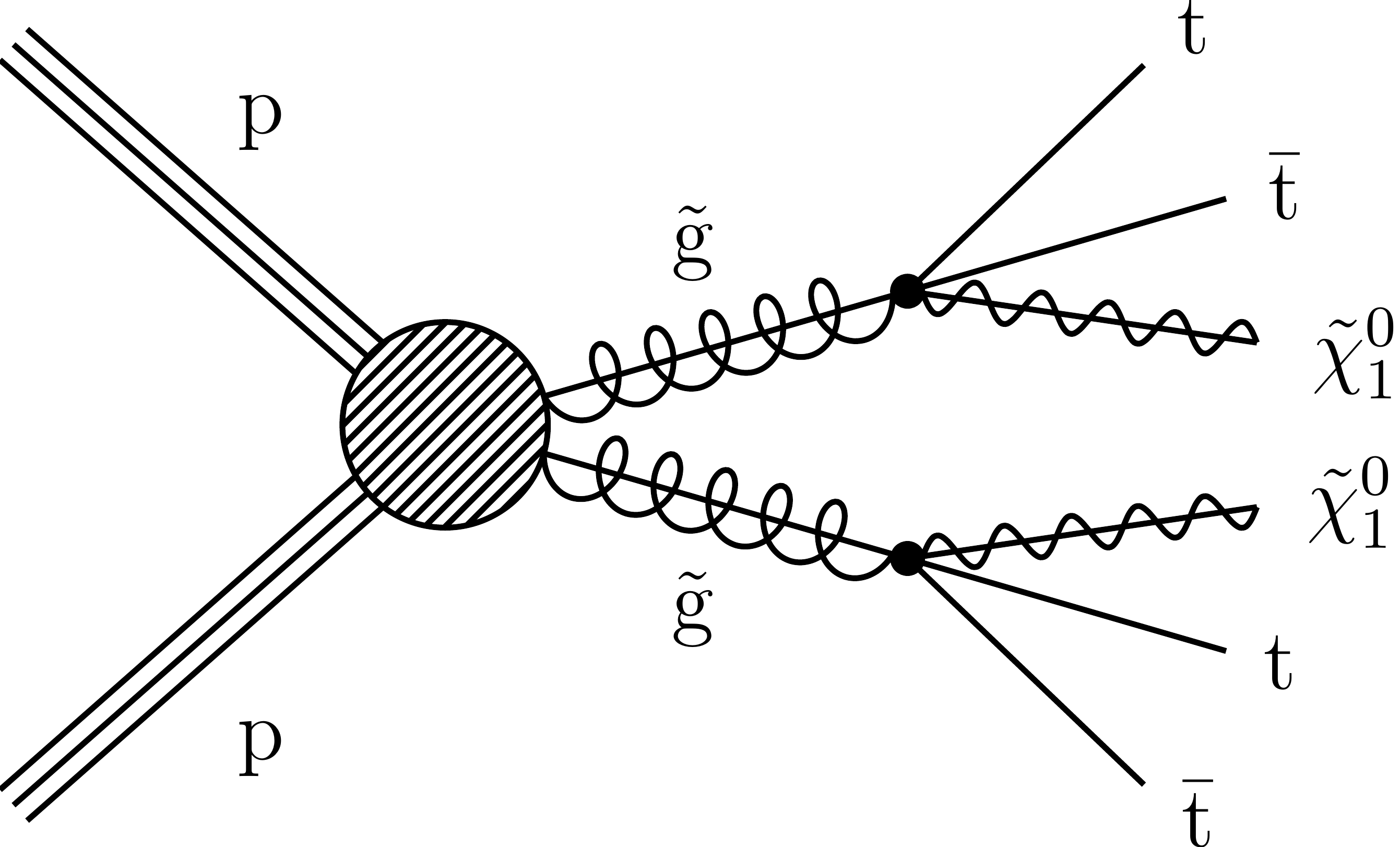}
\includegraphics[width=0.45\linewidth]{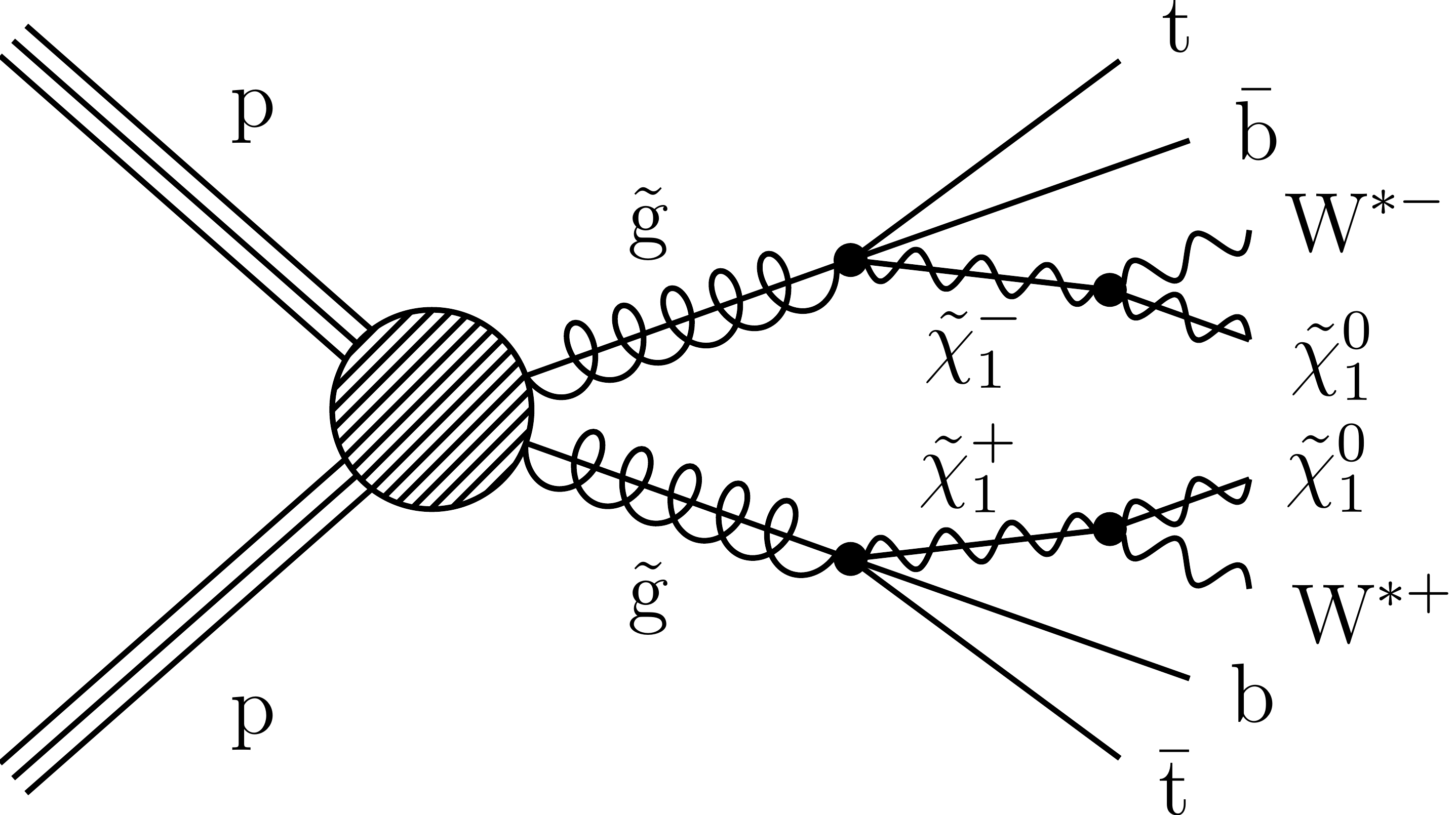}
\includegraphics[width=0.45\linewidth]{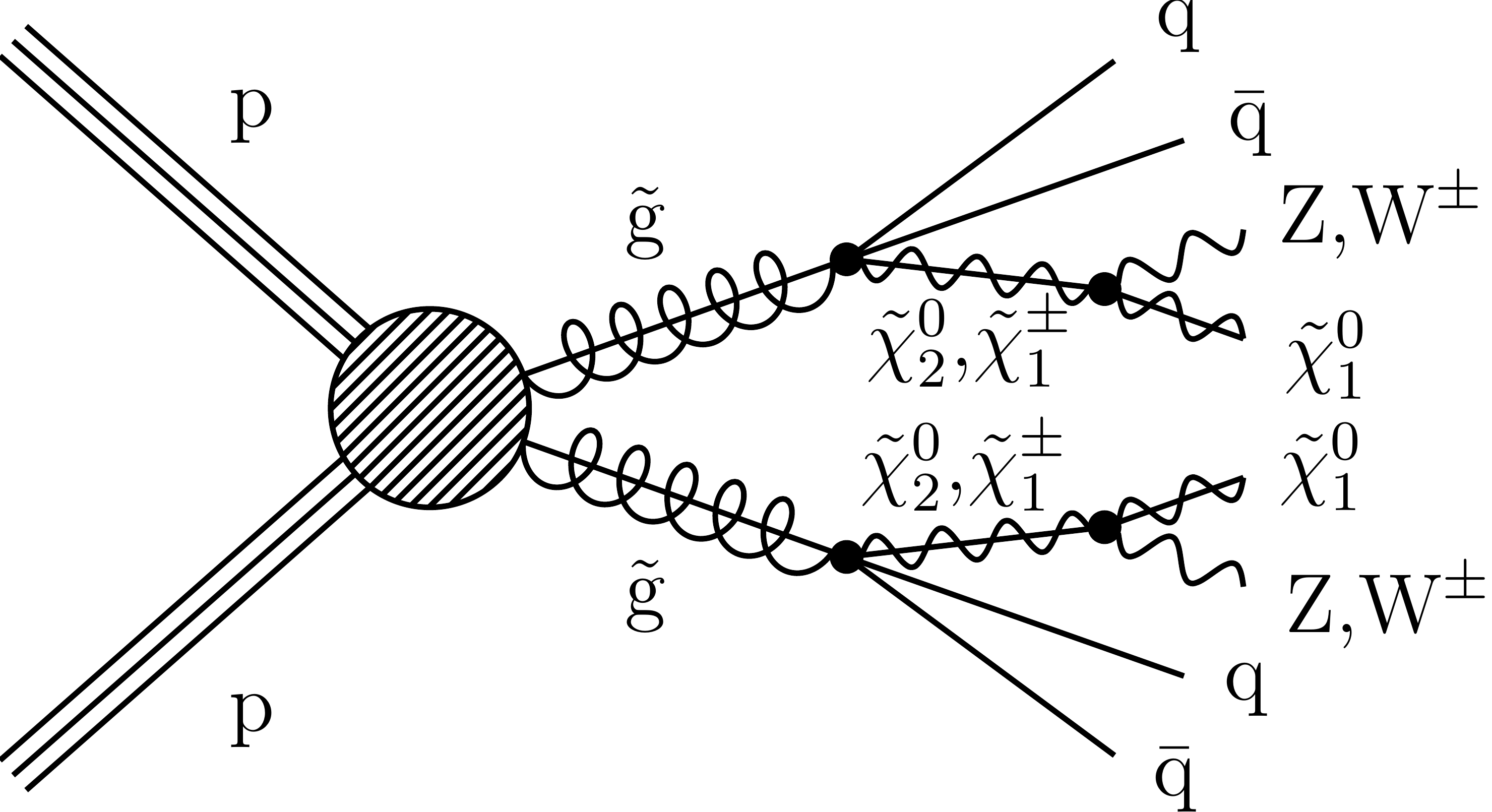}
\includegraphics[width=0.45\linewidth]{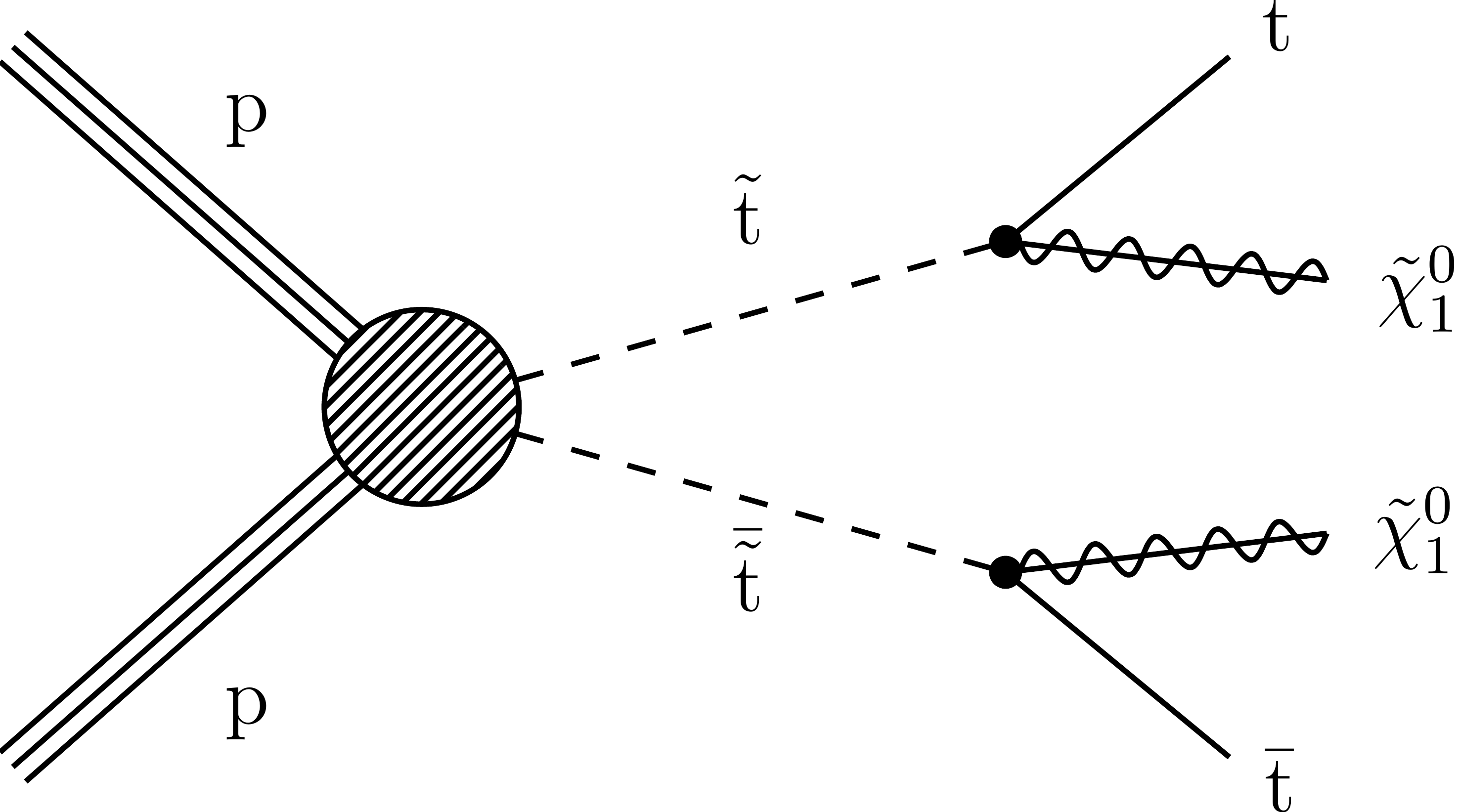}
\caption{
Example Feynman diagrams for the simplified model signal scenarios
considered in this study:
the (upper left) T1tttt,
(upper right) T1tbtb,
(lower left) T5qqqqVV,
and (lower right) T2tt
scenarios.
In the T5qqqqVV model,
the flavors of the quark {\cPq} and antiquark {\cPaq}
differ from each other if the gluino {\PSg} decays as
$\PSg\to\cPq\cPaq\PSGcpm_1$,
where $\PSGcpm_1$ is the lightest chargino.
}
\label{fig:event-diagrams}
\end{figure*}

Gluino and squark pair production are
studied in the context of simplified
models~\cite{bib-sms-1,bib-sms-2,bib-sms-3,bib-sms-4}.
For all models considered,
the lightest neutralino {\PSGczDo} is the LSP.
For gluino pair production,
the T1tttt, T1bbbb, T1qqqq, T1tbtb,
and T5qqqqVV~\cite{Chatrchyan:2013sza}
simplified models are considered,
defined as follows.
In the T1tttt scenario [Fig.~\ref{fig:event-diagrams} (upper left)],
each gluino decays to a top quark-antiquark (\ttbar) pair and the {\PSGczDo}.
The T1bbbb and T1qqqq scenarios are the same as the T1tttt scenario
except with the \ttbar pairs replaced by
bottom quark-antiquark (\bbbar) or
light-flavored (\cPqu, \cPqd, \cPqs, \cPqc) quark-antiquark (\qqbar) pairs,
respectively.
In the T1tbtb scenario
[Fig.~\ref{fig:event-diagrams} (upper right)],
each gluino decays either as $\PSg\to\PAQt\PQb\PSGcp_1$
or as its charge conjugate,
each with 50\% probability,
where $\PSGcpm_1$ denotes the lightest chargino.
The $\PSGcpm_1$ is assumed to be nearly degenerate in mass
with the \PSGczDo,
representing the expected situation should the
$\PSGcpm_1$ and $\PSGczDo$
appear within the same SU(2) multiplet~\cite{bib-sms-4}.
The chargino subsequently decays to the \PSGczDo
and to an off-shell {\PW} boson ($\PW^*$).
In the T5qqqqVV scenario [Fig.~\ref{fig:event-diagrams} (lower left)],
each gluino decays to a light-flavored \qqbar pair
and either to the next-to-lightest neutralino \PSGczDt
or to the $\PSGcpm_1$.
The probability for the decay to proceed via the \PSGczDt,
$\PSGcpDo$, or $\PSGc_1^-$ is 1/3 for each possibility.
The \PSGczDt ($\PSGcpm_1$) subsequently decays to the \PSGczDo
and to an on- or off-shell {\cPZ} ({$\PW^\pm$}) boson.

We also consider models in which more than one of the decays
$\PSg\to\ttbar\PSGczDo$,
$\PSg\to\bbbar\PSGczDo$,
and $\PSg\to\PAQt\PQb\PSGcp_1$
(or its charge conjugate)
can occur~\cite{bib-sms-4}.
Taken together,
these scenarios reduce the model dependence of the assumptions for
gluino decay to third-generation particles.
Specifically, we consider the following three mixed scenarios,
with the respective branching fractions in parentheses:
\begin{itemize}
\item
  $\PSg\to\cPaqt\cPqb\PSGcp_1$ (25\%), $\PSg\to\cPqt\cPaqb\PSGcm_1$ (25\%),
   $\PSg\to\ttbar\PSGczDo$ (50\%);
\item
  $\PSg\to\cPaqt\cPqb\PSGcp_1$ (25\%), $\PSg\to\cPqt\cPaqb\PSGcm_1$ (25\%),
   $\PSg \to \bbbar \PSGczDo$ (50\%);
\item
  $\PSg\to\cPaqt\cPqb\PSGcp_1$ (25\%), $\PSg\to\cPqt\cPaqb\PSGcm_1$ (25\%),
   $\PSg\to\ttbar\PSGczDo$ (25\%), $\PSg\to\bbbar\PSGczDo$ (25\%).
\end{itemize}

For squark-antisquark production,
three simplified models are considered,
denoted T2tt, T2bb, and T2qq.
In the T2tt scenario [Fig.~\ref{fig:event-diagrams} (lower right)],
top squark-antisquark production is followed by the decay of
each squark to a top quark and the \PSGczDo.
The T2bb and T2qq scenarios are the same as the T2tt scenario
except with bottom squarks and quarks,
or light-flavored squarks and quarks,
respectively,
in place of the top squarks and quarks.

Supersymmetric particles not participating in the
respective reaction are assumed to have infinite mass.
All considered SUSY particles are taken to decay promptly.

Background from SM processes
arises from events with a top quark
(either \ttbar events or events with a single top quark),
events with jets and an on- or off-shell {\PW} or {\cPZ} boson
(\wjets and \zjets events, respectively),
and QCD events.
Top quark and {\PW}+jets events can exhibit
significant \MHT and thus contribute to the background
if a {\PW} boson decays to a neutrino and an
undetected or out-of-acceptance charged lepton.
Similarly,
\zjets events can exhibit significant \MHT if
the {\cPZ} boson decays to two neutrinos.
Significant \MHT in QCD events is mostly the consequence
of mismeasured jet \pt,
but it can also arise if an event contains a charm or bottom quark
that decays semileptonically.
Note that \ttbar events in which both top quarks decay hadronically
are indistinguishable in our analysis from QCD events
and are accounted for in the evaluation of the QCD background.
Because the cross section is small compared to that for QCD events,
all-hadronic \ttbar events
comprise only a small (sub-percent level) component
of the evaluated QCD background.

\section{Detector and trigger}
\label{sec:detector}

A detailed description of the CMS detector,
along with a definition of the coordinate system and
pertinent kinematic variables,
is given in Ref.~\cite{Chatrchyan:2008aa}.
Briefly,
a cylindrical superconducting solenoid with an inner diameter of 6\unit{m}
provides a 3.8\unit{T} axial magnetic field.
Within the cylindrical volume
are a silicon pixel and strip tracker,
a lead tungstate crystal electromagnetic calorimeter (ECAL),
and a brass and scintillator hadron calorimeter (HCAL).
The tracking detectors cover the pseudorapidity range $\abs{\eta}<2.5$.
The ECAL and HCAL,
each composed of a barrel and two endcap sections,
cover $\abs{\eta}<3.0$.
Forward calorimeters extend the coverage to $3.0<\abs{\eta}<5.0$.
Muons are measured within $\abs{\eta}<2.4$ by gas-ionization detectors
embedded in the steel flux-return yoke outside the solenoid.
The detector is nearly hermetic,
permitting accurate measurements of~\MHT.

The CMS trigger is described in Ref.~\cite{Khachatryan:2016bia}.
For this analysis,
signal event candidates were recorded
by requiring \MHT at the trigger level
to exceed a threshold that varied between
100 and 120\GeV depending on the LHC instantaneous luminosity.
The efficiency of this trigger,
which exceeds 98\% following application of the event selection
criteria described below,
is measured in data and is taken into account in the analysis.
Additional triggers,
requiring the presence of charged leptons,
photons, or minimum values of \HT,
are used to select samples employed in the evaluation of backgrounds,
as described below.

\section{Event reconstruction}
\label{sec:reconstruction}

Individual particles are reconstructed
with the CMS particle-flow (PF)
algorithm~\cite{CMS-PRF-14-001},
which identifies them as photons,
charged hadrons, neutral hadrons, electrons, or muons.
To improve the quality of electron candidates~\cite{Khachatryan:2015hwa},
additional criteria are imposed on the ECAL shower
shape and on the ratio of associated energies in the HCAL and ECAL.
Analogously, for muon candidates~\cite{Chatrchyan:2013sba},
more stringent requirements are imposed on the matching between
silicon-tracker and muon-detector track segments.
Electron and muon candidates are restricted to $\abs{\eta}<2.5$ and $<2.4$,
respectively.

The reconstructed vertex with the largest value of
summed physics-object $\pt^2$ is taken to be the primary
$\Pp\Pp$ interaction vertex.
The physics objects are the objects returned by a jet finding
algorithm~\cite{Cacciari:2008gp,Cacciari:2011ma} applied to
all charged tracks associated with the vertex,
plus the corresponding associated missing transverse momentum.
The primary vertex is required to lie within 24\unit{cm} of the
center of the detector in the direction along the beam axis
and within 2\unit{cm} in the plane transverse to that axis.
Charged-particle tracks associated with vertices other than
the primary vertex are removed.

To suppress jets erroneously identified
as leptons and genuine leptons from hadron decays,
electron and muon candidates are subjected to an
isolation requirement.
The isolation criterion is based on the variable~$\imini$,
which is the scalar \pt sum of charged hadron,
neutral hadron, and photon PF candidates within a cone
of radius $\sqrt{\smash[b]{(\Delta\phi)^2+(\Delta\eta)^2}}$
around the lepton direction,
divided by the lepton~\pt,
where $\phi$ is the azimuthal angle.
The expected contributions of neutral particles from
extraneous $\Pp\Pp$ interactions (pileup)
are subtracted~\cite{Cacciari:2007fd}.
The radius of the cone is 0.2 for lepton $\pt<50\GeV$,
$10\GeV/\pt$ for $50\leq\pt\leq 200\GeV$,
and 0.05 for $\pt>200\GeV$.
The decrease in cone size with increasing lepton \pt
accounts for the increased collimation of the
decay products from the lepton's parent particle
as the Lorentz boost of the parent particle
increases~\cite{Rehermann:2010vq}.
The isolation requirement is $\imini<0.1$ (0.2) for electrons (muons).

Charged-particle tracks not identified as an isolated electron or muon,
including PF electrons and muons not so identified,
are subjected to a track isolation requirement.
To be identified as an isolated track,
the scalar \pt sum of all other charged-particle tracks
within a cone of radius 0.3 around the track direction,
divided by the track~\pt,
must be less than 0.2 if the track is identified
as a PF electron or muon
and less than 0.1 otherwise.
Isolated tracks are required to satisfy $\abs{\eta}<2.4$.

Jets are defined by clustering PF candidates
using the anti-\kt jet algorithm~\cite{Cacciari:2008gp,Cacciari:2011ma}
with a distance parameter of~0.4.
Jet quality criteria~\cite{cms-pas-jme-10-003}
are imposed to eliminate jets from
spurious sources such as electronics noise.
The jet energies are corrected for the nonlinear response of the
detector~\cite{Khachatryan:2016kdb}
and to account for the expected contributions of neutral
particles from pileup~\cite{Cacciari:2007fd}.
Jets are required to have $\pt>30\GeV$.

The identification of bottom quark jets ({\cPqb} jets)
is performed by applying the combined secondary vertex
algorithm (CSVv2) at the medium working point~\cite{CMS-PAS-BTV-15-001}
to the selected jet sample.
The signal efficiency for {\cPqb} jets with $\pt\approx30\GeV$ is 55\%.
The corresponding misidentification probability for gluon and
light-flavored (charm) quark jets is 1.6 (12)\%.

\section{Event selection and search regions}
\label{sec:event-selection}

Events considered as signal candidates are required to satisfy:
\begin{itemize}
\item $\njets\geq 2$, where jets must appear within $\abs{\eta}<2.4$;
\item $\HT>300\GeV$, with \HT the scalar \pt sum of jets with $\abs{\eta}<2.4$;
\item $\MHT>300\GeV$, where \MHT is the magnitude of \htvecmiss,
  the negative of the vector \pt sum of jets with $\abs{\eta}<5$;
  an extended $\eta$ range is used to calculate \MHT
  so that it better represents
  the total missing transverse momentum in an event;
\item no identified, isolated electron or muon candidate with $\pt>10\GeV$;
\item no isolated track with $\mt<100\GeV$
  and $\pt>10\GeV$
  ($\pt>5\GeV$ if the track is identified as a PF electron or muon),
  where \mt is the transverse mass~\cite{Arnison:1983rp}
  formed from the \ptvecmiss and isolated-track \pt vector,
  with \ptvecmiss the
  negative of the vector \pt sum of all PF objects;
\item
  $\dphimht>0.5$ for the two highest \pt jets j$_1$ and j$_2$,
  with \dphimht the azimuthal angle between \htvecmiss
  and the \pt vector of jet j$_{i}$;
  if $\njets\geq3$, then, in addition, $\dpmht3>0.3$
  for the third highest \pt jet j$_3$;
  if $\njets\geq4$, then, yet in addition, $\dpmht4>0.3$
  for the fourth highest \pt jet j$_4$;
  all considered jets must have $\abs{\eta}<2.4$.
\end{itemize}
In addition,
anomalous events with reconstruction failures or
that arise from noise or beam halo interactions
are removed~\cite{CMS-PAS-JME-16-004}.
A breakdown of the efficiency at different stages of the selection
process for representative signal models is given in
Tables~\ref{tab:sel-eff-gg} and~\ref{tab:sel-eff-qq}
of \cmsAppendix\ref{sec:cutflow}.

The isolated-track veto requirement suppresses events with
a hadronically decaying $\tau$ lepton,
or with an isolated electron or muon not identified as such;
the \mt requirement restricts the isolated-track veto to
situations consistent with {\PW} boson decay.
The selection criteria on \dphimht suppress background
from QCD events,
for which \htvecmiss is usually aligned along a jet direction.

\begin{figure*}[tbp]
\centering
    \includegraphics[width=0.65\textwidth]{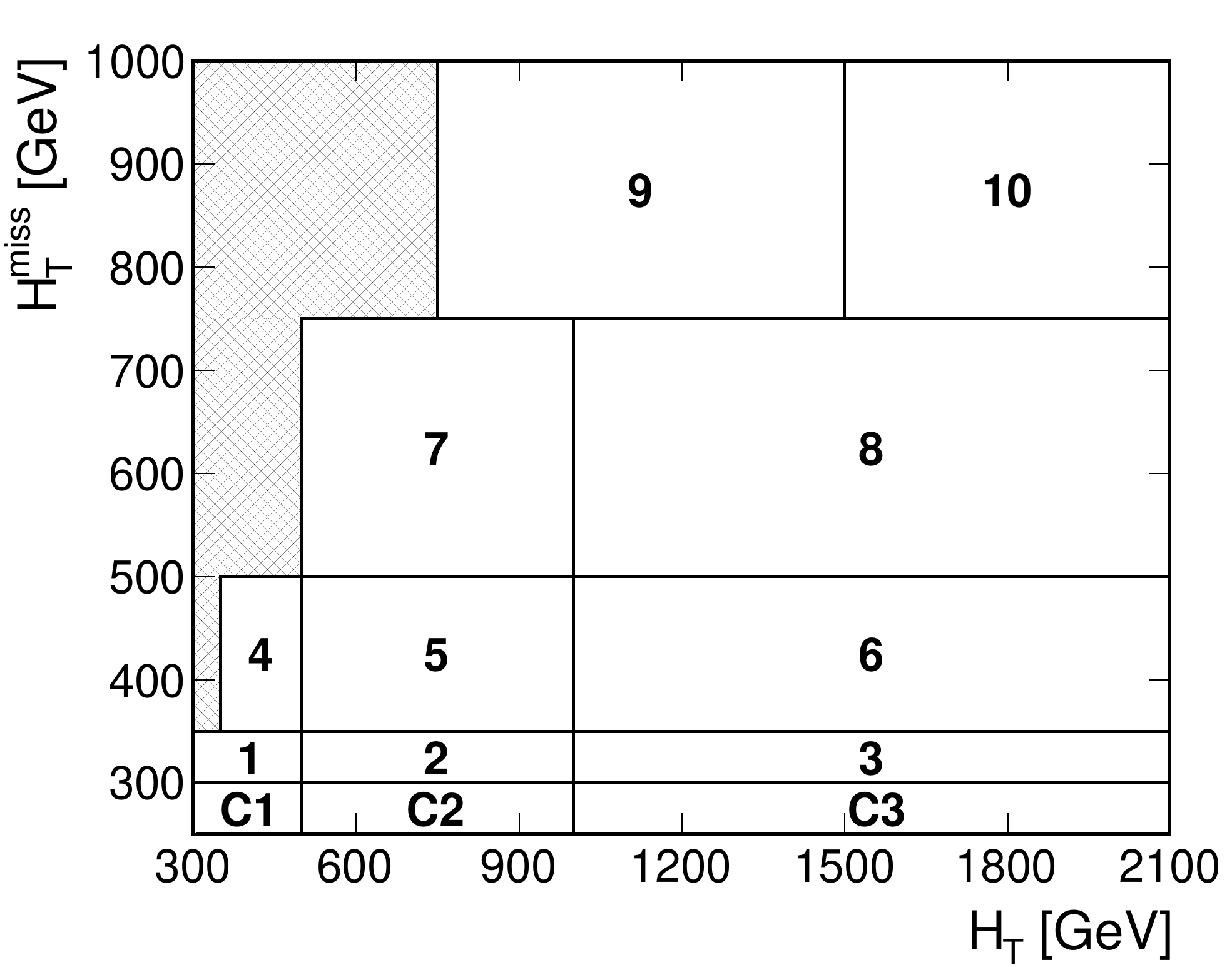}
    \caption{
      Schematic illustration of the 10 kinematic search intervals
      in the \MHT versus \HT plane.
      Intervals 1 and 4 are discarded for $\njets\geq7$.
      The intervals labeled C1, C2, and C3 are control regions
      used to evaluate the QCD background.
      The rightmost and topmost bins are unbounded,
      extending to
      $\HT=\infty$ and $\MHT=\infty$, respectively.
    }
    \label{fig:HT-MHT}
\end{figure*}

The search is performed in four-dimensional exclusive regions of
\njets, \nbjets, \HT, and \MHT.
The search intervals in \njets and \nbjets are:
\begin{itemize}
\item \njets: 2, 3--4, 5--6, 7--8, $\geq$9;
\item \nbjets: 0, 1, 2, $\geq$3.
\end{itemize}
Intervals with $\nbjets\geq3$ and $\njets=2$
are discarded since there are no entries.
For \HT and \MHT,
10 kinematic intervals are defined,
as specified in Table~\ref{tab:kine-bins}
and illustrated in Fig.~\ref{fig:HT-MHT}.
Events with both small \HT and large \MHT are not considered
(see the hatched area in Fig.~\ref{fig:HT-MHT})
because such events are likely to arise from mismeasurement.
For $\njets\geq7$,
the kinematic intervals labeled 1 and 4 are discarded
because of the small number of events.
The total number of search regions is~174.

\begin{table}[tbp]
\centering
\topcaption{
Definition of the search intervals in
the \MHT and~\HT variables.
Intervals 1 and 4 are discarded for $\njets\geq7$.
}
\label{tab:kine-bins}
\begin{scotch}{ccc}
Interval & \MHT [\GeVns{}] & \HT [\GeVns{}]  \\ \hline
1 & 300--350 & 300--500 \\
2 & 300--350 & 500--1000 \\
3 & 300--350 & $>$1000 \\
4 & 350--500 & 350--500  \\
5 & 350--500 & 500--1000  \\
6 & 350--500 & $>$1000 \\
7 & 500--750 & 500--1000 \\
8 & 500--750 & $>$1000 \\
9 & $>$750 & 750--1500 \\
10 & $>$750 & $>$1500 \\
\end{scotch}
\end{table}

The intervals labeled C1, C2, and C3 in Fig.~\ref{fig:HT-MHT}
are control regions defined by $250<\MHT<300\GeV$,
with the same boundaries in \HT as kinematic intervals 1, 2, and 3,
respectively.
These regions are used in the method
to estimate the QCD background
described in Section~\ref{sec:qcd-dphi}.

\section{Simulated event samples}
\label{sec:mc}

To evaluate the background,
we mostly rely on data control regions,
as discussed in Section~\ref{sec:background}.
Samples of simulated SM events are used to validate
the analysis procedures and
for some secondary aspects of the background estimation.
The SM production of \ttbar, \wjets, \zjets, \gjets,
and QCD events is simulated using the
{\MADGRAPH}5{\textunderscore}a{\MCATNLO} 2.2.2~\cite{Alwall:2014hca,Alwall:2007fs}
event generator at leading order (LO).
The \ttbar events are generated with
up to three additional partons in the matrix element calculations,
while up to four additional partons can be present for
\wjets, \zjets, and \gjets events.
Single top quark events produced through the $s$ channel,
diboson events such as
$\PW\PW$, $\cPZ\cPZ$, and $\cPZ\PH$ production,
where $\PH$ is a Higgs boson,
and rare events
such as $\ttbar\PW$, $\ttbar\cPZ$, and $\PW\PW\cPZ$ production,
are generated with this same program~\cite{Alwall:2014hca,Frederix:2012ps}
at next-to-leading (NLO) order,
except that $\PW\PW$ events in which both {\PW} bosons decay leptonically
are generated using the \POWHEG
v2.0~\cite{Nason:2004rx,Frixione:2007vw,Alioli:2010xd,Alioli:2009je,Re:2010bp}
program at NLO.
The same \POWHEG generator is used to describe
single top quark events produced through the $t$ and $\cPqt\PW$ channels.
The detector response is modeled with
the \GEANTfour~\cite{Agostinelli:2002hh} suite of programs.
Normalization of the simulated background samples is performed using
the most accurate cross section calculations
available~\cite{Alioli:2009je,Re:2010bp,Alwall:2014hca,Melia:2011tj,Beneke:2011mq,
Cacciari:2011hy,Baernreuther:2012ws,Czakon:2012zr,Czakon:2012pz,Czakon:2013goa,
Gavin:2012sy,Gavin:2010az},
which generally correspond to NLO or next-to-NLO precision.

Samples of simulated signal events are generated at LO using the
{\MADGRAPH}5{\textunderscore}a{\MCATNLO} program.
Up to two additional partons are included in the
matrix element calculation.
The production cross sections are
determined with NLO plus next-to-leading logarithmic (NLL)
accuracy~\cite{bib-nlo-nll-01,bib-nlo-nll-02,bib-nlo-nll-03,bib-nlo-nll-04,bib-nlo-nll-05}.
Events with gluino (squark) pair production are generated
for a range of gluino \mgluino (squark \msquark) and LSP \mlsp mass values,
with $\mlsp<\mgluino$ ($\mlsp<\msquark$).
The ranges of mass considered vary according to the model but
are generally from around 600 to 2200\GeV for \mgluino,
200 to 1700\GeV for \msquark,
and 0 to 1200\GeV for~\mlsp
(see the results shown in Section~\ref{sec:results} for more detail).
For the T1tbtb model,
the mass of the intermediate $\PSGcpm_1$ state
is taken to be $\mlsp+5\GeV$,
while for the T5qqqqVV model,
the masses of the intermediate \PSGczDt and $\PSGcpm_1$
are given by the mean of \mlsp and $\mgluino$.
The gluinos and squarks decay according to
phase space~\cite{Sjostrand:2014zea}.
To render the computational requirements manageable,
the detector response is described using the CMS fast
simulation~\cite{Orbaker:2010zz,bib-cms-fastsim-02},
which yields consistent results with the
{\GEANTfour}-based simulation,
except that we apply a correction of 1\% to account for
differences in the efficiency of the
jet quality requirements~\cite{cms-pas-jme-10-003},
corrections of 5--12\% to account for differences
in the {\cPqb} jet tagging efficiency,
and corrections of 0--14\% to account for differences
in the modeling of \HT and \MHT.

For simulated samples generated at LO (NLO),
the NNPDF3.0LO~\cite{Ball:2014uwa} (NNPDF3.0NLO~\cite{Ball:2014uwa})
parton distribution functions (PDFs) are used.
Parton showering and hadronization are described by the
\PYTHIA 8.205~\cite{Sjostrand:2014zea} program for all samples.

To improve the description of initial-state radiation (ISR),
we compare the {\MADGRAPH} prediction to data in a control
region enriched in \ttbar events:
two leptons ($\Pe\Pe$, $\Pgm\Pgm$, or $\Pe\Pgm$)
and two tagged {\cPqb} jets are required.
The number of all other jets in the event is denoted \njetsisr.
The correction factor is derived as a function of \njetsisr,
with a central value ranging from 0.92 for
$\njetsisr = 1$ to 0.51 for $\njetsisr \geq 6$.
These corrections are applied to simulated \ttbar and signal events.
From studies with a single-lepton data control sample,
dominated by \ttbar events,
the associated systematic uncertainty is taken to be 20\%
of the correction for \ttbar events
and 50\% of the correction for signal events,
where the larger uncertainty in the latter case accounts
for possible differences between \ttbar and signal event production.

\section{Signal systematic uncertainties}
\label{sec:systematic}

\begin{table*}[tb]
\topcaption{
Systematic uncertainties in the yield of signal events,
averaged over all search regions.
The variations correspond to different signal
models and choices for the SUSY particle masses.
Results reported as 0.0 correspond to values less than~0.05\%.
``Mixed T1'' refers to the mixed models of gluino
decays to heavy squarks described in the introduction.
}
\centering
\begin{scotch}{lc}
Item & Relative uncertainty (\%) \\
\hline
Trigger efficiency                              & $0.2-2.8$ \\
Jet quality requirements                        & 1.0 \\
Initial-state radiation                         & $0.0-14$ \\
Renormalization and factorization scales        & $0.0-6.2$ \\
Jet energy scale                                & $0.0-7.7$ \\
Jet energy resolution                           & $0.0-4.2$ \\
Statistical uncertainty of MC samples           & $1.5-30$ \\
\HT and \MHT modeling                           & $0.0-13$ \\
Pileup                                          & $0.2-5.5$ \\
Isolated-lepton \& isolated-track vetoes        & 2.0 \\
\hspace*{5mm} (T1tttt, T1tbtb, mixed T1, T5qqqqVV, and T2tt models) & \\
Integrated luminosity                           & 2.5 \\
\hline
Total                                           & $3.9-34$ \\
\end{scotch}
\label{tab:sig-syst}
\end{table*}

Systematic uncertainties in the
signal event yield are listed in Table~\ref{tab:sig-syst}.
To evaluate the uncertainty associated with the
renormalization (\rscale) and factorization (\fscale) scales,
each scale is varied independently
by a factor of 2.0 and 0.5~\cite{Catani:2003zt,Cacciari:2003fi}.
The uncertainties associated with \rscale, \fscale, and ISR,
integrated over all search regions, typically lie below 0.1\%
but can be as large as the maximum values noted in Table~\ref{tab:sig-syst}
for $\dmass\approx 0$,
where \dmass is the difference between the gluino or squark mass
and the sum of the masses of the particles into which it decays.
For example,
for the T1tttt model,
\dmass is given by $\dmass=\mgluino - (\mlsp+2\mtop)$,
with \mtop the top quark mass.
The uncertainties associated with the jet energy scale and jet energy resolution
are evaluated as a function of jet \pt and~$\eta$.
An uncertainty in the event yield
associated with pileup is evaluated
based on the observed distribution of the number \nvtx of reconstructed vertices,
and on the selection efficiency and its uncertainty
determined from simulation as a function of \nvtx.
The isolated-lepton and isolated-track vetoes have a
minimal impact on the T1bbbb, T1qqqq, T2bb, and T2qq models because events in
these models rarely contain an isolated lepton.
Thus, the associated uncertainty is negligible ($\lesssim 0.1\%$).
The systematic uncertainty in the determination of the
integrated luminosity is~2.5\%~\cite{CMS-PAS-LUM-17-001}.

Systematic uncertainties in the signal predictions
associated with the {\cPqb} jet tagging and misidentification efficiencies
are also evaluated.
These uncertainties do not affect the signal yield
but can potentially alter the shape of signal distributions.
The systematic uncertainties associated with the trigger,
\rscale, \fscale, ISR,
jet energy scale, jet energy resolution,
statistical precision in the event samples, and \MHT modeling
can also affect the shapes of the signal distributions.
We account for these potential changes in shape,
i.e., migration of events between search regions,
in the limit-setting procedure
described in Section~\ref{sec:results}.

\section{Background evaluation}
\label{sec:background}

The evaluation of background
is primarily based on data control regions (CRs).
Signal events, if present, could populate the CRs,
an effect known as signal contamination.
The impact of signal contamination is evaluated as
described in Section~\ref{sec:results}.
Signal contamination
is negligible for all CRs except those used to evaluate the
top quark and \wjets background (Section~\ref{sec:ttbar}).
It is nonnegligible only for the models that can produce an
isolated track or lepton,
viz., the T1tttt, T1tbtb, T5qqqqVV, and T2tt models,
and the mixed models of gluino decays to heavy squarks described in the introduction.

\subsection{Background from top quark and \texorpdfstring{\wjets}{W+jets} events}
\label{sec:ttbar}

The background from the SM production of
\ttbar, single top quark, and \wjets events
originates from {\PW} bosons that decay leptonically
to yield a neutrino and a charged lepton.
If the charged lepton is an electron or muon,
including those from $\tau$ lepton decay,
it is called a ``lost'' lepton.
A lost lepton arises if an electron or muon lies outside
the analysis acceptance,
is not reconstructed,
or is not isolated,
and thus is not vetoed by the requirements of
Section~\ref{sec:event-selection}.
The other possibility is that the charged lepton is
a hadronically decaying $\tau$ lepton,
denoted ``{\tauh}.''

\subsubsection{Lost-lepton background}
\label{sec:ttbar-ll}

The procedure used to evaluate the lost-lepton background
is described in Ref.~\cite{Khachatryan:2016kdk}
(see also Refs.~\cite{Chatrchyan:2014lfa,Collaboration:2011ida,Chatrchyan:2012lia}).
Briefly,
single-lepton CRs are selected using the standard trigger
and selection criteria,
except with the electron and muon vetoes inverted
and the isolated-track veto not applied.
Exactly one isolated electron or muon must be present.
In addition,
the transverse mass \mt formed from the \ptvecmiss and
lepton $\vec{p}_{\text{T}}$ is required to satisfy $\mt<100\GeV$:
this requirement is effective at identifying SM events,
while reducing potential signal contamination.
The T1tttt (T1tbtb, T5qqqqVV, T2tt)
signal contamination in the resulting CRs
is generally negligible ($\lesssim 0.1\%$),
but it can be as large as 30--50\% (25--60\%, 2--15\%, 5--50\%)
for large values of \njets, \nbjets, \HT, and/or \MHT,
depending on \mgluino or \msquark and~\mlsp.
Similar results to the T1tbtb model are obtained
for the mixed models of gluino decay to heavy squarks.

Each CR event is entered into one of the 174 search regions
with a weight that represents
the probability for a lost-lepton event to appear with
the corresponding values of \HT, \MHT, \njets, and \nbjets.
The weights are determined from the \ttbar, \wjets, single top quark,
and rare process simulations through evaluation of the efficiency
of the lepton acceptance, lepton reconstruction, lepton isolation,
isolated-track, and \mt requirements.
Corrections are applied to account for the purity of the CR,
the contributions of dilepton events to the signal regions and CR,
and efficiency differences with respect to data.
More details are provided in Ref.~\cite{Khachatryan:2016kdk}.
The efficiencies are determined as a function of \HT, \MHT, \njets, \nbjets,
lepton \pt and $\eta$, and other kinematic variables.
Improvements relative to Ref.~\cite{Khachatryan:2016kdk}
are that we now use $\nbjets$ and lepton $\eta$ to help characterize
the efficiencies,
and the efficiency of the isolated-track veto is now determined
separately for lost-lepton events that fail the acceptance,
reconstruction, or isolation requirements.
Previously, only a single overall isolated-track veto
efficiency was evaluated (as a function of search region)
when constructing the weights.

\begin{figure}[htp]
  \centering
  \includegraphics[width=\cmsFigWidth]{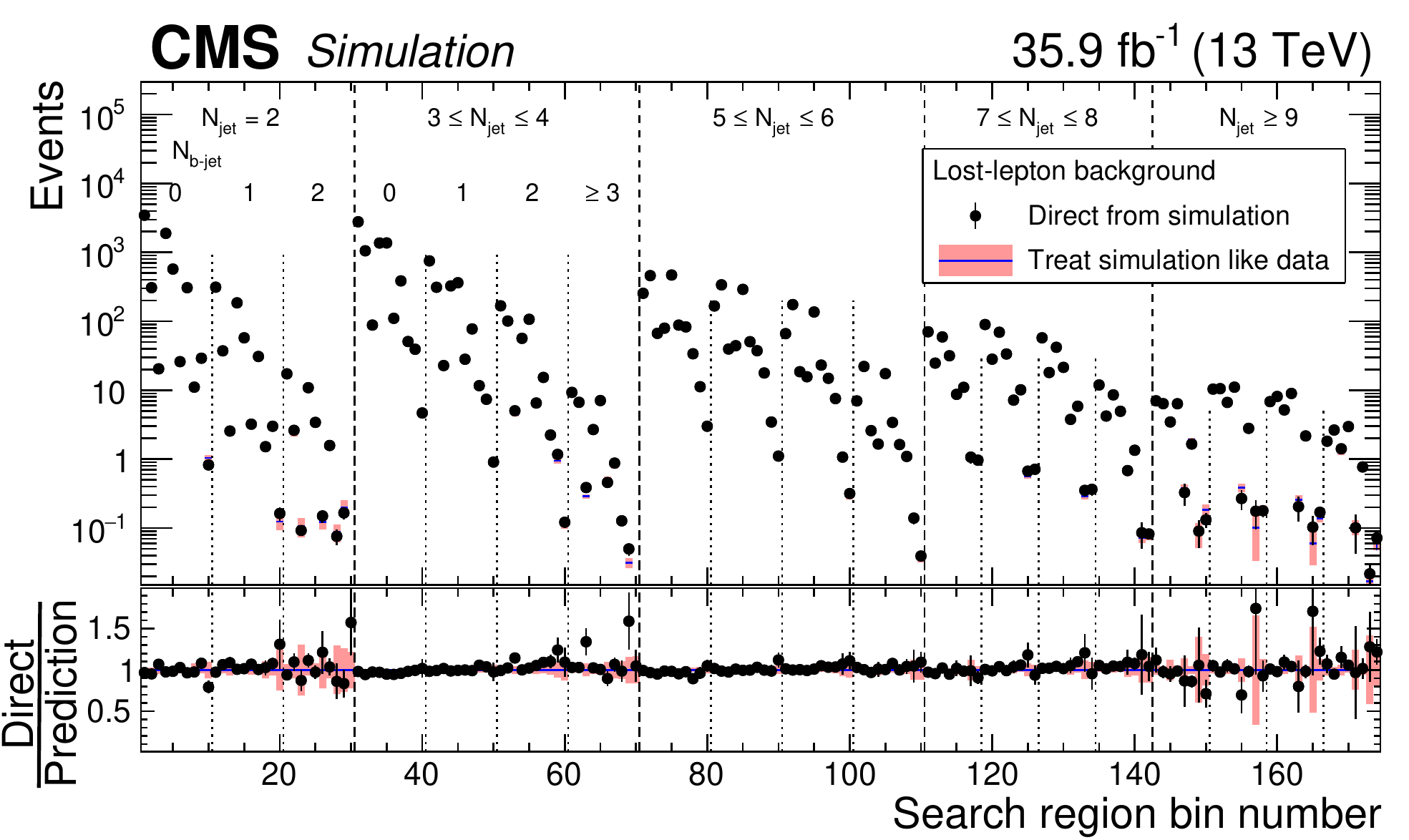}
  \caption{The lost-lepton background in the 174 search regions
    of the analysis as determined directly from \ttbar,
    single top quark, \wjets, diboson, and rare-event simulation
    (points, with statistical uncertainties) and as predicted by applying the
    lost-lepton background determination procedure to simulated
    electron and muon control samples
    (histograms, with statistical uncertainties).
    The results in the lower panel are obtained through bin-by-bin
    division of the results
    in the upper panel, including the uncertainties,
    by the central values of the ``predicted'' results.
    The 10 results (8 results for $\njets\geq7$)
    within each region delineated by vertical dashed lines
    correspond sequentially to the 10
    (8) kinematic intervals of \HT and \MHT
    indicated in Table~\ref{tab:kine-bins} and Fig.~\ref{fig:HT-MHT}.
  }
  \label{fig:lost-lepton-closure}
\end{figure}

The weighted distributions of the search variables,
summed over the events in the CRs,
define the lost-lepton background prediction.
The procedure
is performed separately for the single-electron and single-muon CRs,
both of which are used to predict the total lost-lepton background,
\ie, the background due both to lost electrons and to lost muons.
The two predictions yield consistent results and are averaged,
with correlations in the uncertainties taken into account,
to obtain the final lost-lepton background estimate.
The method is checked with a closure test,
namely by determining the ability of the method,
applied to simulated event samples,
to predict correctly the true number of background events.
The results of this test are shown
in Fig.~\ref{fig:lost-lepton-closure}.

The dominant uncertainty in the lost-lepton background
prediction is statistical,
due to the limited number of CR events.
As a systematic uncertainty,
we take the larger of the observed nonclosure
and the statistical uncertainty in the nonclosure,
for each search region,
where ``nonclosure'' refers to the bin-by-bin difference between
the solid points and histogram in Fig.~\ref{fig:lost-lepton-closure}.
Additional systematic uncertainties are evaluated as
described in Ref.~\cite{Khachatryan:2016kdk}
and account for potential differences between the data and
simulation for the
lepton acceptance,
lepton reconstruction efficiency,
lepton isolation efficiency,
isolated-track efficiency,
\mt selection efficiency,
dilepton contributions,
and purity of the CRs.

\subsubsection{Hadronically decaying \texorpdfstring{$\tau$}{tau} lepton background}
\label{sec:ttbar-hadtau}

To evaluate the top quark and {\wjets} background due to \tauh events,
a CR event sample is selected
using a trigger that requires either
at least one isolated muon candidate with $\pt>24\GeV$,
or at least one isolated muon candidate with $\pt>15\GeV$
in conjunction with $\HT>500\GeV$.
The reason a special trigger is used,
and not the standard one,
is that the \tauh background determination method
requires there not be a selection
requirement on missing transverse momentum,
as is explained below.
The selected events are required to contain exactly
one identified muon with $\abs{\eta}<2.1$.
The \pt of the muon candidate must exceed 20\GeV,
or 25\GeV if $\HT<500\GeV$.
The fraction of T1tttt (T1tbtb, T5qqqqVV, T2tt) events in the CR due to
signal contamination is generally $\lesssim0.1\%$,
but can be as large as 5--22\% (1--20\%, 1--15\%, 1--40\%)
for large values of \njets, \nbjets, \HT, and/or \MHT,
depending on \mgluino or \msquark and~\mlsp,
with similar results to the T1tbtb model
for the mixed models of gluino decay to heavy squarks.

\begin{figure}[ht]
  \centering
  \includegraphics[width=\cmsFigWidth]{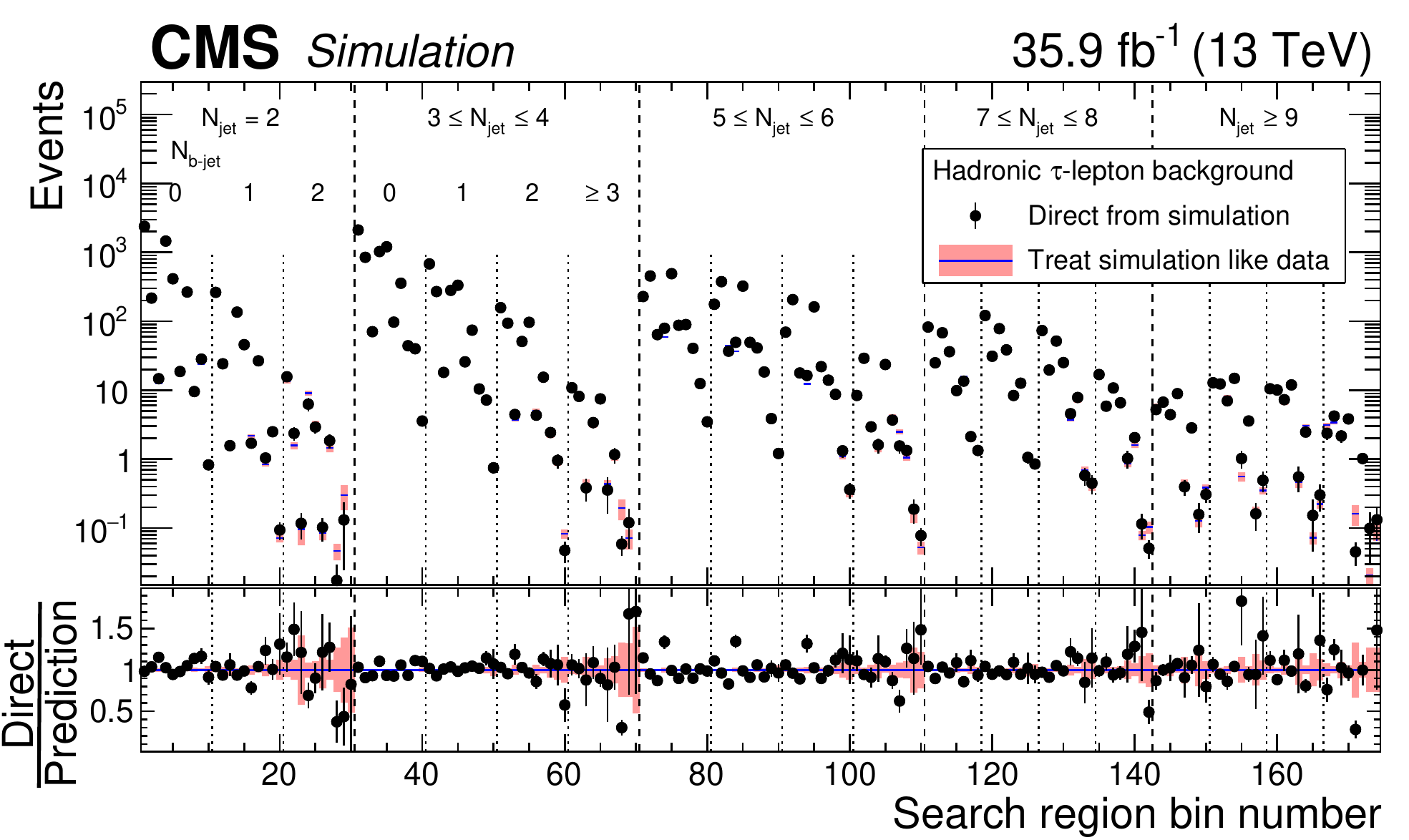}
  \caption{The background from hadronically decaying
    $\tau$ leptons in the 174 search regions
    of the analysis as determined directly from \ttbar,
    single top quark, and \wjets simulation
    (points, with statistical uncertainties) and as predicted by applying the
    hadronically decaying $\tau$ lepton background determination procedure to
    a simulated muon control sample
    (histograms, with statistical uncertainties).
    The results in the lower panel are obtained through bin-by-bin
    division of the results
    in the upper panel, including the uncertainties,
    by the central values of the ``predicted'' results.
    The labeling of the bin numbers is the
    same as in Fig.~\ref{fig:lost-lepton-closure}.
  }
  \label{fig:hadtau-closure}
\end{figure}

The \tauh background is determined
using the method described in Ref.~\cite{Khachatryan:2016kdk}
(see also Refs.~\cite{Chatrchyan:2014lfa,Collaboration:2011ida,Chatrchyan:2012lia}).
It makes use of the similarity between
$\mu$+jets and $\tauh$+jets events
aside from the detector response to the $\mu$ or $\tauh$.
In each CR event, the muon \pt is smeared through random sampling
of \tauh response functions derived from simulation of single
${\PW}\to\tauh\nu_{\tau}$ decay events.
This differs from Ref.~\cite{Khachatryan:2016kdk},
in which $\PW\to\tauh\nu_{\tau}$ decays in simulated
\ttbar and \wjets events were used to derive the response functions.
The change was made in order to reduce the risk of contamination
in the response functions from nearby non-{\tauh}-related particles;
note that the CR already includes the
effects from the underlying event and nearby jets.
The response functions express the expected
visible-\pt distribution of a \tauh candidate
as a function of the true $\tau$ lepton \pt,
taken to be the measured muon~\pt in the CR event.
Following the smearing,
the values of \HT, \MHT, \njets, and \nbjets are calculated for the CR event,
and the selection criteria of Section~\ref{sec:event-selection} are applied.
Note that CR events with relatively low values of \MHT
can be promoted,
after smearing,
to have \MHT values above the nominal threshold,
and thus appear in the \tauh background prediction.
It is for this reason that the CR is selected using a trigger
without a requirement on missing transverse momentum:
to avoid possible \MHT bias.
The probability for a \tauh jet to be erroneously
identified as a {\cPqb} jet is taken into account.
Corrections are applied to account for the trigger efficiency,
the acceptance and efficiency of the $\mu$ selection,
and the ratio of branching fractions
${\mathcal{B}}(\PW\to\tauh\nu)/{\mathcal{B}}(\PW\to\mu\nu) = 0.65$~\cite{PDG2016}.
The resulting event yield provides the \tauh background estimate.
The method is validated with a closure test,
whose results are shown in Fig.~\ref{fig:hadtau-closure}.

Systematic uncertainties are assigned based on the level of nonclosure,
as described for the lost-lepton background.
In addition,
systematic uncertainties
are evaluated for the muon reconstruction, isolation, and acceptance efficiencies,
for the response functions,
and for the misidentification rate of \tauh jets as {\cPqb}~jets.
The dominant source of uncertainty,
as for the lost-lepton background,
is from the limited statistical precision of the CR sample.

\subsection{Background from \texorpdfstring{\znn}{Z -> nu nu} events}
\label{sec:znn}

The evaluation of background from SM \zjets events with \znn is based on
CR samples of \gjets events,
and of \zjets events with \zll ($\ell=\Pe$, $\Pgm$).
The photon in the \gjets events
and the $\ell^+\ell^-$ pair in the \zll events are removed
from the event
in order to emulate missing transverse momentum.
The \gjets and \zll events are then subjected to the same selection
criteria as in the standard analysis,
with corrections applied to account for differences in
acceptance with respect to the \znnjets process.
The use of \gjets events exploits the similarity between {\cPZ} boson
and direct photon production in $\Pp\Pp$ collisions,
where ``direct'' refers to a photon produced through the Compton
scattering ($\cPq\cPg\to\cPq\gamma$)
or annihilation ($\cPq\cPaq\to\cPg\gamma$) process.

The method is an extension of that described in Ref.~\cite{Khachatryan:2016kdk}.
Briefly,
the relatively copious \gjets events are used to evaluate the background
in the 46 search regions with $\nbjets=0$.
We do not use \gjets events for the $\nbjets>0$ search regions
to avoid reliance on the theoretical modeling
of \gjets versus \zjets production with bottom quarks.
The less abundant \zll events are used to validate
and calibrate the $\nbjets=0$ results,
as described below,
and to extrapolate to the $\nbjets>0$ search regions.
For this extrapolation,
the \zll data are integrated over \HT and \MHT
because of the limited number of events.

The \zll CR sample is selected using a combination of
triggers that requires either
i)~at least one isolated electron or muon with $\pt>15\GeV$,
and either $\HT>350$ or 400\GeV depending on the
LHC instantaneous luminosity,
ii)~at least one electron with either $\pt>105$ or 115\GeV
depending on the instantaneous luminosity,
iii)~at least one muon with $\pt>50\GeV$,
or
iv)~at least one isolated electron (muon) with $\pt>27$ (24)\GeV.
The events are required to contain exactly one \EE or one \MM pair
with an invariant mass within 15\GeV of the nominal {\cPZ} boson mass,
with the constituents of the pair identified using the same
criteria for isolated electrons and muons
as in the standard analysis.
The \pt of the lepton pair must exceed 200\GeV.
To ensure that the \zll and \gjets CRs are independent,
a veto is applied to events containing an identified photon.

The \gjets CR sample is selected with a trigger
that requires a photon candidate with $\pt>175\GeV$.
Events are retained if they contain exactly one
well-identified isolated photon with $\pt>200\GeV$.
The photon isolation criteria require the pileup-corrected energy
within a cone of radius 0.3 around the photon direction,
excluding the energy carried by the photon candidate itself,
to satisfy upper bounds that depend on the \pt and $\eta$ of the photon,
and are determined separately for the contributions of
electromagnetic, charged hadronic, and neutral hadronic energy.
About 85\% of the events in the resulting sample are
estimated to contain a direct photon,
while the remaining events either contain a fragmentation photon,
\ie, emitted as initial- or final-state radiation
or during the hadronization process,
or a nonprompt photon,
i.e., from unstable hadron decay.
A fit to the photon isolation variable is performed
as a function of \MHT to determine
the photon purity $\beta_\gamma$,
defined as the fraction of events in the \gjets CR
with a direct or fragmentation photon
(these two types of photons are experimentally indistinguishable
and together are referred to as ``prompt'').

The estimated number \nznn of \znnjets background events
contributing to each $\nbjets=0$ search region is given by:
\begin{equation}
  \left.\nznn\right|_{\nbjets=0}
     = \rho {\rznnsim} \Fdir \beta_\gamma \ngdata \,/\, \SFdmcg,
\label{eq:gjet1}
\end{equation}
where \ngdata is the number of events in the corresponding
\njets, \HT, and \MHT bin of the \gjets CR,
$\beta_\gamma$ is the fraction that are prompt,
\Fdir is the fraction of prompt photons that are also direct
(evaluated from simulation),
and \rznnsim is the ratio from simulation
of the number of \znnjets events to the number of
direct-photon \gjets events,
with the direct photon term obtained from an LO
{\MADGRAPH}5{\textunderscore}a{\MCATNLO} calculation.
The \SFdmcg factors are corrections to the simulation that
account for efficiency differences in photon
reconstruction with respect to data.

The $\rho$ factor in Eq.~(\ref{eq:gjet1}) is determined from \zll data
and is used to account for potential differences between simulation and data
in the \rznn ratio,
such as those that might be present because of missing higher-order corrections
in the simulated \gjets term.
It is given by:
\ifthenelse{\boolean{cms@external}}{
\begin{multline}
\DblR = \frac{\left\langle\mathcal{R}_{\zll/\gamma}^{\text{obs}}\right\rangle}{\left\langle\mathcal{R}_{\zll/\gamma}^{\text{sim}}\right\rangle}\\
 = \frac{\sum{N^{\text{obs}}_{\zll}}}{\sum{N^{\text{sim}}_{\zll}}}
   \frac{\sum{N^{\text{sim}}_{\gamma}}}{\sum{N^{\text{obs}}_\gamma}}
\frac{\left\langle\beta_{\ell\ell}^{\text{data}}\right\rangle}{\left\langle\SFdmcll\right\rangle} \frac{\left\langle\SFdmcg\right\rangle}{\left\langle\Fdir \beta_\gamma\right\rangle},
\label{eq:doubleratio}
\end{multline}
}{
\begin{equation}
\DblR = \frac{\left\langle\mathcal{R}_{\zll/\gamma}^{\text{obs}}\right\rangle}{\left\langle\mathcal{R}_{\zll/\gamma}^{\text{sim}}\right\rangle}
 = \frac{\sum{N^{\text{obs}}_{\zll}}}{\sum{N^{\text{sim}}_{\zll}}}
   \frac{\sum{N^{\text{sim}}_{\gamma}}}{\sum{N^{\text{obs}}_\gamma}}
\frac{\left\langle\beta_{\ell\ell}^{\text{data}}\right\rangle}{\left\langle\SFdmcll\right\rangle} \frac{\left\langle\SFdmcg\right\rangle}{\left\langle\Fdir \beta_\gamma\right\rangle},
\label{eq:doubleratio}
\end{equation}
}
with $N^{\text{obs}}_{\zll}$,
$N^{\text{sim}}_{\zll}$, and $N^{\text{sim}}_{\gamma}$
the numbers of events in the indicated CRs,
with the simulated samples normalized to the integrated luminosity of the data.
The sums and averages span the search regions.
The $\beta_{\ell\ell}^{\text{data}}$ factors represent the purity of the \zll CR,
obtained from fits to the measured lepton-pair mass distributions,
while \SFdmcll are corrections to account for data-versus-simulation
differences in lepton reconstruction efficiencies.
While the \zll sample is too small to allow a meaningful
measurement of $\rho$ in each search region,
we examine the projections of $\rho$ in each dimension.
We find a modest dependence on \HT and on the correlated variable \njets.
Based on the observed empirical result
$\rho(\HT) = 0.91+\left(9.6\times10^{-5}{\GeV}^{-1}\right)\min{(\HT,900\GeV)}$,
we apply a weight to each simulated \gjets event entering
the evaluation of $\rho$ and \rznn.
Following this weighting,
the projections of $\rho$ in the \njets, \HT, and \MHT dimensions
are consistent with a constant value of 1.00,
with uncertainties deduced from linear fits to
the projections that vary with these variables between 2 and 13\%.

For search regions with $\nbjets>0$, the \znn background estimate is:
\begin{equation}
  \left(\nznn\right)_{j,b,k} = \left(\nznn\right)_{j,0,k}\mathcal{F}_{j,b},
  \label{eq:Zbjet}
\end{equation}
where $j$, $b$, and $k$ are bin indices (numbered from zero)
for the \njets, \nbjets, and kinematic
(i.e., \HT and \MHT) variables, respectively.
For example, $j=1$ corresponds to $\njets=3$--4,
$b=3$ to $\nbjets\ge3$,
and $k=0$ to kinematic interval 1 of Table~\ref{tab:kine-bins} and Fig.~\ref{fig:HT-MHT}.
The first term on the right-hand side of Eq.~(\ref{eq:Zbjet}) is
obtained from Eq.~(\ref{eq:gjet1}).

\begin{figure}[tb]
\centering
\includegraphics[width=\cmsFigWidth]{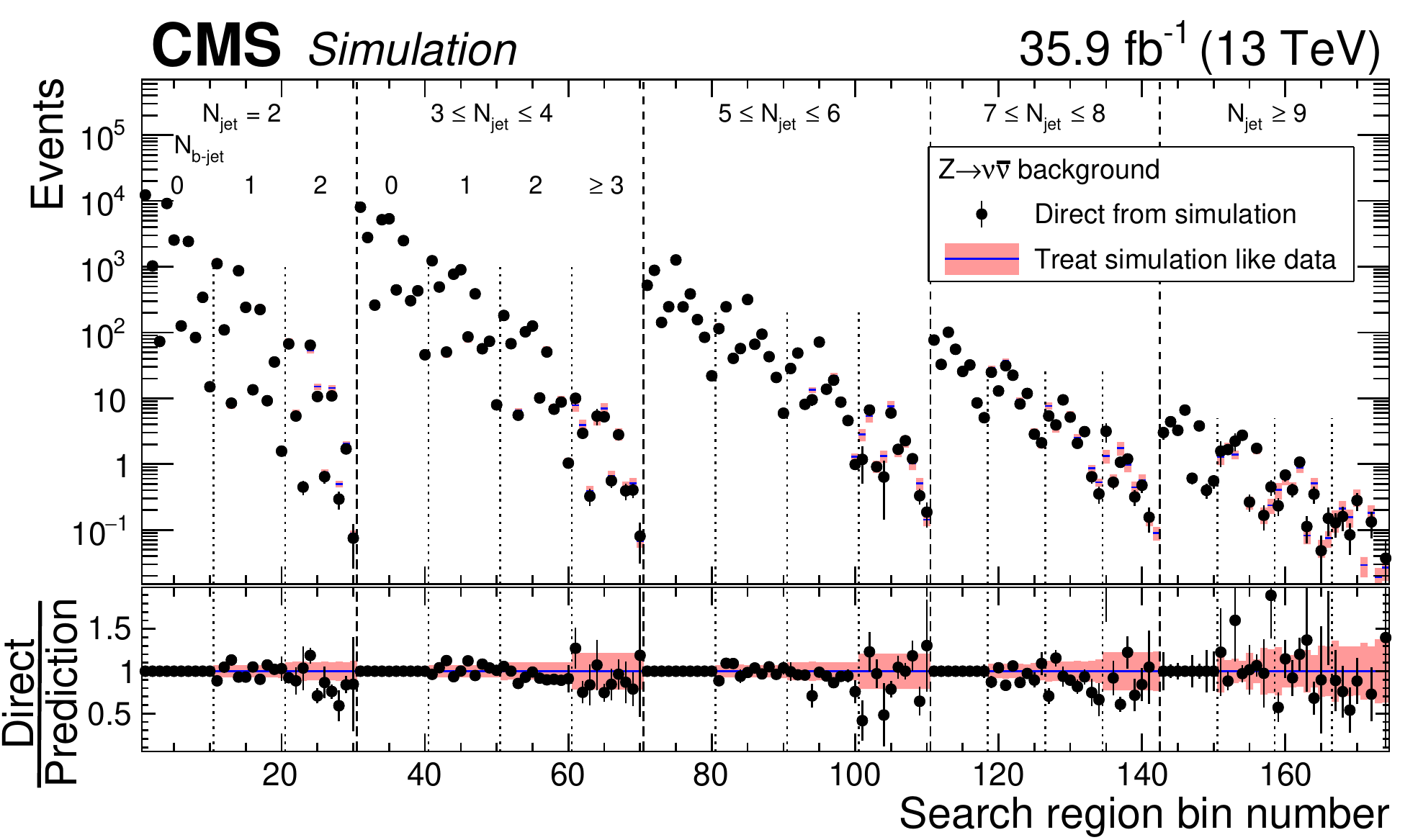}
\caption{
    The \znn background in the 174 search regions of the analysis
    as determined directly from \znnjets
    simulation (points, with statistical uncertainties),
    and as predicted by applying the \znn background determination
    procedure to statistically independent
    \zlljets simulated event samples
    (histogram, with
    shaded regions indicating the quadrature sum of the
    systematic uncertainty associated with the assumption that
    $\mathcal{F}_{j,b}$ is independent of \HT and \MHT,
    and the statistical uncertainty).
    For bins corresponding to $\nbjets=0$,
    the agreement is exact by construction.
    The results in the lower panel are obtained through bin-by-bin
    division of the results
    in the upper panel, including the uncertainties,
    by the central values of the ``predicted'' results.
    The labeling of the bin numbers is the
    same as in Fig.~\ref{fig:lost-lepton-closure}.
}
\label{fig:bjetSF}
\end{figure}

For all but the $\njets\geq9$ bin, corresponding to $j=4$,
the \nbjets extrapolation factor $\mathcal{F}_{j,b}$
is obtained from the fitted \zll data yields,
with data-derived corrections $\beta_{\ell\ell}^{\text{data}}$
to account for the {\nbjets}-dependent purity.
Other efficiencies cancel in the ratio.
Specifically,
\ifthenelse{\boolean{cms@external}}{
\begin{multline}
  \mathcal{F}_{j,b} =
  \left(N^\text{data}_{\zll} \beta_{\ell\ell}^{\text{data}}\right)_{j,b} \big/
  \left(N^\text{data}_{\zll} \beta_{\ell\ell}^{\text{data}}\right)_{j,0}; \\ j=0,1,2,3.
\end{multline}
}{
\begin{equation}
  \mathcal{F}_{j,b} =
  \left(N^\text{data}_{\zll} \beta_{\ell\ell}^{\text{data}}\right)_{j,b} \big/
  \left(N^\text{data}_{\zll} \beta_{\ell\ell}^{\text{data}}\right)_{j,0}; \quad j=0,1,2,3.
\end{equation}
}
For $\njets\geq9$,
there are very few \zll events
and we use the measured results for $\njets=7-8$ (the $j=3$ bin)
multiplied by an \nbjets extrapolation factor from simulation:
\begin{align}
  \mathcal{F}_{4,b} &= \mathcal{F}_{3,b}
    \left(\mathcal{F}_{4,b}^\text{sim}/
    \mathcal{F}_{3,b}^\text{sim} \label{eq:Jbjet}\right).
\end{align}
A systematic uncertainty is
assigned to the ratio of simulated yields in Eq.~(\ref{eq:Jbjet})
based on a lower bound equal to 1.0 and an upper bound determined
using the binomial model of Ref.~\cite{Khachatryan:2016kdk}.
The resulting uncertainty ranges from 7 to 40\%,
depending on \nbjets.

A closure test of the method
is presented in Fig.~\ref{fig:bjetSF}.
The shaded bands represent systematic uncertainties
of 7, 10, and 20\% for $\nbjets=1$, 2, and $\geq$3, respectively,
combined with the statistical uncertainties from the simulation.
The systematic uncertainties account for the assumption
that the $\mathcal{F}_{j,b}$ terms are independent of \HT and~\MHT.

The rare process $\ttbar\cPZ$
and the even more rare processes $\cPZ\cPZ$, $\PW\PW\cPZ$,
$\PW\cPZ\cPZ$, and $\cPZ\cPZ\cPZ$
can contribute to the background.
We add the expectations for these processes,
obtained from simulation,
to the numerator and denominator of Eq.~(\ref{eq:Jbjet}).
Note that processes with a {\cPZ} boson
that have a counterpart with the {\cPZ} boson replaced by a photon
are already accounted for in $N^\text{obs}_\gamma$
and largely cancel in the {\rznn} ratio.
For search regions with $\njets\geq9$ and $\nbjets\ge2$,
the contribution of $\ttbar\cPZ$ events
is comparable to that from \zjets events,
with an uncertainty of ${\approx}50\%$,
consistent with the rate and uncertainty for
$\ttbar\cPZ$ events found in Ref.~\cite{cms_ttZ}.

Besides the uncertainties associated with the \nbjets extrapolation
and the $\rho$ term,
discussed above,
systematic uncertainties associated with
the statistical precision of the simulation,
the photon reconstruction efficiency,
the photon and dilepton purities,
and the $\rznnsim$ term are evaluated.
The principal uncertainty arises from the
limited number of events in the~CRs.

\subsection{Background from QCD events}
\label{sec:qcd}

Background from QCD events is not,
in general,
expected to be large.
Nonetheless,
since \MHT in these events
primarily arises from the mismeasurement of jet \pt
rather than from genuine missing transverse momentum,
it represents a difficult background to model.
We employ two methods,
complementary to each other,
to evaluate the QCD background:
the rebalance-and-smear (\rands)
method~\cite{Collaboration:2011ida,Chatrchyan:2014lfa}
and the low-{\dphi} extrapolation
method~\cite{Chatrchyan:2013wxa,Khachatryan:2016kdk}.
The \rands method is selected as our primary technique
because it is more strongly motivated from first principles
and is less empirical in nature.
Thus the \rands method is used
for the interpretation of the data,
presented in Section~\ref{sec:results}.
The low-\dphi extrapolation method
is used as a cross-check.

\subsubsection{The rebalance-and-smear method}
\label{sec:qcd-rs}

The \rands method utilizes a special CR event sample,
selected using triggers that require
\HT to exceed thresholds ranging from 250 to 800\GeV.

In a first step, called ``rebalance,''
the jet momenta in a CR event are rescaled
to effectively undo the effects of detector response.
This step is performed using Bayesian inference.
The prior probability distribution $\pi$
is derived from the particle-level QCD simulation,
where ``particle level''
corresponds to the level of an event generator,
\ie, without simulation of the detector.
It is given by
\begin{equation}
\pi(\htvecmiss,\jtrans) =
  {\cal P}\left(\MHT\right) \,
  {\cal P}\left(\dphimhtb\right),
\end{equation}
where ${\cal P}(\MHT)$
is the distribution of \MHT,
and ${\cal P}(\dphimhtb)$
the distribution of the azimuthal angle between \htvecmiss and
the highest \pt jet in the event,
or between \htvecmiss and the highest \pt tagged {\cPqb} jet
if \mbox{$\nbjets\geq1$}.
The prior is binned in intervals of \HT and \nbjets.
The prior thus incorporates information about both the
magnitude and direction of the genuine \htvecmiss
expected in QCD events.
This represents a more sophisticated treatment than
the one used in Refs.~\cite{Collaboration:2011ida,Chatrchyan:2014lfa},
where the prior was merely taken to be a Dirac delta function at $\MHT=0$.

The jets in a CR event are then rescaled,
using Bayes' theorem,
to represent the event at the particle level.
Jets with $\pt>15\GeV$ and $\abs{\eta}<5.0$ are included in this procedure.
The expression of Bayes' theorem is:
\begin{equation}
  {\cal P}(\vec{J}_{\text{part}}|\vec{J}_{\text{meas}})
   \sim
  {\cal P}(\vec{J}_{\text{meas}}|\vec{J}_{\text{part}})\,\pi(\htvecmiss,\jtrans).
\end{equation}
The ${\cal P}(\vec{J}_{\text{part}}|\vec{J}_{\text{meas}})$ term is
the posterior probability density,
expressing the probability for a given set
of particle-level jet momenta $\vec{J}_{\text{part}}$
given the measured set $\vec{J}_{\text{meas}}$.
The ${\cal P}(\vec{J}_{\text{meas}}|\vec{J}_{\text{part}})$
term is a likelihood function,
defined by the product over the jets in the event
of the response functions for the individual jets.
The jet response functions,
determined in bins of jet \pt and $\eta$,
are derived from simulation as the
distribution of the ratio of reconstructed jet \pt values
to a given generated value,
corrected with separate scale factors for
the Gaussian cores and non-Gaussian tails
to account for jet energy resolution
differences with respect to data.
The likelihood function is maximized by rescaling the
momenta of the measured jets,
with the respective jet \pt uncertainties as constraints.
The set $\vec{J}_{\text{part}}$ corresponding to the resulting
most-likely posterior probability defines the rebalanced event.

In a second step, denoted ``smear,''
the magnitudes of the jet momenta are rescaled
by \pt- and $\eta$-dependent
factors obtained from random sampling
of the jet response functions.
This sampling is performed numerous times for each
rebalanced event to increase the statistical precision
of the resulting sample.
Each event is weighted with a factor inversely proportional
to the number of times it is sampled.

Application of the \rands procedure
produces an event sample that closely resembles the
original sample of CR events,
except the contributions of events with genuine \MHT,
viz.,
top quark, {\wjets}, {\zjets}, and possible signal events,
are effectively eliminated~\cite{Collaboration:2011ida}.
The rebalanced and smeared events are subjected
to the standard event selection criteria of
Section~\ref{sec:event-selection} to obtain the predictions for the
QCD background in each search region.

The principal uncertainty in the \rands QCD background
prediction is systematic,
associated with the uncertainty in the shape of the
jet response functions.
This uncertainty is evaluated by varying the
jet energy resolution scale factors within their uncertainties,
resulting in uncertainties in the prediction that
range from 20--80\% depending on the search region.
Smaller uncertainties related to the trigger, the prior,
and the statistical uncertainties are also evaluated.

\begin{figure}[!htbp]
\centering
\includegraphics[width=\cmsFigWidth]{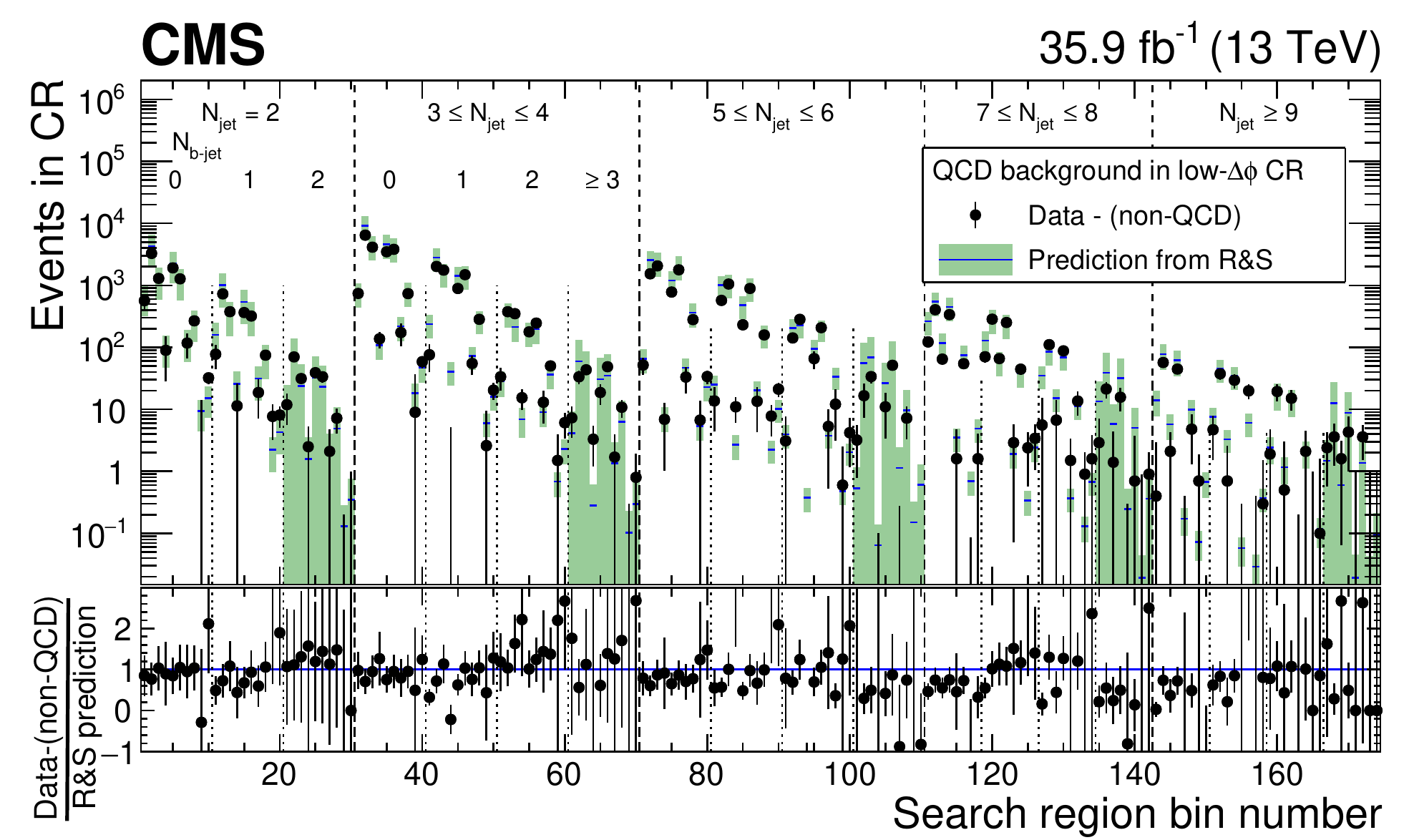}
\caption{
The QCD background in the low-\dphi control region (CR)
as predicted by the rebalance-and-smear (\rands) method
(histograms, with statistical and systematic uncertainties added in quadrature),
compared to the corresponding data from which the expected contributions of
top quark, \wjets, and \zjets events
have been subtracted
(points, with statistical uncertainties).
The lower panel shows the ratio of the measured to the predicted results
and its propagated uncertainty.
The labeling of the bin numbers is the same
as in Fig.~\ref{fig:lost-lepton-closure}.
}
\label{fig:qcd-rands}
\end{figure}

As a test of the method,
we determine the \rands prediction for the QCD contribution
to a QCD-dominated CR
selected with the standard trigger
and event selection,
except for the \dphimht requirements
of Section~\ref{sec:event-selection},
which are inverted.
Specifically, at least one of the two (for $\njets=2$),
three (for $\njets=3$),
or four (for $\njets\geq4$) highest \pt jets in an event must fail
a \dphimht selection criterion.
The resulting QCD-dominated
sample is called the low-\dphi CR.
The \rands prediction
for the QCD background in the low-\dphi CR is shown
in Fig.~\ref{fig:qcd-rands}
in comparison to the corresponding measured results,
following subtraction from the data of the contributions from
top quark, \wjets, and \zjets events,
evaluated as described in the previous sections.
Note that because of this subtraction,
the resulting difference is sometimes negative.
The prediction from the \rands method is seen to agree with the data
within the uncertainties.

\subsubsection{The low-\texorpdfstring{\dphi}{dphi} extrapolation method}
\label{sec:qcd-dphi}

In the low-\dphi extrapolation method,
the QCD background in each search region is evaluated
by multiplying the observed event yield in the corresponding region
of the low-\dphi CR (Section~\ref{sec:qcd-rs}),
after accounting for the contributions of non-QCD SM events,
by a factor \rqcd determined primarily from data.
The \rqcd terms express the ratio of the
expected QCD background in the
corresponding signal and low-\dphi regions.

The \rqcd term is empirically observed to have a
negligible dependence on \nbjets for a given value of \njets.
The functional dependence of \rqcd can therefore be expressed
in terms of \HT, \MHT, and \njets alone.
The \rqcd term is modeled as:
\begin{equation}
  \label{eqn:rqcd}
  \rqcd_{i,j,k} = \khtnj \,\smhtsim\,,
\end{equation}
with $i$, $j$, and $k$ the \HT, \njets, and \MHT bin indices, respectively.
In Ref.~\cite{Khachatryan:2016kdk}
we used a model in which the \HT, \MHT, and \njets dependencies in \rqcd
factorized.
For the $\njets=2$ search regions,
introduced for the present study,
this factorization is found to be less well justified and
we adopt the parameterization of Eq.~(\ref{eqn:rqcd}).

The \khtnj factors are determined from a maximum likelihood fit to
data in a sideband region defined by $250<\MHT<300\GeV$
(regions C1, C2, and C3 in Fig.~\ref{fig:HT-MHT}).
They are the ratio of the number of QCD events
in the high-\dphi region to that in the low-\dphi region,
where ``high \dphi'' refers to events selected with
the standard (noninverted) \dphimht requirements.
The fit accounts for the contributions
of top quark, \wjets, and \zjets events
using the results of the methods described in the preceding sections.
Uncertainties in \khtnj are determined
from the covariance matrix of the fit.
The \smhtsim terms,
taken from the QCD simulation,
represent corrections to account for
the \MHT dependence of \rqcd.
Based on studies of the differing contributions of
events in which the jet with the largest \pt mismeasurement
is or is not amongst the two (for $\njets=2$),
three (for $\njets=3$),
or four (for $\njets\geq4$) highest \pt jets,
uncertainties between 14 and 100\% are assigned to the
\smhtsim terms to account for
potential differences between data and simulation.
The total uncertainties in \smhtsim are defined by the
sum in quadrature of the systematic uncertainties
and the statistical uncertainties from the simulation.

\begin{figure}[!htb]
\centering
\includegraphics[width=\cmsFigWidth]{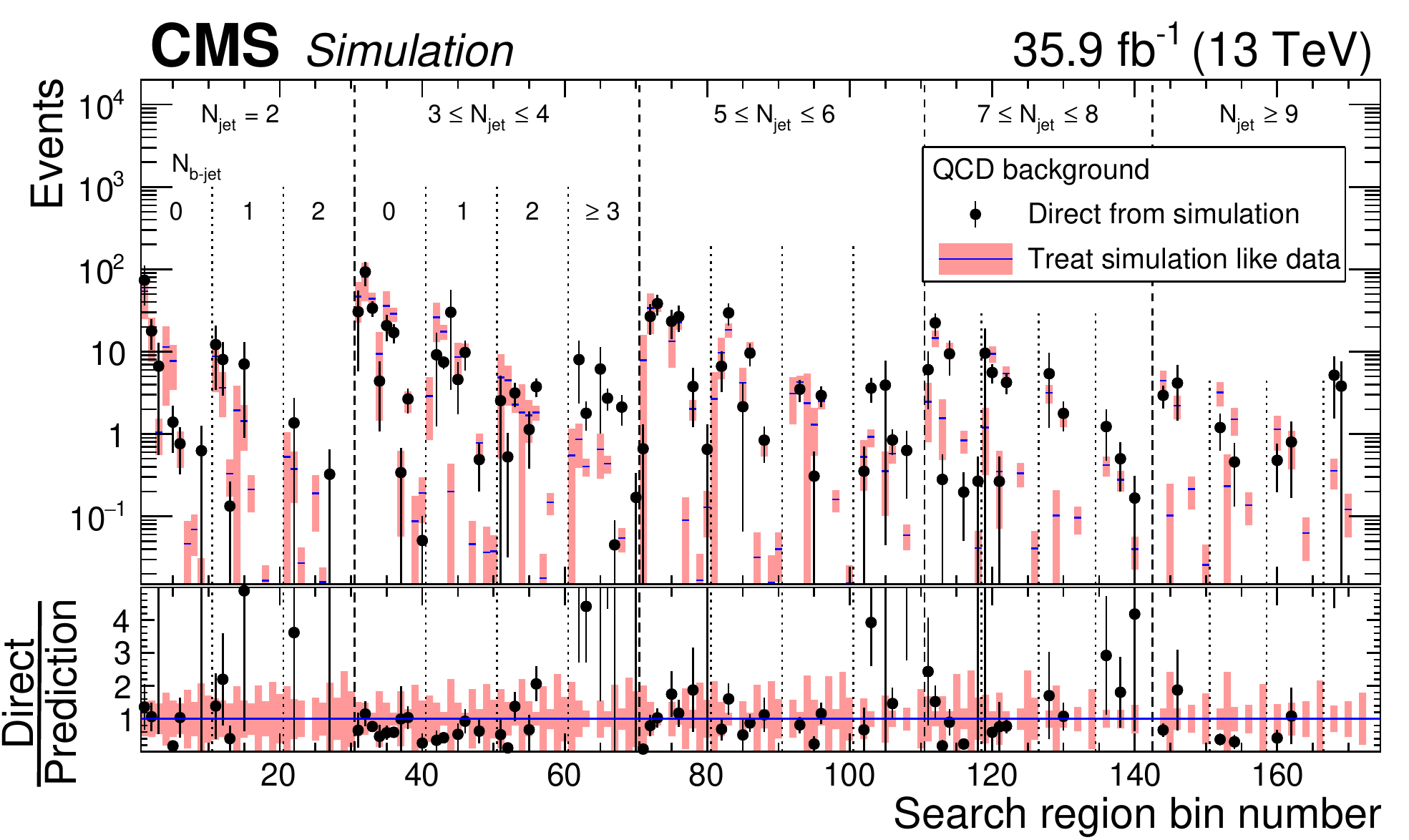}
\caption{
The QCD background
in the 174 search regions of the analysis
as determined directly from QCD simulation
(points, with statistical uncertainties)
and as predicted by applying the low-\dphi extrapolation
QCD background determination
procedure to simulated event samples
(histograms, with statistical and systematic uncertainties added in quadrature).
Bins without a point have no simulated QCD events in the search region,
while bins without a histogram have no simulated QCD events in
the corresponding control region.
The results in the lower panel are obtained through bin-by-bin
division of the results
in the upper panel, including the uncertainties,
by the central values of the ``predicted'' results.
No result is given in the lower panel if the value of
the prediction is zero.
The labeling of the bin numbers is the same
as in Fig.~\ref{fig:lost-lepton-closure}.
}
\label{fig:qcd-lowdphi}
\end{figure}

\begin{figure}[tbh]
\centering
\includegraphics[width=\cmsFigWidth]{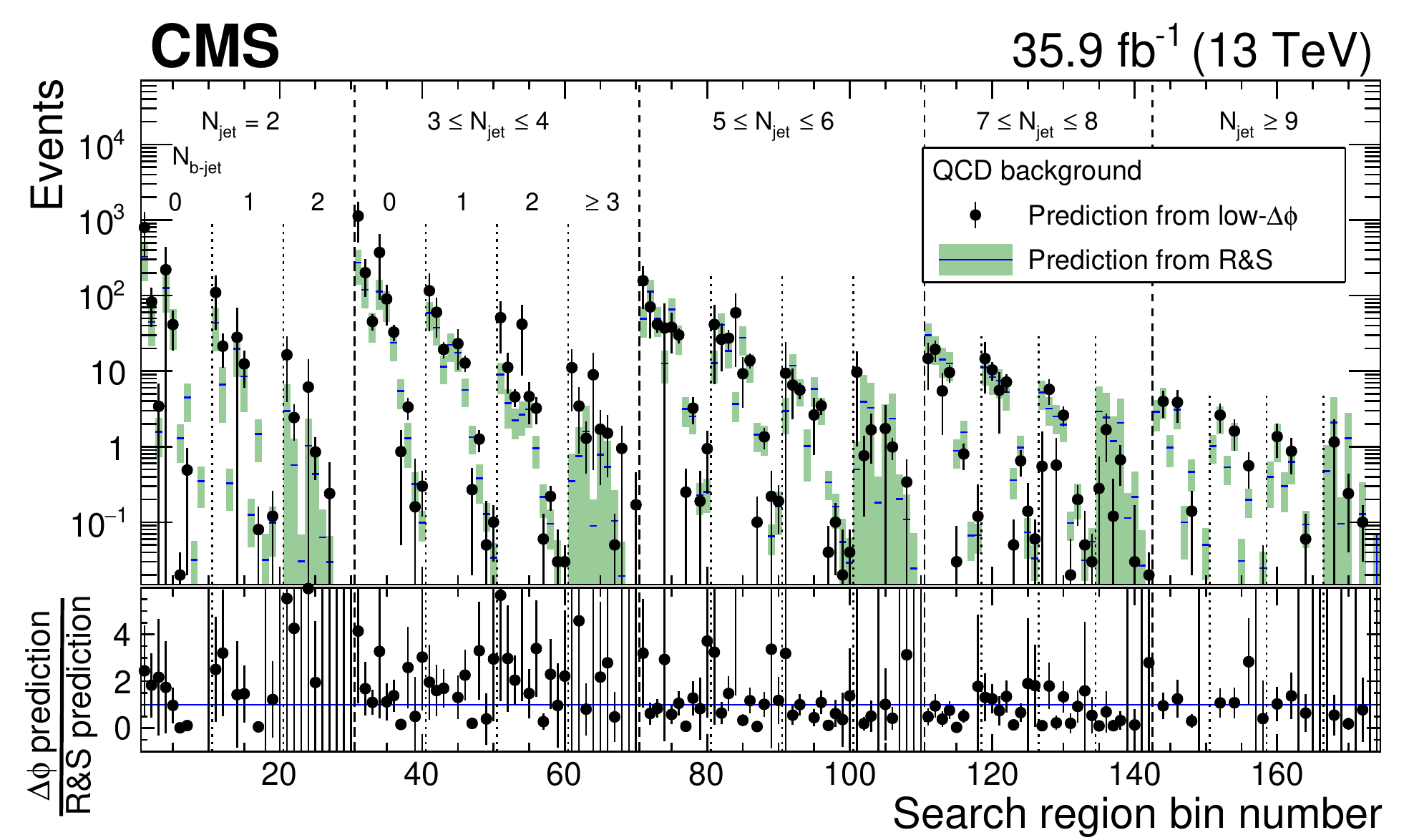}
\caption{
Comparison between the predictions for the number
of QCD events in the 174 search regions of the analysis
as determined from the rebalance-and-smear (\rands, histograms)
and low-\dphi extrapolation
(points)
methods.
For both methods,
the error bars indicate the combined
statistical and systematic uncertainties.
The lower panel shows the ratio of the low-\dphi extrapolation
to the \rands results
and its propagated uncertainty.
The labeling of the bin numbers is the same
as in Fig.~\ref{fig:lost-lepton-closure}.
}
\label{fig:qcd-compare}
\end{figure}

\begin{figure}[tbh]
\centering
\includegraphics[width=\cmsFigWidth]{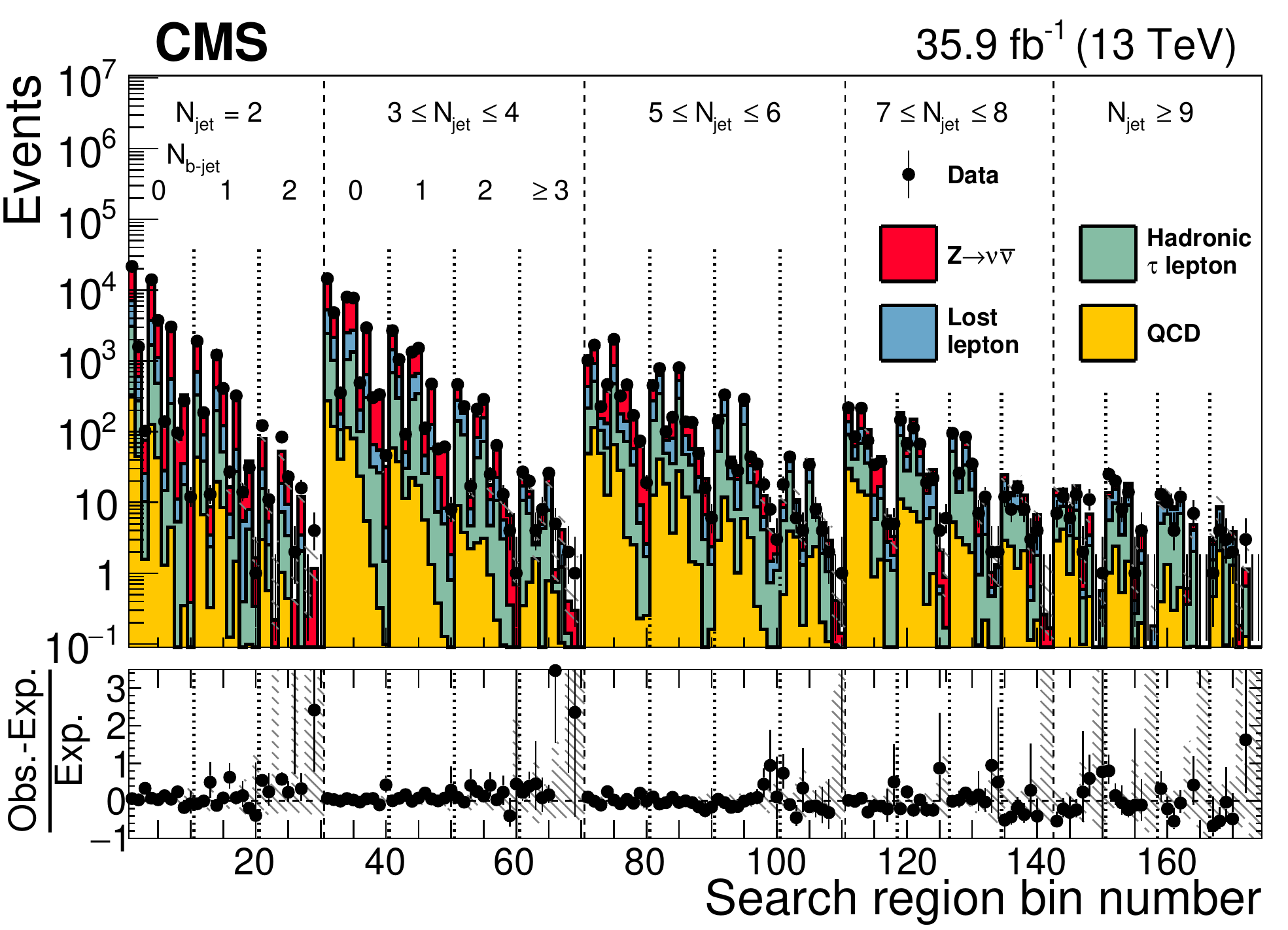}
\caption{
  The observed numbers of events and prefit SM background predictions
  in the 174 search regions of the analysis,
  where ``prefit'' means there is no constraint from the likelihood fit.
  Numerical values are given in
  Tables~\ref{tab:pre-fit-results-nj0}--\ref{tab:pre-fit-results-nj4}.
  The hatching indicates the total uncertainty in the background
  predictions.
  The lower panel displays the fractional differences
  between the data and SM predictions.
  The labeling of the bin numbers is the same as in
  Fig.~\ref{fig:lost-lepton-closure}.
}
\label{fig:fit-results}
\end{figure}

Figure~\ref{fig:qcd-lowdphi} presents
a closure test for the method.
An additional systematic uncertainty is included in \rqcd
to account for the level of nonclosure.
Figure~\ref{fig:qcd-compare}
shows a comparison between the predictions of the \rands and \dphi methods,
which are seen to be consistent.
Residual differences between the results from the two methods
are negligible compared to the overall uncertainties.

\section{Results}
\label{sec:results}

Figure~\ref{fig:fit-results} presents
the observed numbers of events in the 174 search regions.
The data are shown
in comparison with the summed predictions for the SM backgrounds.
Numerical values are given in
Tables~\ref{tab:pre-fit-results-nj0}--\ref{tab:pre-fit-results-nj4}
of \cmsAppendix\ref{sec:prefit}.
Signal region 126 exhibits a difference of 3.5 standard deviations
with respect to the SM expectation.
Signal regions 74, 114, and 151 exhibit differences
between 2 and 3 standard deviations.
The differences for all other signal regions lie below 2 standard deviations.
Thus, the evaluated SM background is found to be statistically
compatible with the data
and we do not obtain evidence for supersymmetry.

\begin{figure}[htb]
\centering
\includegraphics[width=\cmsFigWidth]{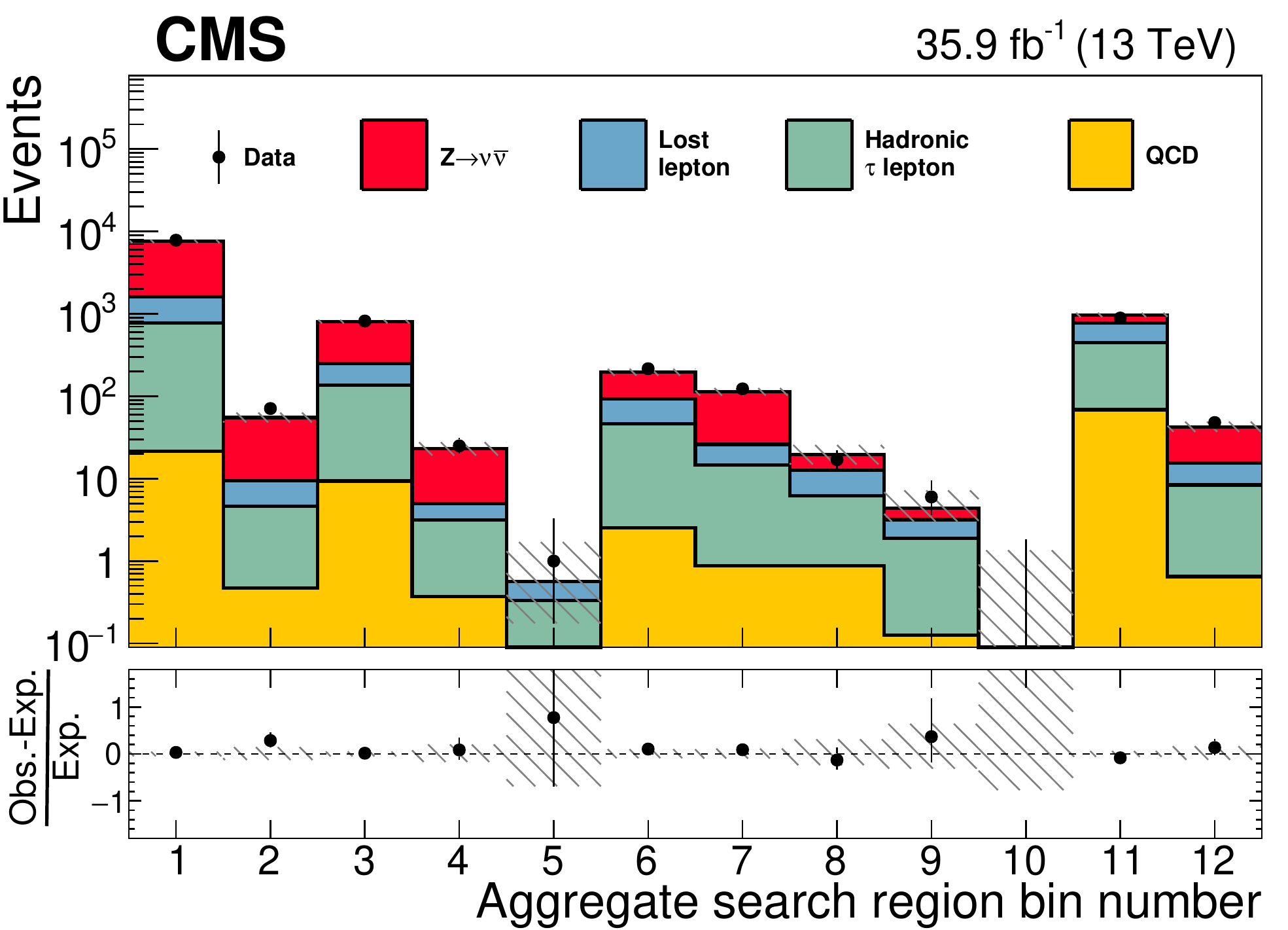}
\caption{
  The observed numbers of events and prefit
  SM background predictions in the 12 aggregate search regions,
  with fractional differences displayed in the lower panel,
  where ``prefit'' means there is no constraint from the likelihood fit.
  The hatching indicates the total uncertainty in the background
  predictions.
  The numerical values are given in Table~\ref{tab:pre-fit-results-asr}.
 }
\label{fig:agg}
\end{figure}

In addition to the finely segmented search regions of Fig.~\ref{fig:fit-results},
we evaluate the background predictions in 12 aggregate regions,
determined by summing the results from the nominal search regions
while accounting for correlations.
The aggregate regions are intended to represent
12 potentially interesting signal topologies.
For representative values of the SUSY particle masses,
the cross section upper limits from individual aggregate
signal regions are found to be around 50--300\% larger than
those presented below for the full 174 bin fit,
with a typical difference of about 100\%.
Nonetheless, 
the limits on SUSY particle masses derived using
the aggregate regions are generally no more than
around 10\% lower than those found using
the fit based on the 174 regions.
While the aggregate regions do not provide as much sensitivity
to the presence of new physics as the full set of search regions,
they allow our data to be used in a simpler manner
for the investigation of signal scenarios not examined in this paper.
The aggregate regions,
and the signal topologies they are intended to help probe,
are specified in Table~\ref{tab:agg}.
The aggregate regions are characterized by their heavy flavor
(top or bottom quark) content,
parton multiplicity,
and the mass difference \dmass
discussed in Section~\ref{sec:systematic}.
Aggregate regions 11 and 12 target models
with direct top squark production.
The results for the aggregate regions are presented in Fig.~\ref{fig:agg},
with numerical values provided in
Table~\ref{tab:pre-fit-results-asr} of \cmsAppendix\ref{sec:prefit}.

\begin{table*}[tbh]
  \centering
  \topcaption{Definition of the aggregate search regions.
Note that the cross-hatched region in Fig.~\ref{fig:HT-MHT},
corresponding to large \MHT relative to \HT,
is excluded from the definition of the aggregate regions.
}
\label{tab:agg}
   \begin{scotch}{c|cccc|ccc}
Region  & \njets & \nbjets & \HT [\GeVns{}]	& \MHT [\GeVns{}] &
   Parton multiplicity & Heavy flavor ? & $\Delta m$\\
    \hline
1  & $\geq$2 & 0 & $\geq$500 & $\geq$500  & Low & No & Small \\
2  & $\geq$3 & 0 & $\geq$1500 & $\geq$750 & Low & No & Large \\
3  & $\geq$5 & 0 & $\geq$500 & $\geq$500  & Medium & No & Small\\
4  & $\geq$5 & 0 & $\geq$1500 & $\geq$750 & Medium & No & Large\\
5  & $\geq$9 & 0 & $\geq$1500 & $\geq$750 & High & No & All\\
6  & $\geq$2 & $\geq$2 & $\geq$500 & $\geq$500 & Low & Yes & Small\\
7  & $\geq$3 & $\geq$1 & $\geq$750 & $\geq$750 & Low & Yes & Large\\
8  & $\geq$5 & $\geq$3 & $\geq$500 & $\geq$500 & Medium & Yes & Small\\
9  & $\geq$5 & $\geq$2 & $\geq$1500 & $\geq$750 & Medium & Yes & Large\\
10 & $\geq$9 & $\geq$3 & $\geq$750 & $\geq$750 & High & Yes & All\\
11 & $\geq$7 & $\geq$1 & $\geq$300 & $\geq$300 & Medium high & Yes & Small\\
12 & $\geq$5 & $\geq$1 & $\geq$750 & $\geq$750 & Medium & Yes & Large\\
  \end{scotch}
\end{table*}

\begin{figure*}[htbp]
\centering
\includegraphics[width=0.40\textwidth]{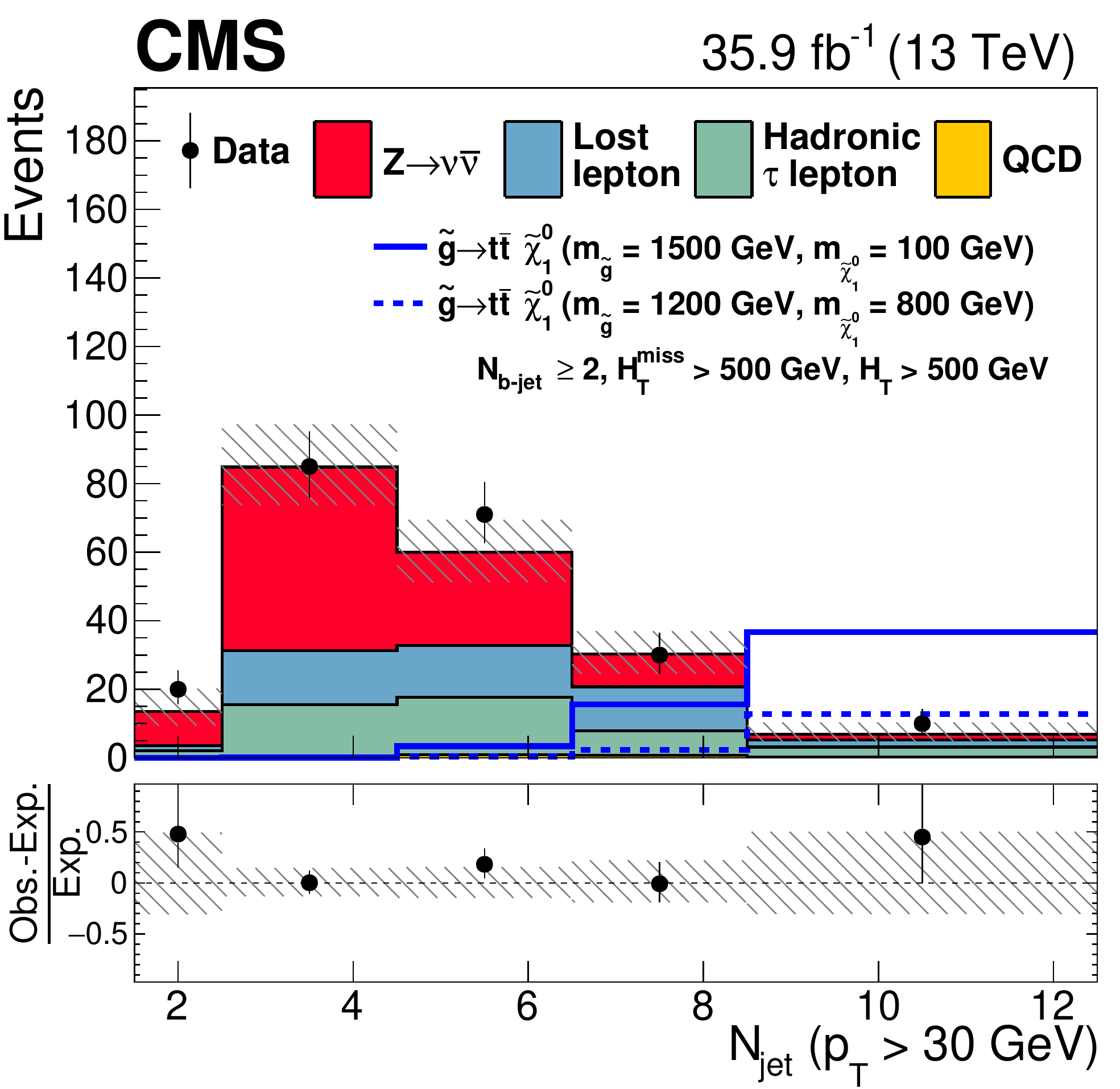}
\includegraphics[width=0.40\textwidth]{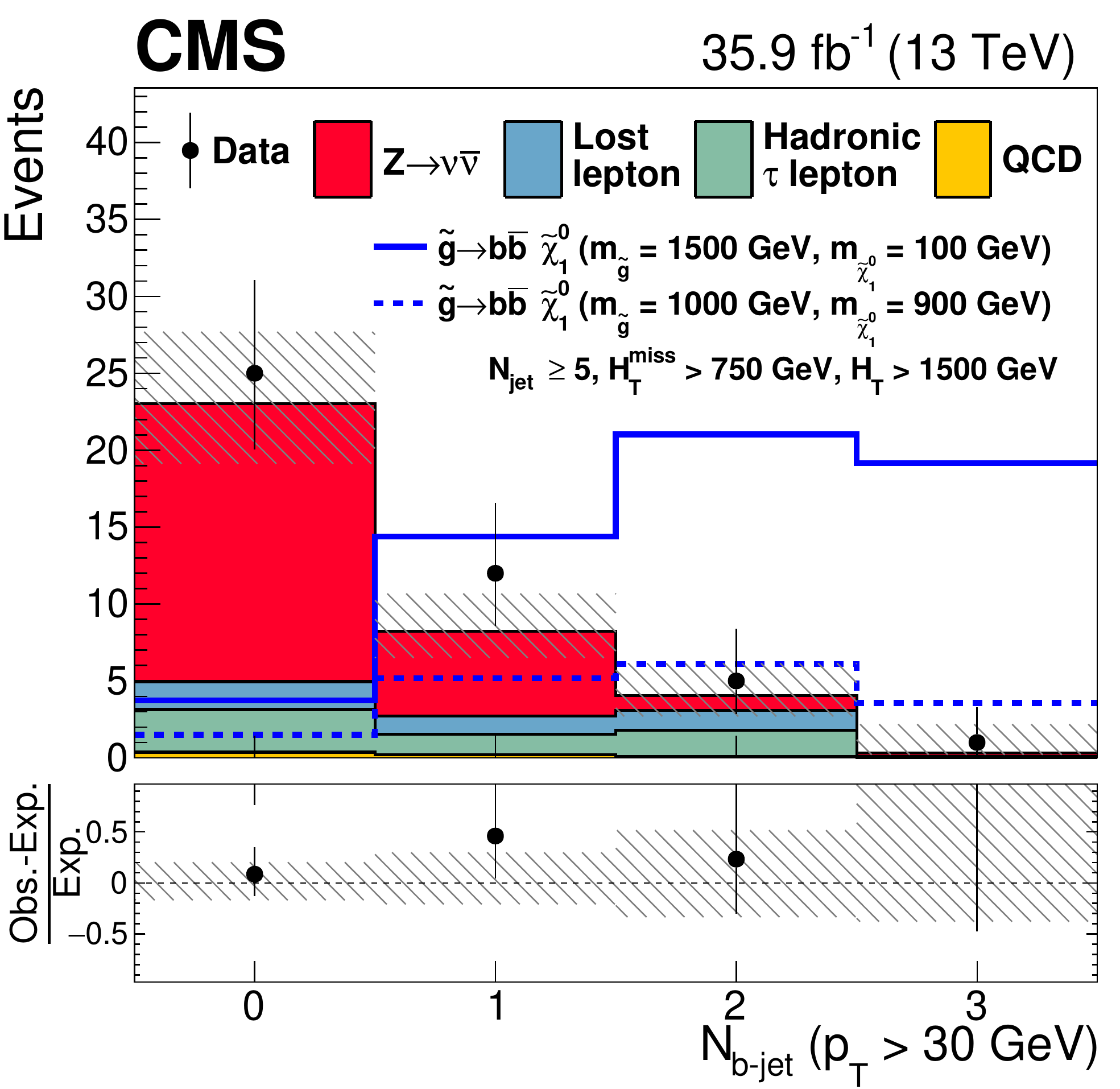}\\
\includegraphics[width=0.40\textwidth]{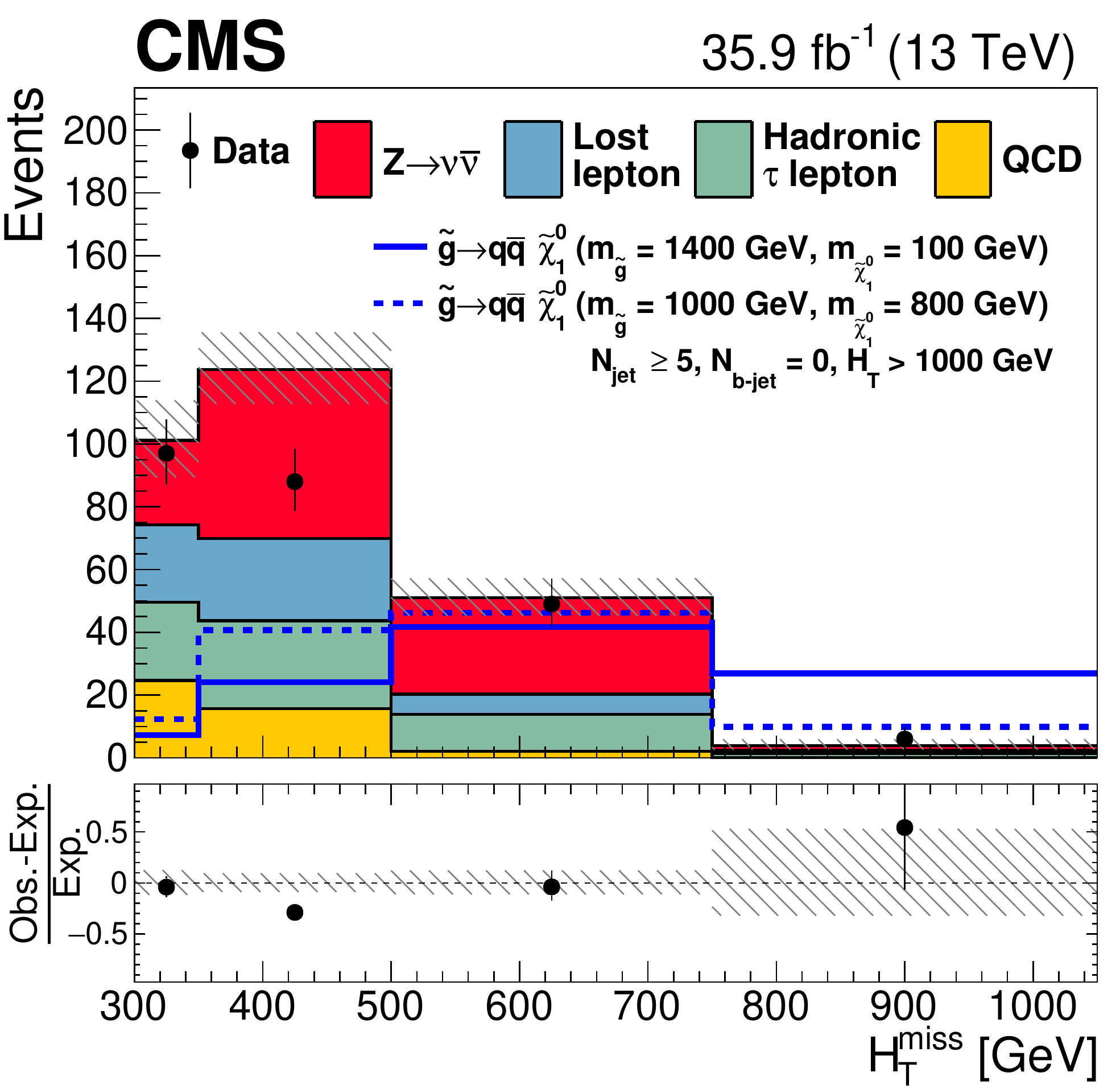}
\includegraphics[width=0.40\textwidth]{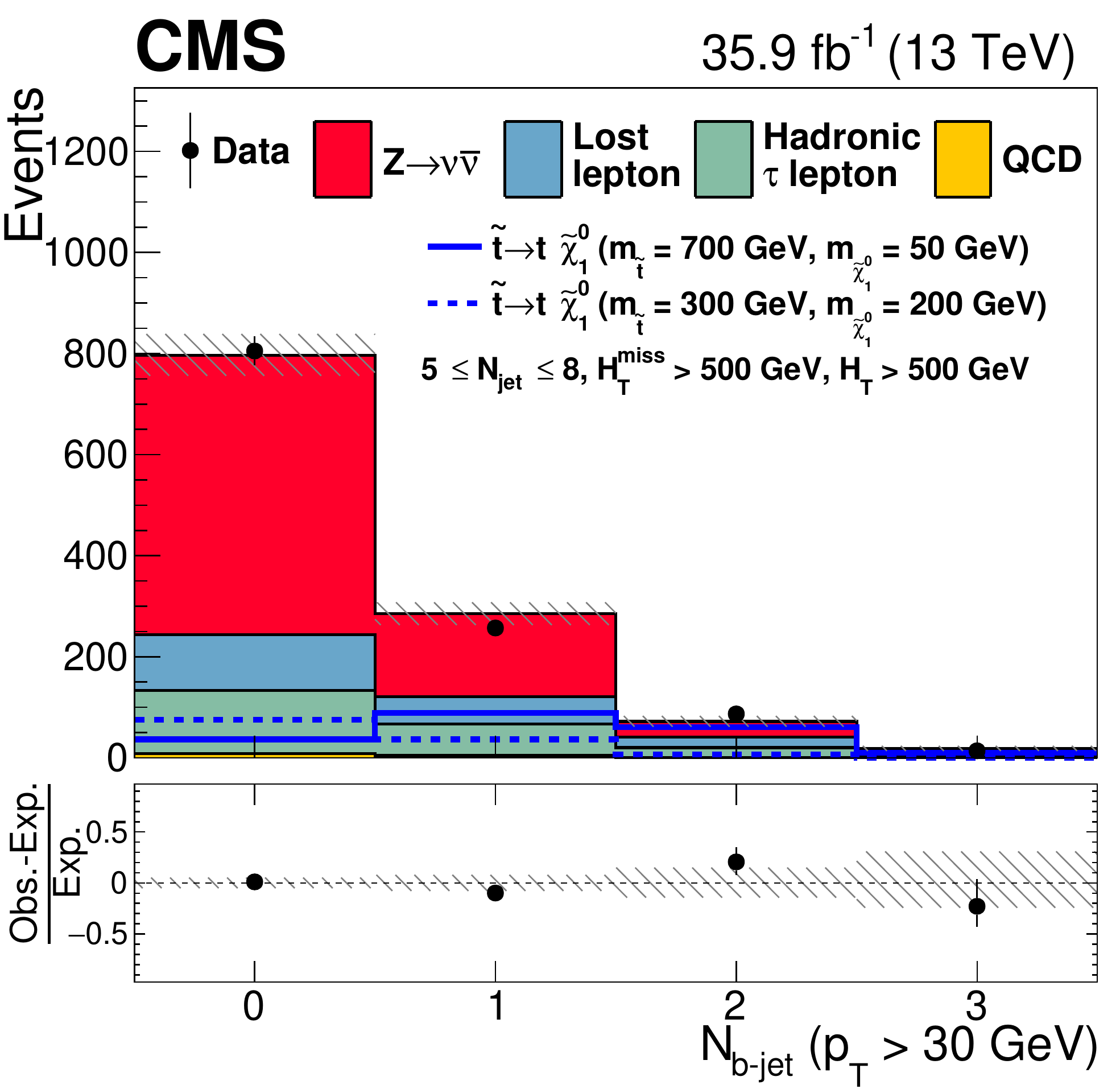}\\
\includegraphics[width=0.40\textwidth]{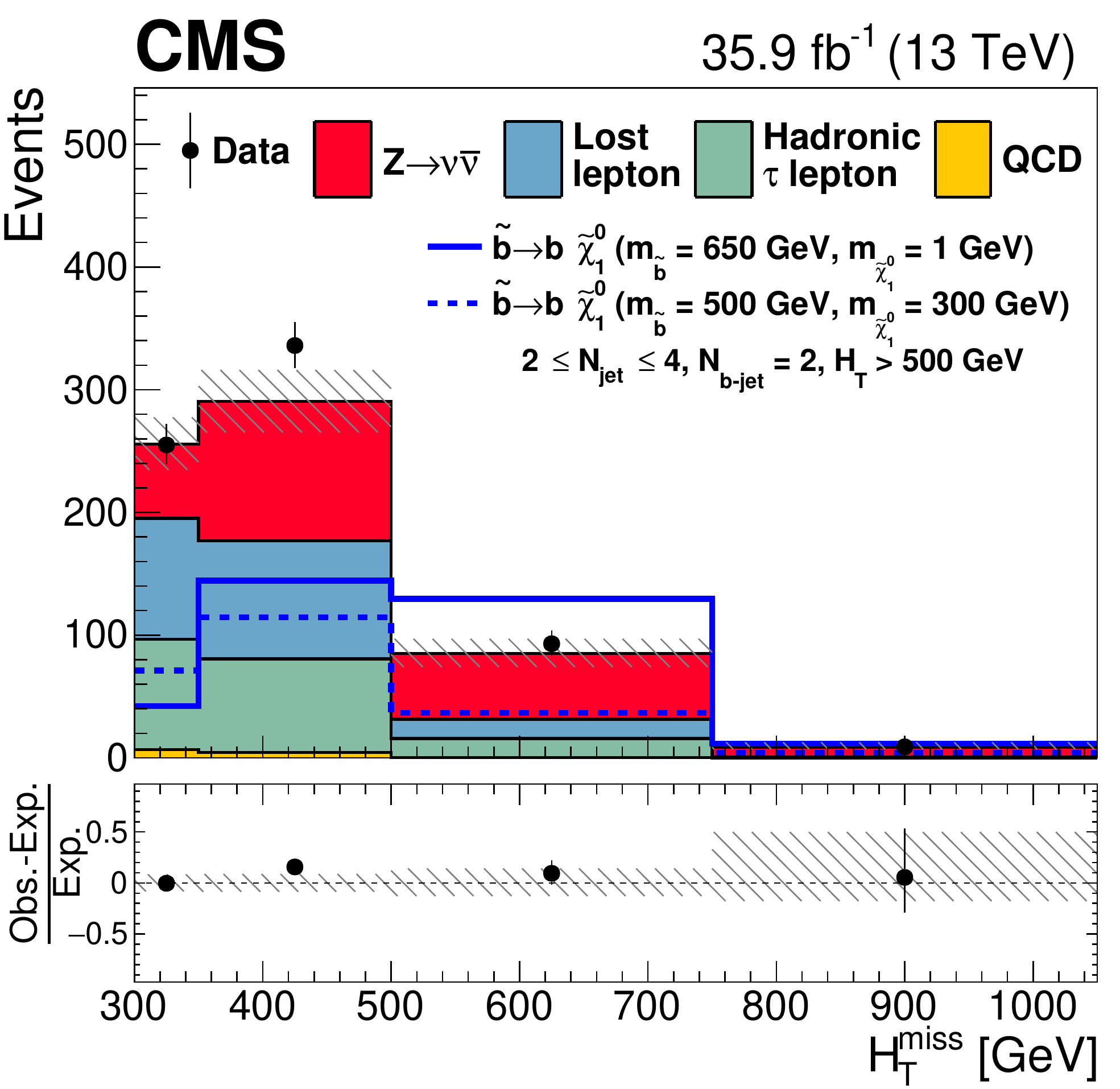}
\includegraphics[width=0.40\textwidth]{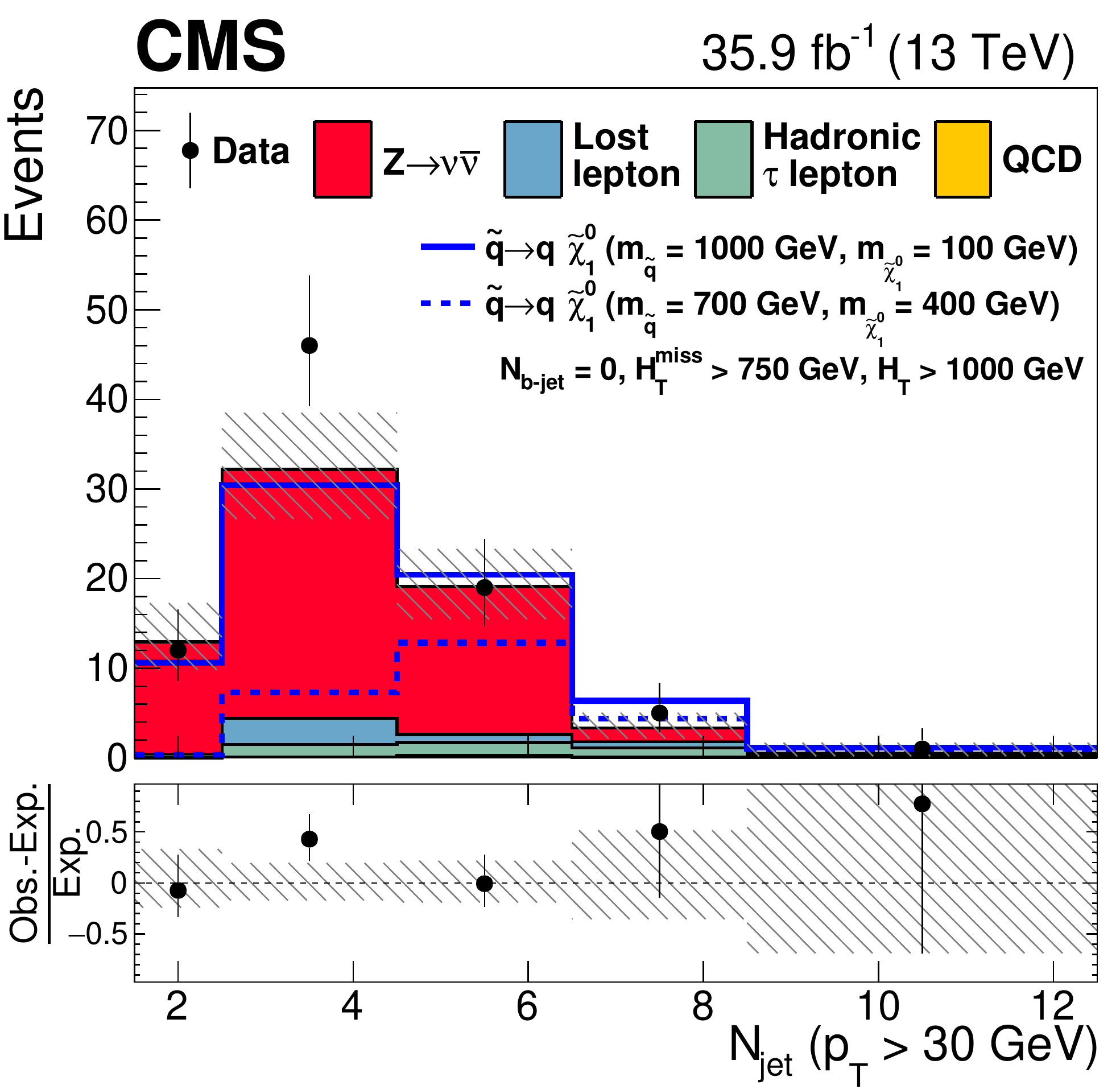}
\caption{
  The observed numbers of events and SM background predictions
  for regions in the search region parameter space
  particularly sensitive to the production of events in the
  (upper left) T1tttt,
  (upper right) T1bbbb,
  (middle left) T1qqqq,
  (middle right) T2tt,
  (lower left) T2bb, and
  (lower right) T2qq scenarios.
  The selection requirements are given in the figure legends.
  The hatched regions indicate the total uncertainties in the background
  predictions.
  The (unstacked) results for two example signal scenarios are shown in each instance,
  one with $\dmass\gg 0$ and the other with $\dmass\approx 0$,
  where \dmass is the difference between the gluino or squark mass
  and the sum of the masses of the particles into which it decays.
 }
\label{fig:projections}
\end{figure*}

In Fig.~\ref{fig:projections},
for purposes of illustration,
we present one-dimensional projections of
the data and SM predictions in either the
\MHT, \njets, or \nbjets variable
after imposing criteria, indicated in the legends,
to enhance the expected contributions of
T1tttt, T1bbbb, T1qqqq, T2tt, T2bb, or T2qq events.
In each case,
two example signal distributions are shown:
one with $\dmass\gg 0$,
and one with $\dmass\approx 0$,
where both example scenarios lie well within the
parameter space excluded by the present study.

Limits are evaluated for the production cross sections of the
signal scenarios using a likelihood fit,
with the SUSY signal strength,
the yields of the four classes of background
shown in Fig.~\ref{fig:fit-results},
and various nuisance parameters as fitted parameters,
where a nuisance parameter refers to a variable of little physical interest,
such as a scale factor in a background determination procedure.
The nuisances are constrained in the fit. 
For the models of gluino (squark) pair production,
the limits are derived as a function of \mgluino (\msquark) and \mlsp.
All 174 search regions are used for each
choice of the SUSY particle masses.
The likelihood function is given by the product of
Poisson probability density functions,
one for each search region,
and constraints that account for
uncertainties in the background predictions and signal yields.
These uncertainties are treated as nuisance parameters
with log-normal probability density functions.
Correlations are taken into account.
The signal yield uncertainties associated with
the renormalization and factorization scales, ISR,
jet energy scale,
{\cPqb} jet tagging,
pileup,
and statistical fluctuations
are evaluated as a function of \mgluino and~\mlsp,
or \msquark and \mlsp.
The test statistic is
$q_\mu =  - 2 \ln \left( \mathcal{L}_\mu/\mathcal{L}_\text{max} \right)$,
where $\mathcal{L}_\text{max}$ is the maximum likelihood
determined by allowing all parameters including the
SUSY signal strength $\mu$ to vary,
and $\mathcal{L}_\mu$ is the maximum likelihood for a fixed signal strength.
To set limits,
asymptotic results for the test statistic~\cite{Cowan:2010js} are used,
in conjunction with the CL$_\mathrm{s}$
criterion described in Refs.~\cite{Junk1999,bib-cls}.

\begin{figure*}[htbp]
\centering
    \includegraphics[width=0.40\textwidth]{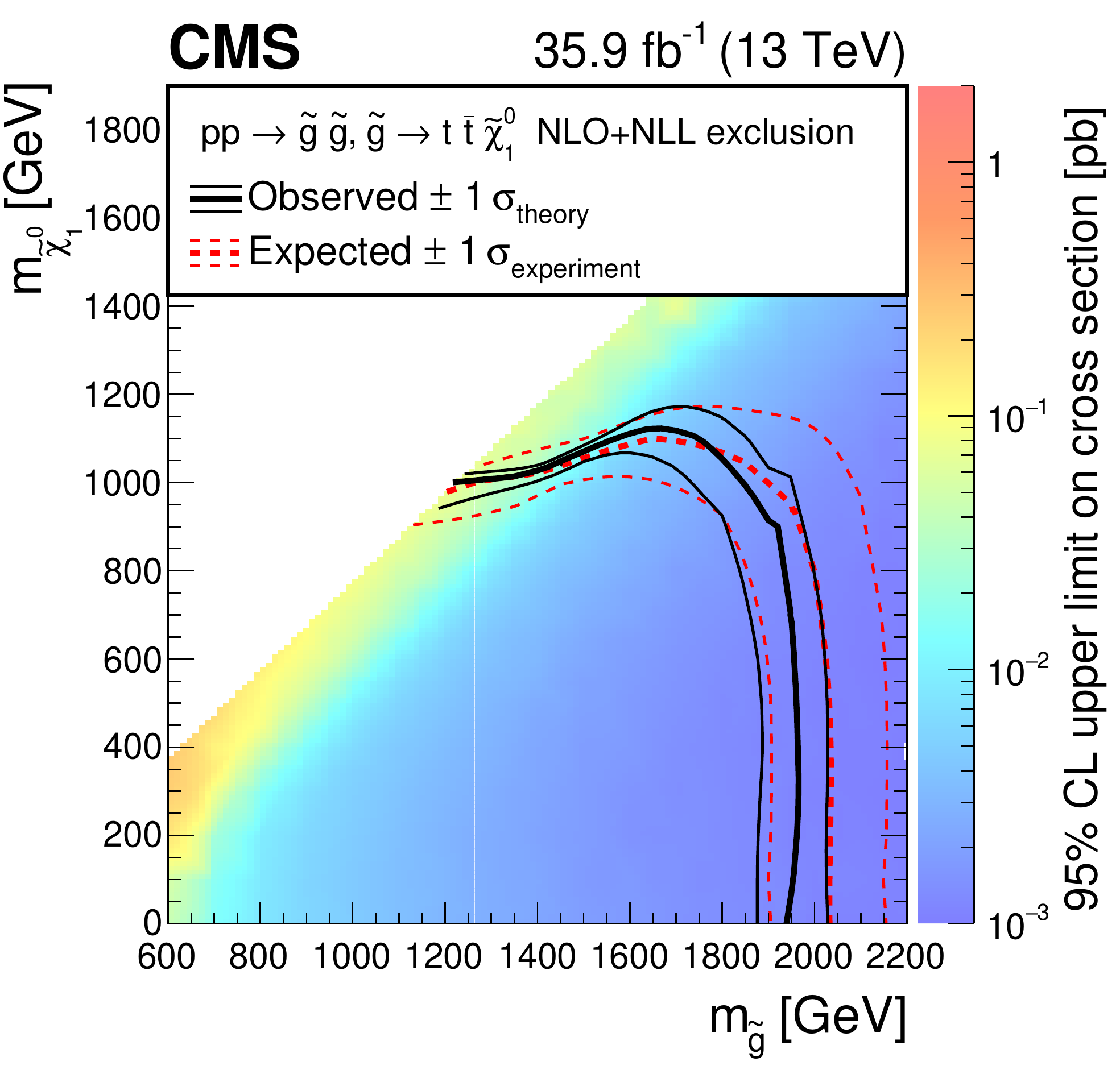}
    \includegraphics[width=0.40\textwidth]{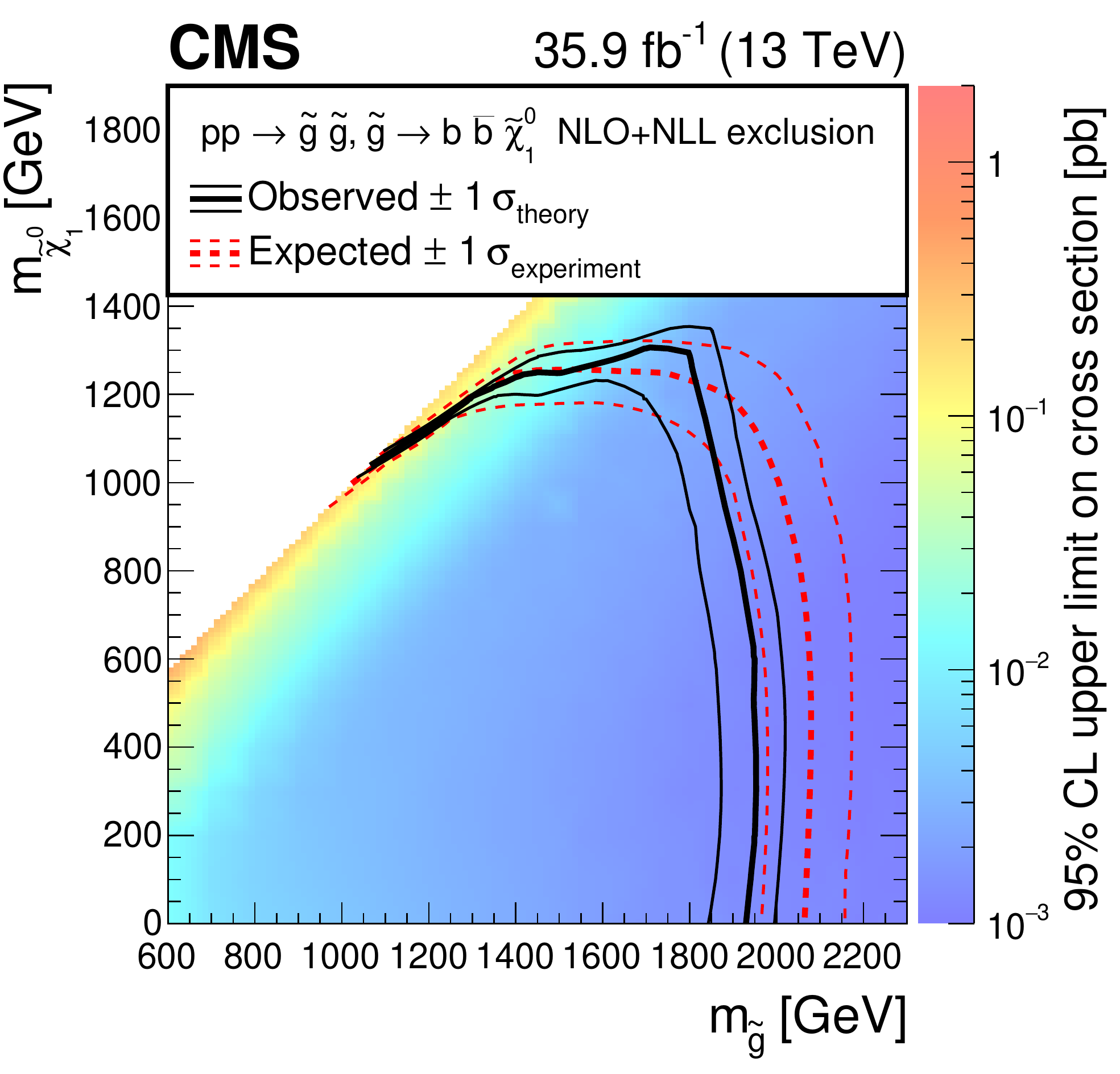}
    \includegraphics[width=0.40\textwidth]{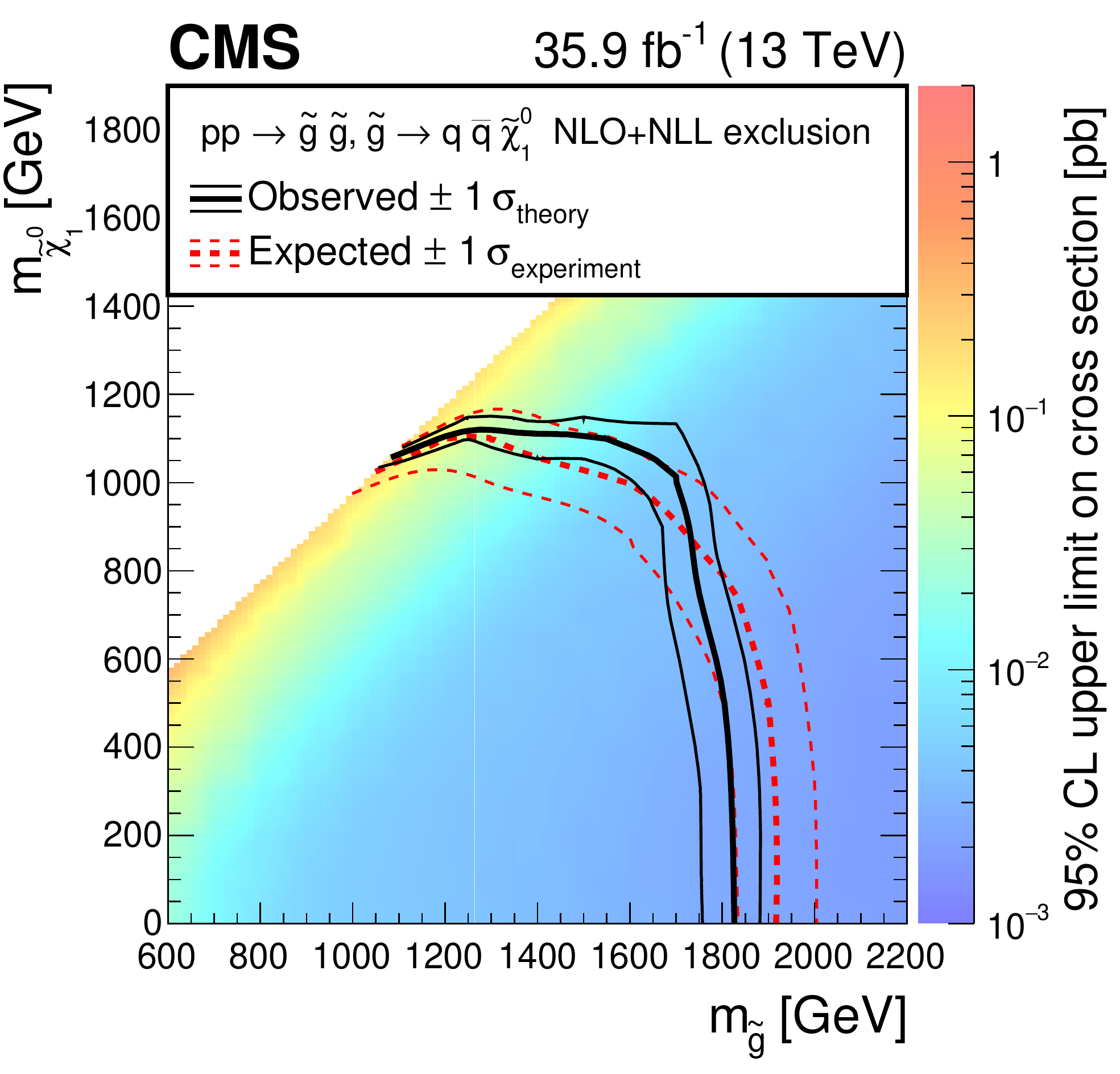}
    \includegraphics[width=0.40\textwidth]{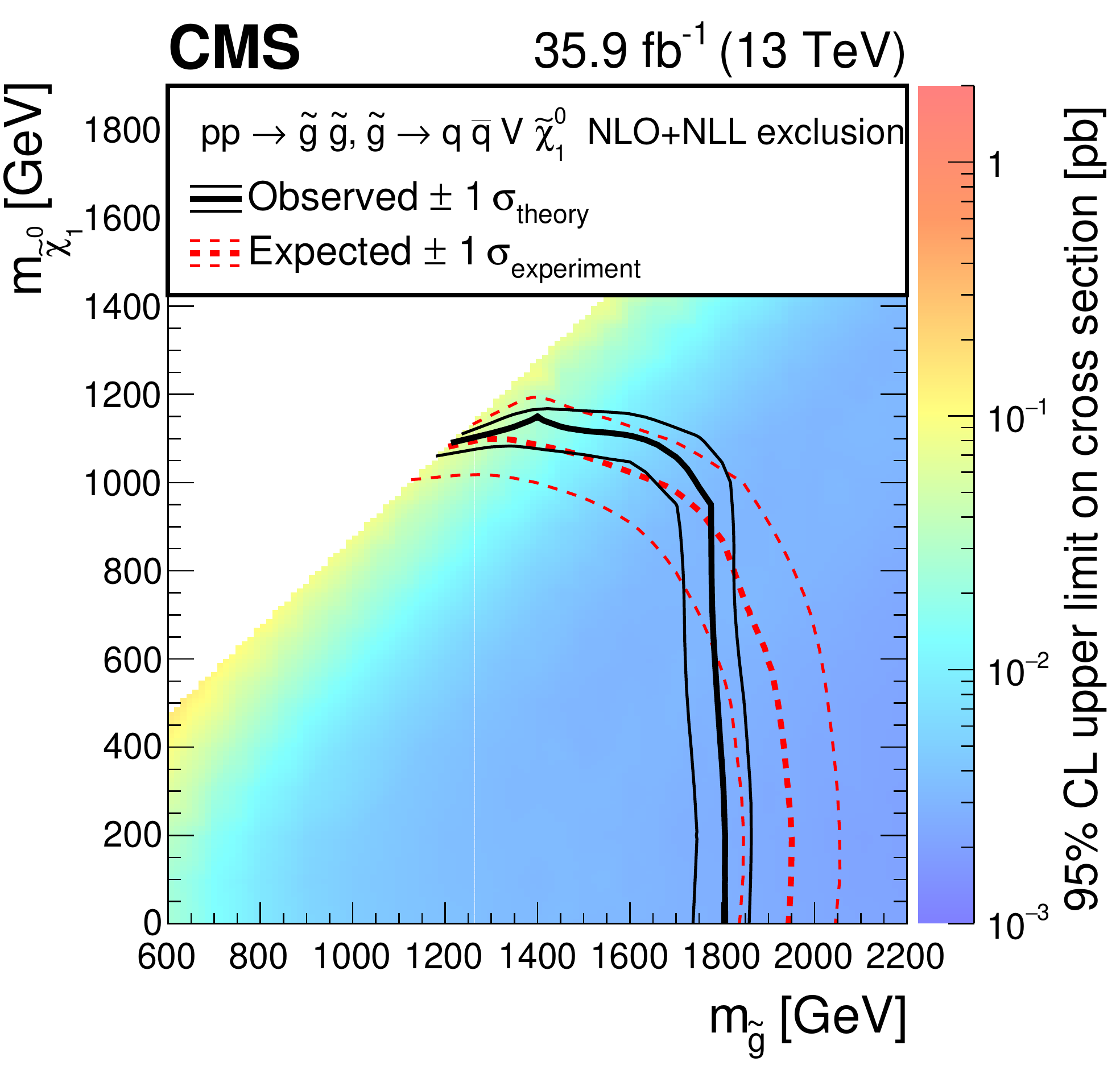}
    \includegraphics[width=0.40\textwidth]{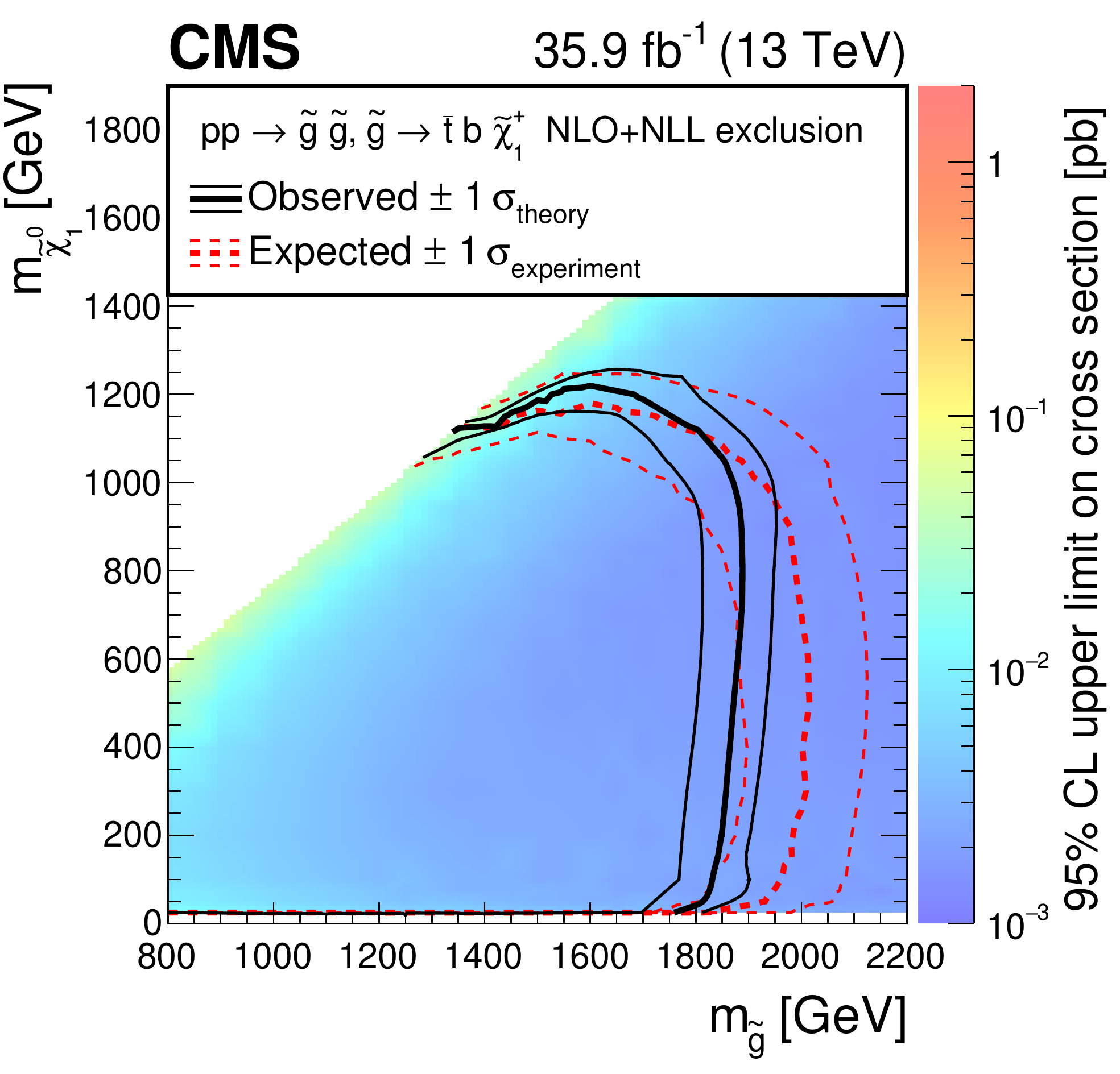}
    \includegraphics[width=0.40\textwidth]{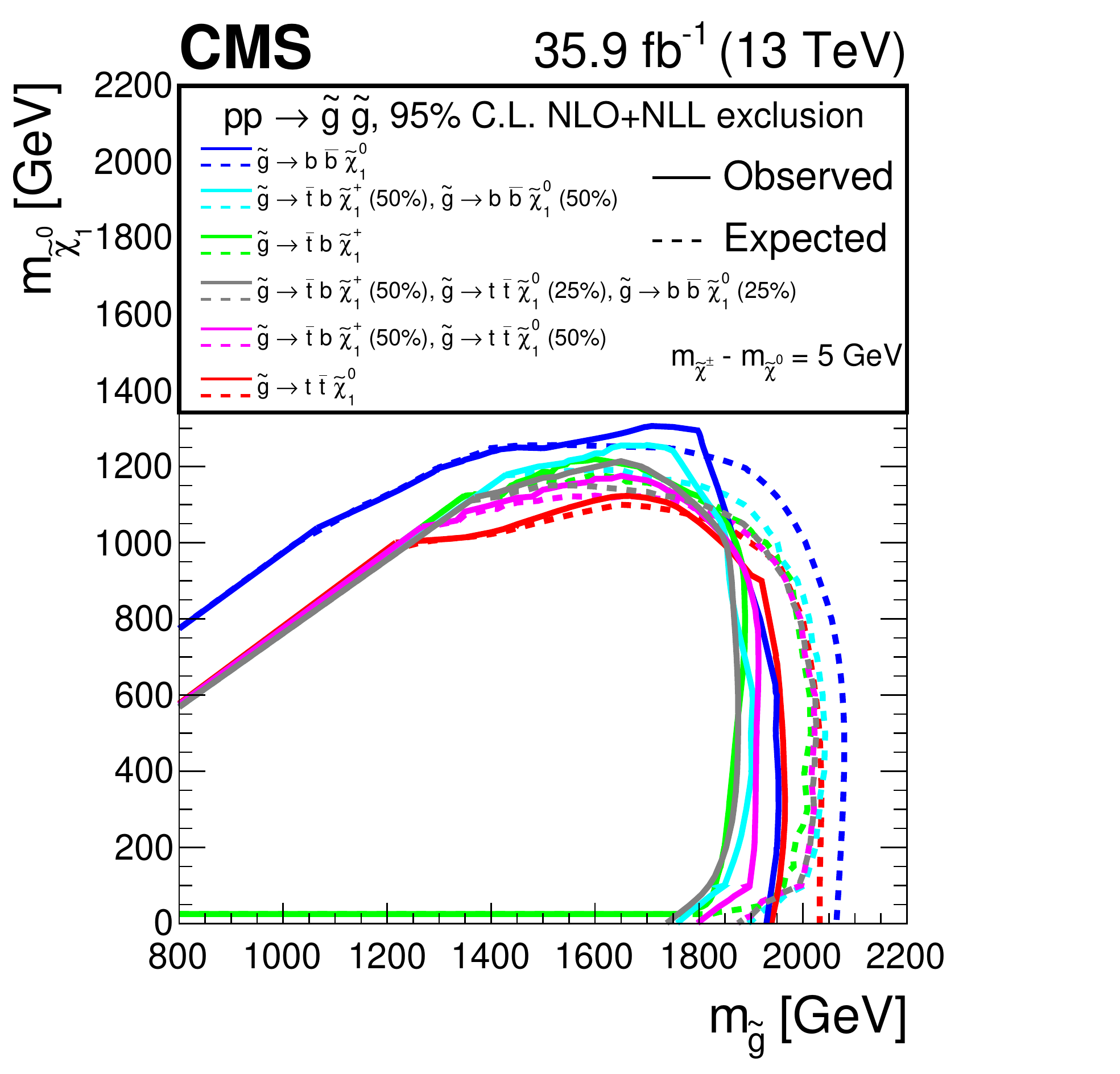}
    \caption{
      The 95\% CL upper limits on the production
      cross sections for the (upper left) T1tttt,
      (upper right) T1bbbb,
      (middle left) T1qqqq,
      (middle right) T5qqqqVV,
      and (lower left) T1tbtb simplified models
      as a function of the gluino and LSP masses \mgluino and~\mlsp.
      The thick solid (black) curves show the observed exclusion limits
      assuming the NLO+NLL cross
      sections~\cite{bib-nlo-nll-01,bib-nlo-nll-02,bib-nlo-nll-03,
      bib-nlo-nll-04,bib-nlo-nll-05} and
      the thin solid (black) curves the change in these limits
      due to variation of the signal cross sections
      within their theoretical uncertainties~\cite{Borschensky:2014cia}.
      The thick dashed (red) curves present the expected limits
      under the background-only hypothesis,
      while the thin dotted (red) curves indicate the region containing 68\%
      of the distribution of limits expected under this hypothesis.
      (Lower right)
      The corresponding 95\% NLO+NLL exclusion curves
      for the mixed models of gluino decays
      to heavy squarks.
      For the T1tbtb model,
      the results are restricted to $\mlsp>25\GeV$
      for the reason stated in the text.
    }
    \label{fig:limits-gluinos}
\end{figure*}

We evaluate 95\% confidence level (CL) upper limits
on the signal cross sections.
The NLO+NLL cross section is used
to determine corresponding exclusion curves.
When computing the limits,
the signal yields are corrected to account for
possible signal contamination in the CRs.
Beyond the observed exclusion limits,
we derive expected exclusion limits
by using the expected Poisson fluctuations around the predicted
numbers of background events when evaluating the test statistic.

The results for the T1tttt, T1bbbb, T1qqqq, and T5qqqqVV
models are shown in the upper and middle rows
of Fig.~\ref{fig:limits-gluinos}.
Depending on the value of \mlsp,
and using the NLO+NLL cross sections,
gluinos with masses as large as
1960, 1950, 1825, and 1800\GeV, respectively,
are excluded.
These results significantly extend those
of our previous study~\cite{Khachatryan:2016kdk},
for which the corresponding limits
vary between 1440 and 1600\GeV.

The corresponding results for the T1tbtb model
and for the mixed models of gluino decay to heavy squarks
are shown in the lower row of Fig.~\ref{fig:limits-gluinos}.
In this case gluinos with masses as large as
1850 to 1880\GeV are excluded,
extending the limits
of between 1550 and 1600\GeV presented in Ref.~\cite{Khachatryan:2016epu}.
Note that for the T1tbtb model,
the acceptance is small for $\mlsp\lesssim 25\GeV$
and we are unable to exclude the scenario.
The reason is that as \mlsp approaches zero,
the mass of the nearly mass-degenerate $\PSGcpm_1$
parent particle also becomes small.
The $\PSGcpm_1$ becomes highly Lorentz boosted,
and more of the momentum from the parent $\PSGcpm_1$
is carried by the daughter off-shell {\PW} boson
[see Fig.~\ref{fig:event-diagrams} (upper right)]
and less by the daughter {\PSGczDo}.
The net effect is that
the \MHT spectrum becomes softer
for hadronic $\PW^*$ decays,
leading to reduced signal acceptance,
while the charged-lepton or isolated-track \pt spectrum
becomes harder for leptonic $\PW^*$ decays,
increasing the probability for the event to be vetoed
and thus also leading to reduced signal acceptance.
Furthermore,
jets arising from the $\PW^*$ decay tend to be aligned with
the missing transverse momentum from the {\PSGczDo}.
When these jets become harder,
as \mlsp becomes small,
they are more likely to appear amongst the highest \pt jets
in the event,
causing the event to be rejected by the \dphimht
requirements.
Because of the small signal acceptance for $\mlsp\to 0$,
the relative contribution of signal contamination in this region
becomes comparable to the true signal content,
and a precise determination of the search sensitivity becomes difficult.
Therefore,
for the T1tbtb model,
we limit our determination of the cross section upper limit
to $\mlsp>25\GeV$.

\begin{figure*}[htbp]
\centering
    \includegraphics[width=0.48\textwidth]{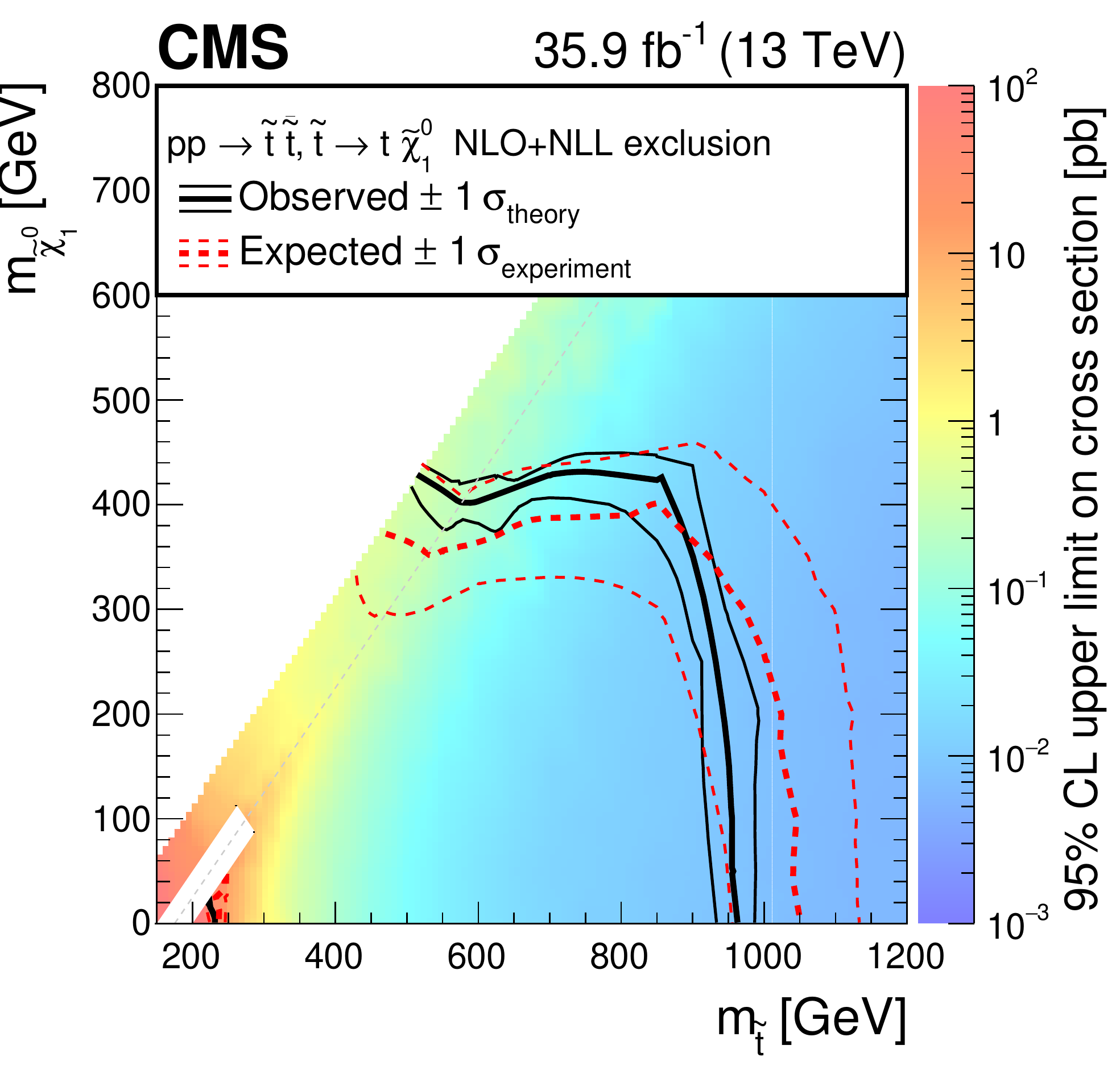}
    \includegraphics[width=0.48\textwidth]{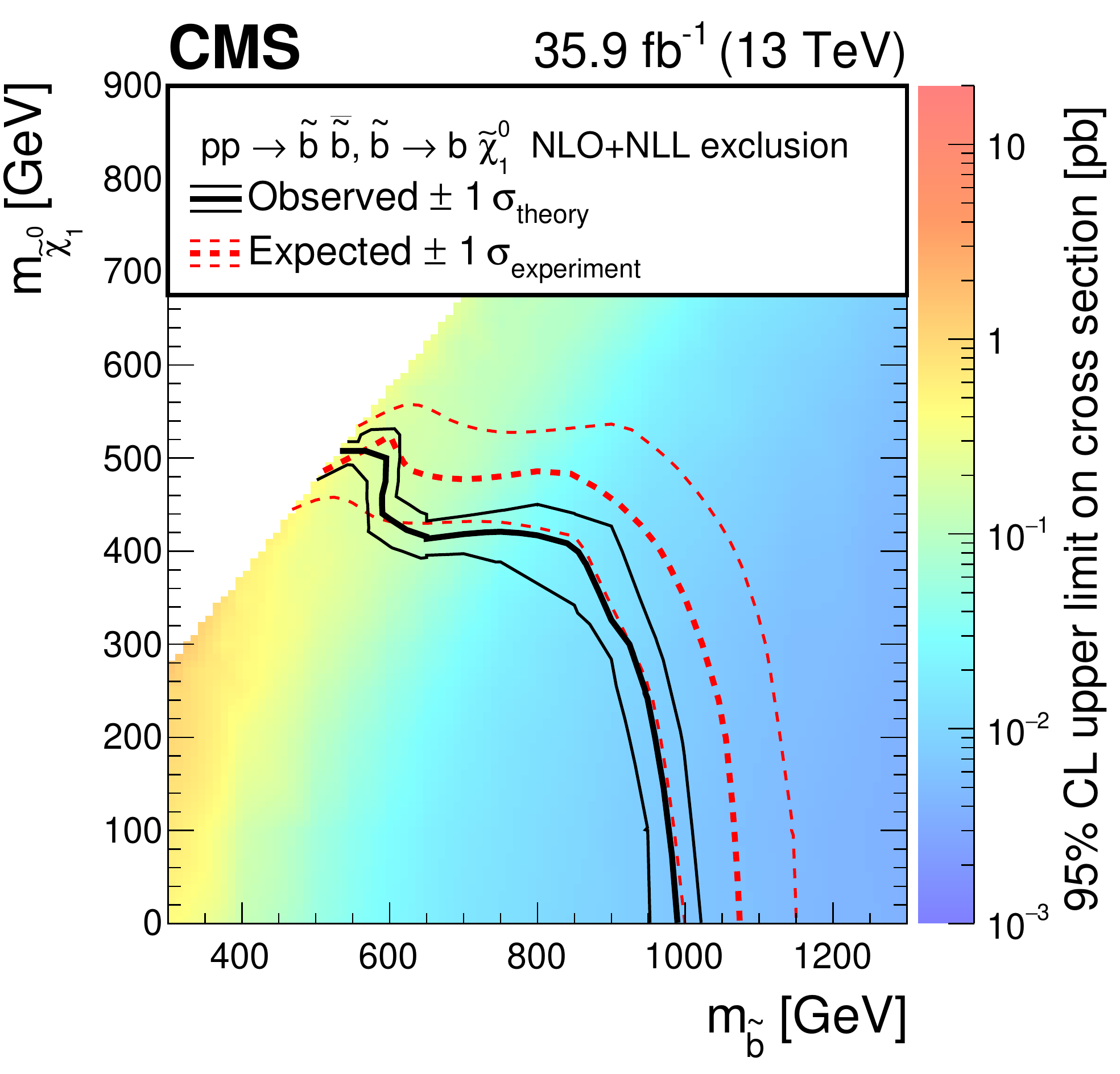}
    \includegraphics[width=0.48\textwidth]{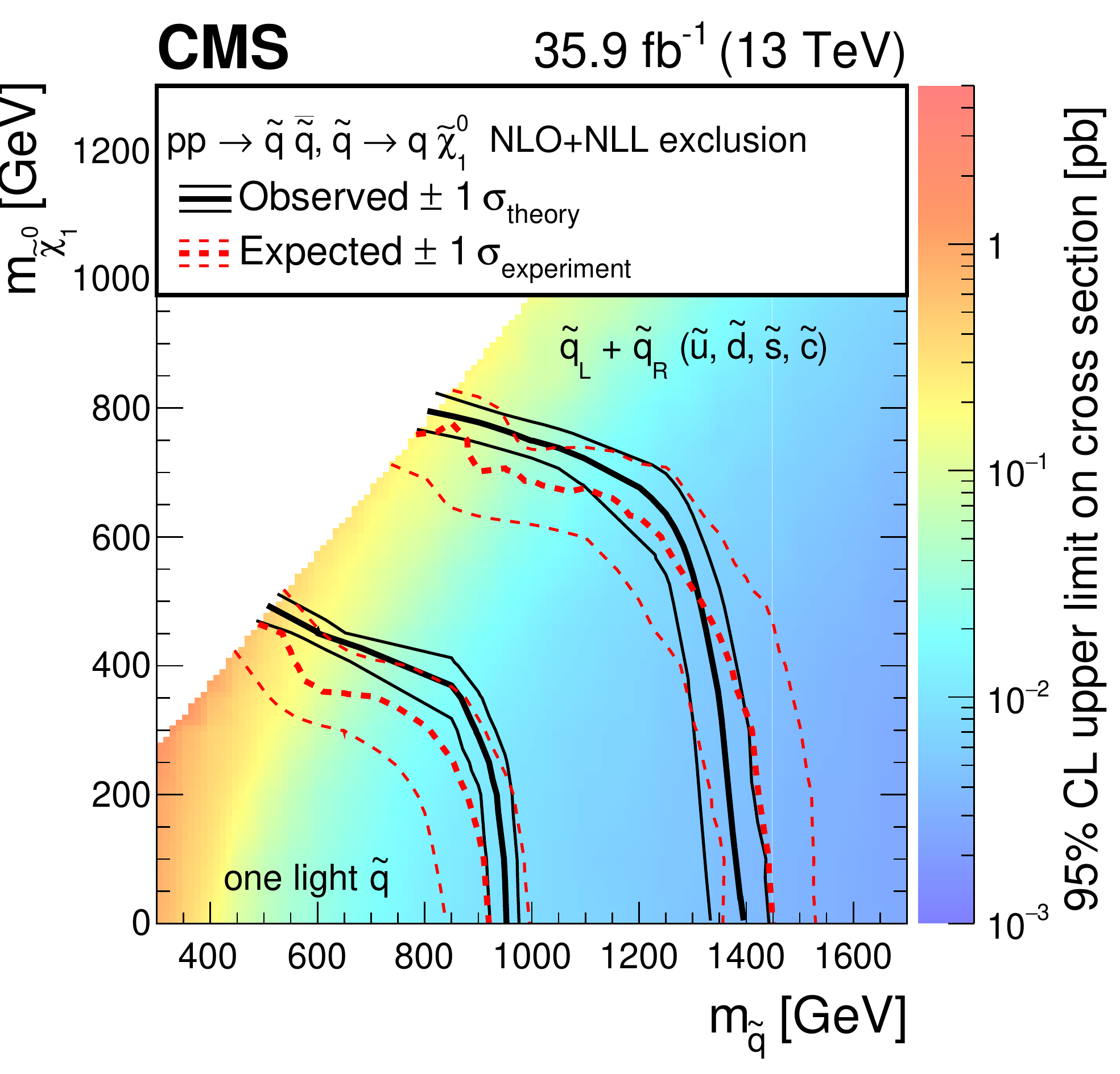}
    \caption{
      (Left) The 95\% CL upper limits on the production
      cross section for the (upper left) T2tt,
      (upper right) T2bb, and (lower) T2qq simplified models
      as a function of the squark and LSP masses \msquark and~\mlsp.
      The diagonal dotted line shown for the T2tt model corresponds to
      $\msquark-\mlsp=\mtop$.
      Note that for the T2tt model
      we do not present cross section upper limits
      in the unshaded diagonal region at low \mlsp
      for the reasons discussed in the text,
      and that  there is a small region corresponding to
      $\mstop\lesssim 230\GeV$ and $\mlsp\lesssim 20\GeV$
      that is not included in the NLO+NLL exclusion region.
      The results labeled ``one light \PSQ'' for the T2qq model are
      discussed in the text.
      The meaning of the curves is described in the
      Fig.~\ref{fig:limits-gluinos} caption.
    }
    \label{fig:limits-squarks}
\end{figure*}

Finally,
Fig.~\ref{fig:limits-squarks} shows
the results for the T2tt, T2bb, and T2qq models.
Based on the NLO+NLL cross sections,
squarks with masses up to 960, 990, and 1390\GeV,
respectively,
are excluded.
Note that for the T2tt model
we do not present cross section upper limits for
small values of \mlsp if
$\msquark-\mlsp\approx \mtop$,
corresponding to the unshaded diagonal region at low \mlsp
visible in Fig.~\ref{fig:limits-squarks} (upper left).
The reason for this is that signal events are essentially
indistinguishable from SM \ttbar events in this region,
rendering the signal event acceptance difficult to model.
Note also for the T2tt model that there is a small region
corresponding to
$\mstop\lesssim 230\GeV$ and $\mlsp\lesssim 20\GeV$
that is not excluded by the data.

In addition to the main T2qq model,
with four mass-degenerate squark flavors
(up, down, strange, and charm),
each arising from two different quark spin states,
Fig.~\ref{fig:limits-squarks} (lower) shows the results
should only one of these eight states
(``one light~\PSQ'')
be accessible at the LHC.
In this case,
the upper limit on the squark mass
based on the NLO+NLL cross section
is reduced to 950\GeV.

\section{Summary}
\label{sec:summary}

A search for gluino and squark pair production is presented
based on a sample of proton-proton collisions collected
at a center-of-mass energy of 13\TeV
with the CMS detector.
The search is performed in the multijet channel,
i.e., the visible reconstructed final state consists solely of jets.
The data correspond to an integrated luminosity of 35.9\fbinv.
Events are required to have at least two jets,
$\HT>300\GeV$, and $\MHT>300\GeV$,
where \HT is the scalar sum of jet transverse momenta~\pt.
The \MHT variable, used as a measure of missing transverse momentum,
is the magnitude of the vector \pt sum of jets.
Jets are required to have $\pt>30\GeV$
and to appear in the pseudorapidity range $\abs{\eta}<2.4$.

The data are examined in 174 exclusive four-dimensional
search regions defined by
the number of jets,
the number of tagged bottom quark jets,
\HT, and \MHT.
Background from standard model processes
is evaluated using control samples in the data.
We also provide results for 12 aggregated search regions,
to simplify use of our data by others.
The estimates of the standard model background are found to agree
with the observed numbers of events for all regions.

The results are interpreted in the context of simplified models.
We consider models in which pair-produced gluinos
each decay to a \ttbar pair and an undetected,
stable, lightest-supersymmetric-particle (LSP) neutralino {\PSGczDo}
(T1tttt model);
to a \bbbar pair and the {\PSGczDo}
(T1bbbb model);
to a light-flavored \qqbar pair and the {\PSGczDo}
(T1qqqq model);
to a light-flavored quark and antiquark and either the
second-lightest neutralino \PSGczDt
or the lightest chargino $\PSGcpm_1$,
followed by decay of the \PSGczDt ($\PSGcpm_1$)
to the \PSGczDo and an on- or off-shell
{\cPZ} ({$\PW^\pm$}) boson (T5qqqqVV model);
or to $\PAQt\PQb\PSGcp_1$
or $\PQt\PAQb\PSGcm_1$,
followed by the decay of the $\PSGcpm_1$ to the {\PSGczDo}
and an off-shell {\PW} boson (T1tbtb model).
To provide more model independence,
we also consider mixed scenarios in which a gluino
can decay to
$\ttbar\PSGczDo$,
$\bbbar\PSGczDo$,
$\PAQt\PQb\PSGcp_1$,
or
$\PQt\PAQb\PSGcm_1$
with various probabilities.
Beyond the models for gluino production,
we examine models for direct squark pair production.
We consider scenarios in which each
squark decays to a top quark and the {\PSGczDo}
(T2tt model);
to a bottom quark and the {\PSGczDo}
(T2bb model);
or to a light-flavored
(\cPqu, \cPqd, \cPqs, \cPqc)
quark and the {\PSGczDo}
(T2qq model).
We derive
upper limits at 95\% confidence level on
the model cross sections as a function of
the gluino and LSP masses,
or of the squark and LSP masses.

Using the predicted cross sections with next-to-leading-order plus
next-to-leading-logarithm accuracy as a reference,
95\% confidence level lower limits on the gluino mass
as large as 1800 to 1960\GeV are derived,
depending on the scenario.
The corresponding limits on the mass of directly produced
squarks range from 960 to 1390\GeV.
These results extend those from previous searches.

\begin{acknowledgments}
\hyphenation{Bundes-ministerium Forschungs-gemeinschaft Forschungs-zentren Rachada-pisek} We congratulate our colleagues in the CERN accelerator departments for the excellent performance of the LHC and thank the technical and administrative staffs at CERN and at other CMS institutes for their contributions to the success of the CMS effort. In addition, we gratefully acknowledge the computing centers and personnel of the Worldwide LHC Computing Grid for delivering so effectively the computing infrastructure essential to our analyses. Finally, we acknowledge the enduring support for the construction and operation of the LHC and the CMS detector provided by the following funding agencies: the Austrian Federal Ministry of Science, Research and Economy and the Austrian Science Fund; the Belgian Fonds de la Recherche Scientifique, and Fonds voor Wetenschappelijk Onderzoek; the Brazilian Funding Agencies (CNPq, CAPES, FAPERJ, and FAPESP); the Bulgarian Ministry of Education and Science; CERN; the Chinese Academy of Sciences, Ministry of Science and Technology, and National Natural Science Foundation of China; the Colombian Funding Agency (COLCIENCIAS); the Croatian Ministry of Science, Education and Sport, and the Croatian Science Foundation; the Research Promotion Foundation, Cyprus; the Secretariat for Higher Education, Science, Technology and Innovation, Ecuador; the Ministry of Education and Research, Estonian Research Council via IUT23-4 and IUT23-6 and European Regional Development Fund, Estonia; the Academy of Finland, Finnish Ministry of Education and Culture, and Helsinki Institute of Physics; the Institut National de Physique Nucl\'eaire et de Physique des Particules~/~CNRS, and Commissariat \`a l'\'Energie Atomique et aux \'Energies Alternatives~/~CEA, France; the Bundesministerium f\"ur Bildung und Forschung, Deutsche Forschungsgemeinschaft, and Helmholtz-Gemeinschaft Deutscher Forschungszentren, Germany; the General Secretariat for Research and Technology, Greece; the National Scientific Research Foundation, and National Innovation Office, Hungary; the Department of Atomic Energy and the Department of Science and Technology, India; the Institute for Studies in Theoretical Physics and Mathematics, Iran; the Science Foundation, Ireland; the Istituto Nazionale di Fisica Nucleare, Italy; the Ministry of Science, ICT and Future Planning, and National Research Foundation (NRF), Republic of Korea; the Lithuanian Academy of Sciences; the Ministry of Education, and University of Malaya (Malaysia); the Mexican Funding Agencies (BUAP, CINVESTAV, CONACYT, LNS, SEP, and UASLP-FAI); the Ministry of Business, Innovation and Employment, New Zealand; the Pakistan Atomic Energy Commission; the Ministry of Science and Higher Education and the National Science Centre, Poland; the Funda\c{c}\~ao para a Ci\^encia e a Tecnologia, Portugal; JINR, Dubna; the Ministry of Education and Science of the Russian Federation, the Federal Agency of Atomic Energy of the Russian Federation, Russian Academy of Sciences, the Russian Foundation for Basic Research and the Russian Competitiveness Program of NRNU ``MEPhI"; the Ministry of Education, Science and Technological Development of Serbia; the Secretar\'{\i}a de Estado de Investigaci\'on, Desarrollo e Innovaci\'on, Programa Consolider-Ingenio 2010, Plan de Ciencia, Tecnolog\'{i}a e Innovaci\'on 2013-2017 del Principado de Asturias and Fondo Europeo de Desarrollo Regional, Spain; the Swiss Funding Agencies (ETH Board, ETH Zurich, PSI, SNF, UniZH, Canton Zurich, and SER); the Ministry of Science and Technology, Taipei; the Thailand Center of Excellence in Physics, the Institute for the Promotion of Teaching Science and Technology of Thailand, Special Task Force for Activating Research and the National Science and Technology Development Agency of Thailand; the Scientific and Technical Research Council of Turkey, and Turkish Atomic Energy Authority; the National Academy of Sciences of Ukraine, and State Fund for Fundamental Researches, Ukraine; the Science and Technology Facilities Council, UK; the US Department of Energy, and the US National Science Foundation.

Individuals have received support from the Marie-Curie program and the European Research Council and EPLANET (European Union); the Leventis Foundation; the A. P. Sloan Foundation; the Alexander von Humboldt Foundation; the Belgian Federal Science Policy Office; the Fonds pour la Formation \`a la Recherche dans l'Industrie et dans l'Agriculture (FRIA-Belgium); the Agentschap voor Innovatie door Wetenschap en Technologie (IWT-Belgium); the Ministry of Education, Youth and Sports (MEYS) of the Czech Republic; the Council of Scientific and Industrial Research, India; the HOMING PLUS program of the Foundation for Polish Science, cofinanced from European Union, Regional Development Fund, the Mobility Plus program of the Ministry of Science and Higher Education, the National Science Center (Poland), contracts Harmonia 2014/14/M/ST2/00428, Opus 2014/13/B/ST2/02543, 2014/15/B/ST2/03998, and 2015/19/B/ST2/02861, Sonata-bis 2012/07/E/ST2/01406; the National Priorities Research Program by Qatar National Research Fund; the Programa Clar\'in-COFUND del Principado de Asturias; the Thalis and Aristeia programs cofinanced by EU-ESF and the Greek NSRF; the Rachadapisek Sompot Fund for Postdoctoral Fellowship, Chulalongkorn University and the Chulalongkorn Academic into Its 2nd Century Project Advancement Project (Thailand); and the Welch Foundation, contract C-1845.
\end{acknowledgments}

\clearpage
\bibliography{auto_generated}

\clearpage
\numberwithin{table}{section}
\appendix

\section{Selection efficiency for representative signal models}
\label{sec:cutflow}

Tables~\ref{tab:sel-eff-gg} and~\ref{tab:sel-eff-qq} present
cumulative selection efficiencies for representative
simplified models of gluino and squark pair production,
respectively.

\begin{table*}[hbt]
\topcaption{
Absolute cumulative efficiencies in \% for each step
of the event selection process
for representative models of gluino pair production.
The uncertainties are statistical.
Uncertainties reported as 0.0 correspond to values less than 0.05\%.
}
\centering
\begin{scotch}{>{$}l<{$}xxx}
\multicolumn{1}{c}{Selection} & \multicolumn{1}{r}{$\Pp\Pp\to\PSg\PSg, \PSg\to\ttbar\PSGczDo$}
     & \multicolumn{1}{r}{$\Pp\Pp\to\PSg\PSg, \PSg\to\bbbar\PSGczDo$}
     & \multicolumn{1}{r}{$\Pp\Pp\to\PSg\PSg, \PSg\to\qqbar\PSGczDo$} \\
\multicolumn{1}{c}{} & \multicolumn{1}{r}{$m_{\PSg}=1500\GeV$} & \multicolumn{1}{r}{$m_{\PSg}=1500\GeV$} & \multicolumn{1}{r}{$m_{\PSg}=1400\GeV$} \\
\multicolumn{1}{c}{} & \multicolumn{1}{r}{$m_{\PSGczDo}=100\GeV$} & \multicolumn{1}{r}{$m_{\PSGczDo}=100\GeV$} & \multicolumn{1}{r}{$m_{\PSGczDo}=100\GeV$} \\
\hline
\njets  \geq2 & 100.0 , 0.0 &100.0,  0.0 &100.0,  0.0 \\
\HT   >300\GeV & 100.0 , 0.0 &100.0,  0.0 &100.0,  0.0 \\
\MHT  >300\GeV & 76.7 , 0.3 &80.3,  0.4 &80.0,  0.3 \\
\nmuons =0 & 48.6 , 0.4 &79.8,  0.4 &80.0,  0.3 \\
\nisomuons=0 & 47.8 , 0.4 &79.6,  0.4 &79.9,  0.3 \\
\neles =0  & 30.7 , 0.3 &79.2,  0.4 &79.5,  0.3 \\
\nisoeles=0  & 29.7 , 0.3 &78.7,  0.4 &79.1,  0.3 \\
\nisohads =0 & 28.3 , 0.3 &78.0,  0.4 &78.3,  0.3 \\
\dpmht1 >0.5 & 27.7 , 0.3 &76.7,  0.4 &76.9,  0.3 \\
\dpmht2 >0.5 & 25.2 , 0.3 &69.2,  0.5 &69.8,  0.3 \\
\dpmht3 >0.3 & 23.7 , 0.3 &63.9,  0.5 &64.4,  0.3 \\
\dpmht4>0.3 & 22.1 , 0.3 &58.6,  0.5 &59.4,  0.3 \\
\multicolumn{1}{c}{Event quality filter} & 21.8 , 0.3 &57.7,  0.5 &58.7,  0.3 \\
\hline
\multicolumn{1}{c}{Selection} & \multicolumn{1}{r}{$\Pp\Pp\to\PSg\PSg, \PSg\to\ttbar\PSGczDo$}
     & \multicolumn{1}{r}{$\Pp\Pp\to\PSg\PSg, \PSg\to\bbbar\PSGczDo$}
     & \multicolumn{1}{r}{$\Pp\Pp\to\PSg\PSg, \PSg\to\qqbar\PSGczDo$} \\
\multicolumn{1}{c}{} & \multicolumn{1}{r}{$m_{\PSg}=1200\GeV$} & \multicolumn{1}{r}{$m_{\PSg}=1000\GeV$} & \multicolumn{1}{r}{$m_{\PSg}=1000\GeV$} \\
\multicolumn{1}{c}{} & \multicolumn{1}{r}{$m_{\PSGczDo}=800\GeV$} & \multicolumn{1}{r}{$m_{\PSGczDo}=900\GeV$} & \multicolumn{1}{r}{$m_{\PSGczDo}=800\GeV$} \\
\hline
\njets      \geq2 & 100.0 , 0.0 &92.5,  0.1 &99.6,  0.0 \\
\HT         >300\GeV & 99.0 , 0.0 &38.6,  0.1 &81.3,  0.1 \\
\MHT        >300\GeV & 14.9 , 0.1 &14.1,  0.1 &19.1,  0.1 \\
\nmuons     =0 & 9.6 , 0.1 &13.9,  0.1 &19.1,  0.1 \\
\nisomuons  =0 & 9.2 , 0.1 &13.6,  0.1 &19.1,  0.1 \\
\neles      =0 & 6.2 , 0.1 &13.4,  0.1 &19.0,  0.1 \\
\nisoeles   =0 & 5.8 , 0.1 &13.1,  0.1 &18.8,  0.1 \\
\nisohads   =0 & 5.3 , 0.1 &12.8,  0.1 &18.4,  0.1 \\
\dpmht1     >0.5 & 5.3 , 0.1 &12.8,  0.1 &18.4,  0.1 \\
\dpmht2     >0.5 & 4.5 , 0.1 &11.4,  0.1 &16.9,  0.1 \\
\dpmht3     >0.3 & 4.0 , 0.1 &10.4,  0.1 &15.8,  0.1 \\
\dpmht4     >0.3 & 3.6 , 0.1 &9.6,  0.1 &14.8,  0.1 \\
\multicolumn{1}{c}{Event quality filter} & 3.5 , 0.1 &9.4,  0.1 &14.6,  0.1 \\
\end{scotch}
\label{tab:sel-eff-gg}
\end{table*}

\begin{table*}[htb]
\topcaption{
Absolute cumulative efficiencies in \% for each step
of the event selection process
for representative models of squark pair production.
The uncertainties are statistical.
Uncertainties reported as 0.0 correspond to values less than 0.05\%.
}
\centering
\begin{scotch}{>{$}l<{$}xxx}
\multicolumn{1}{c}{Selection} & \multicolumn{1}{r}{$\Pp\Pp\to\sTop\PASQt,\sTop\to\cPqt\PSGczDo$}
     & \multicolumn{1}{r}{$\Pp\Pp\to\sBot\PASQb, \sBot\to\cPqb\PSGczDo$}
     & \multicolumn{1}{r}{$\Pp\Pp\to\sQua\PASQ, \sQua\to\cPq\PSGczDo$} \\
\multicolumn{1}{c}{} & \multicolumn{1}{r}{$m_{\sTop}=700\GeV$} & \multicolumn{1}{r}{$m_{\sBot}=650\GeV$} & \multicolumn{1}{r}{$m_{\sQua}=1000\GeV$} \\
\multicolumn{1}{c}{} & \multicolumn{1}{r}{$m_{\PSGczDo}=50\GeV$} & \multicolumn{1}{r}{$m_{\PSGczDo}=1\GeV$} & \multicolumn{1}{r}{$m_{\PSGczDo}=100\GeV$} \\
\hline
\njets      \geq2 &99.8,  0.0 &98.2,  0.1 &98.9,  0.1 \\
\HT         >300\GeV &96.4,  0.1 &95.4,  0.1 &98.6,  0.1 \\
\MHT        >300\GeV &57.8,  0.3 &59.8,  0.2 &80.0,  0.3 \\
\nmuons     =0 &46.6,  0.3 &59.6,  0.2 &79.9,  0.3 \\
\nisomuons  =0 &46.1,  0.3 &59.5,  0.2 &79.8,  0.3 \\
\neles      =0 &37.4,  0.3 &59.2,  0.2 &79.6,  0.3 \\
\nisoeles   =0 &36.9,  0.3 &59.0,  0.2 &79.3,  0.3 \\
\nisohads   =0 &35.8,  0.3 &58.5,  0.2 &78.7,  0.3 \\
\dpmht1     >0.5 &35.7,  0.3 &58.4,  0.2 &78.6,  0.3 \\
\dpmht2     >0.5 &34.0,  0.3 &55.7,  0.2 &74.5,  0.3 \\
\dpmht3     >0.3 &33.1,  0.3 &53.3,  0.2 &70.6,  0.3 \\
\dpmht4     >0.3 &31.8,  0.3 &51.6,  0.2 &67.9,  0.3 \\
\multicolumn{1}{c}{Event quality filter} & 31.4,  0.3 &50.8,  0.3 &67.1,  0.3 \\
\hline
\multicolumn{1}{c}{Selection} & \multicolumn{1}{r}{$\Pp\Pp\to\sTop\PASQt, \sTop\to\cPqt\PSGczDo$}
     & \multicolumn{1}{r}{$\Pp\Pp\to\sBot\PASQb, \sBot\to\cPqb\PSGczDo$}
      & \multicolumn{1}{r}{$\Pp\Pp\to\sQua\PASQ, \sQua\to\cPq\PSGczDo$} \\
\multicolumn{1}{c}{} &  \multicolumn{1}{r}{$m_{\sTop}=300\GeV$} & \multicolumn{1}{r}{$m_{\sBot}=500\GeV$} & \multicolumn{1}{r}{$m_{\sQua}=700\GeV$} \\
\multicolumn{1}{c}{} &  \multicolumn{1}{r}{$m_{\PSGczDo}=200\GeV$} & \multicolumn{1}{r}{$m_{\PSGczDo}=300\GeV$} & \multicolumn{1}{r}{$m_{\PSGczDo}=400\GeV$} \\
\hline
\njets      \geq2 &86.9,  0.0 &96.0,  0.1 &98.0,  0.0 \\
\HT         >300\GeV &23.3,  0.0 &68.0,  0.1 &91.3,  0.1 \\
\MHT        >300\GeV &2.84,  0.0 &15.6,  0.1 &43.8,  0.1 \\
\nmuons     =0 &2.16,  0.0 &15.6,  0.1 &43.8,  0.1 \\
\nisomuons  =0 &2.10,  0.0 &15.5,  0.1 &43.7,  0.1 \\
\neles      =0 &1.60,  0.0 &15.4,  0.1 &43.5,  0.1 \\
\nisoeles   =0 &1.52,  0.0 &15.3,  0.1 &43.4,  0.1 \\
\nisohads   =0 &1.41,  0.0 &15.2,  0.1 &43.0,  0.1 \\
\dpmht1   >0.5 &1.40,  0.0 &15.1,  0.1 &42.9,  0.1 \\
\dpmht2   >0.5 &1.03,  0.0 &14.1,  0.1 &41.1,  0.1 \\
\dpmht3   >0.3 &0.85,  0.0 &13.5,  0.1 &39.6,  0.1 \\
\dpmht4   >0.3 &0.73,  0.0 &13.1,  0.1 &38.4,  0.1 \\
\multicolumn{1}{c}{Event quality filter}&  0.72,  0.0 &12.9,  0.1 &37.9,  0.1 \\
\end{scotch}
\label{tab:sel-eff-qq}
\end{table*}

\section{Prefit background predictions}
\label{sec:prefit}

Tables~\ref{tab:pre-fit-results-nj0}--\ref{tab:pre-fit-results-nj4}
present the prefit predictions
for the number of standard model background events
in each of the 174 search regions of the analysis,
along with the observed numbers of events,
where ``prefit'' means there is no constraint from the likelihood fit.
The corresponding information for the 12 aggregate search
regions is presented in Table~\ref{tab:pre-fit-results-asr}.

\begin{table*}[htbp]
\renewcommand{\arraystretch}{1.25}
\centering
\topcaption{
Observed numbers of events and prefit background predictions
in the $\njets=2$ search regions.
The first uncertainty is statistical and second systematic.
}
\label{tab:pre-fit-results-nj0}
\resizebox{\textwidth}{!}{
\begin{scotch}{ ccccccccccc }
Bin & $\MHT$ [\GeVns{}] & $\HT$ [\GeVns{}] & $\njets$ & $\nbjets$ & Lost-$\Pe/\Pgm$ & $\Pgt\to \text{had}$ & $\cPZ\to\Pgn\cPagn$ & QCD & Total pred. & Obs. \\ \hline
1 & 300--350 & 300--500 & 2 & 0 & $4069^{+67+320}_{-67-320}$ & $2744^{+37+510}_{-37-500}$ & $13231^{+67+760}_{-66-740}$ & $326^{+12+170}_{-12-120}$ & $20370^{+120+980}_{-120-960}$ & 21626 \\
2 & 300--350 & 500--1000 & 2 & 0 & $326^{+22+36}_{-22-36}$ & $226^{+11+43}_{-11-42}$ & $944^{+18+55}_{-18-54}$ & $45^{+ 2+24}_{- 2-17}$ & $1541^{+37+82}_{-37-79}$ & 1583 \\
3 & 300--350 & $>$1000 & 2 & 0 & $15.2^{+5.8+2.3}_{-5.1-2.3}$ & $8.7^{+2.1+2.1}_{-2.0-2.1}$ & $50.9^{+4.5+4.4}_{-4.1-3.8}$ & $1.57^{+0.16+0.84}_{-0.16-0.61}$ & $76.3^{+9.1+5.5}_{-8.2-5.0}$ & 102 \\
4 & 350--500 & 350--500 & 2 & 0 & $2049^{+46+160}_{-46-160}$ & $1553^{+27+290}_{-27-290}$ & $9347^{+57+540}_{-57-520}$ & $126^{+ 4+67}_{- 4-48}$ & $13076^{+93+630}_{-93-620}$ & 14019 \\
5 & 350--500 & 500--1000 & 2 & 0 & $631^{+25+54}_{-25-54}$ & $439^{+14+84}_{-14-84}$ & $2502^{+30+150}_{-30-140}$ & $43^{+ 7+22}_{- 7-16}$ & $3615^{+49+180}_{-49-170}$ & 3730 \\ \hline
6 & 350--500 & $>$1000 & 2 & 0 & $13.5^{+4.9+1.9}_{-4.3-1.9}$ & $13.4^{+2.4+2.6}_{-2.3-2.6}$ & $94.0^{+6.2+7.9}_{-5.8-6.9}$ & $1.30^{+0.06+0.68}_{-0.06-0.49}$ & $122.1^{+9.5+8.6}_{-8.8-7.6}$ & 139 \\
7 & 500--750 & 500--1000 & 2 & 0 & $303^{+17+29}_{-17-29}$ & $247^{+10+48}_{-10-47}$ & $2328^{+30+170}_{-29-160}$ & $4.5^{+0.1+2.4}_{-0.1-1.7}$ & $2883^{+40+180}_{-40-170}$ & 3018 \\
8 & 500--750 & $>$1000 & 2 & 0 & $5.8^{+2.7+1.5}_{-2.2-1.5}$ & $5.3^{+1.4+1.3}_{-1.3-1.3}$ & $66.2^{+5.4+5.3}_{-5.0-5.1}$ & $0.03^{+0.02+0.02}_{-0.02-0.01}$ & $77.3^{+6.8+5.7}_{-6.1-5.4}$ & 96 \\
9 & $>$750 & 750--1500 & 2 & 0 & $17.3^{+4.5+3.0}_{-4.1-3.0}$ & $17.4^{+2.5+4.5}_{-2.4-4.5}$ & $295^{+11+41}_{-11-38}$ & $0.35^{+0.06+0.18}_{-0.06-0.13}$ & $330^{+13+42}_{-12-38}$ & 272 \\
10 & $>$750 & $>$1500 & 2 & 0 & $0.0^{+1.8+0.0}_{-0.0-0.0}$ & $0.38^{+0.54+0.09}_{-0.29-0.09}$ & $12.6^{+3.0+2.1}_{-2.4-1.9}$ & $0.01^{+0.01+0.00}_{-0.01-0.00}$ & $13.0^{+3.8+2.1}_{-2.5-1.9}$ & 12 \\ \hline
11 & 300--350 & 300--500 & 2 & 1 & $370^{+21+31}_{-21-31}$ & $288^{+11+63}_{-11-63}$ & $1361^{+ 7+140}_{- 7-140}$ & $44^{+ 6+25}_{- 6-17}$ & $2063^{+33+160}_{-33-160}$ & 1904 \\
12 & 300--350 & 500--1000 & 2 & 1 & $51^{+10+ 7}_{-10- 7}$ & $31.6^{+4.2+7.2}_{-4.2-7.2}$ & $97^{+ 2+10}_{- 2-10}$ & $6.7^{+2.7+3.7}_{-2.7-2.5}$ & $186^{+15+15}_{-14-14}$ & 186 \\
13 & 300--350 & $>$1000 & 2 & 1 & $1.1^{+2.3+0.2}_{-1.1-0.0}$ & $2.0^{+1.1+0.5}_{-1.0-0.5}$ & $5.23^{+0.46+0.63}_{-0.42-0.59}$ & $0.33^{+0.02+0.18}_{-0.02-0.13}$ & $8.7^{+3.4+0.9}_{-2.1-0.8}$ & 13 \\
14 & 350--500 & 350--500 & 2 & 1 & $215^{+16+19}_{-16-19}$ & $179^{+ 9+39}_{- 9-39}$ & $962^{+ 6+99}_{- 6-98}$ & $20^{+ 2+11}_{- 2- 8}$ & $1376^{+26+110}_{-26-110}$ & 1212 \\
15 & 350--500 & 500--1000 & 2 & 1 & $69.8^{+9.9+7.5}_{-9.8-7.5}$ & $43.3^{+4.4+9.7}_{-4.4-9.6}$ & $257^{+ 3+27}_{- 3-26}$ & $8.5^{+3.0+4.8}_{-3.0-3.2}$ & $379^{+15+30}_{-15-29}$ & 409 \\ \hline
16 & 350--500 & $>$1000 & 2 & 1 & $3.7^{+2.5+0.7}_{-1.9-0.7}$ & $3.1^{+1.1+0.9}_{-1.0-0.9}$ & $9.7^{+0.6+1.2}_{-0.6-1.1}$ & $0.13^{+0.04+0.07}_{-0.04-0.05}$ & $16.6^{+3.7+1.6}_{-3.0-1.6}$ & 27 \\
17 & 500--750 & 500--1000 & 2 & 1 & $28.9^{+5.8+3.3}_{-5.6-3.3}$ & $26.0^{+2.9+5.8}_{-2.9-5.8}$ & $240^{+ 3+27}_{- 3-26}$ & $1.48^{+0.18+0.83}_{-0.18-0.56}$ & $296^{+ 9+28}_{- 9-27}$ & 321 \\
18 & 500--750 & $>$1000 & 2 & 1 & $5.1^{+6.2+1.6}_{-4.1-1.6}$ & $0.36^{+0.55+0.12}_{-0.30-0.12}$ & $6.81^{+0.56+0.80}_{-0.52-0.78}$ & $0.03^{+0.03+0.02}_{-0.03-0.00}$ & $12.3^{+6.8+1.8}_{-4.5-1.7}$ & 14 \\
19 & $>$750 & 750--1500 & 2 & 1 & $3.8^{+2.2+0.8}_{-1.7-0.8}$ & $4.1^{+1.5+1.1}_{-1.4-1.1}$ & $30.4^{+1.1+5.0}_{-1.1-4.7}$ & $0.10^{+0.03+0.06}_{-0.03-0.04}$ & $38.4^{+3.9+5.1}_{-3.3-4.8}$ & 31 \\
20 & $>$750 & $>$1500 & 2 & 1 & $0.0^{+1.4+0.0}_{-0.0-0.0}$ & $0.34^{+0.51+0.13}_{-0.22-0.13}$ & $1.29^{+0.31+0.24}_{-0.25-0.23}$ & $0.00^{+0.01+0.00}_{-0.00-0.00}$ & $1.6^{+2.0+0.3}_{-0.3-0.3}$ & 1 \\ \hline
21 & 300--350 & 300--500 & 2 & 2 & $14.1^{+4.5+2.6}_{-4.0-2.6}$ & $12.9^{+2.3+2.8}_{-2.2-2.8}$ & $49^{+ 0+17}_{- 0-17}$ & $3.0^{+0.8+3.6}_{-0.8-2.1}$ & $79^{+ 7+18}_{- 6-18}$ & 122 \\
22 & 300--350 & 500--1000 & 2 & 2 & $2.8^{+2.4+0.9}_{-1.7-0.9}$ & $2.0^{+1.1+1.0}_{-0.9-1.0}$ & $3.5^{+0.1+1.2}_{-0.1-1.2}$ & $0.57^{+0.17+0.69}_{-0.17-0.40}$ & $8.9^{+3.5+2.0}_{-2.6-1.9}$ & 11 \\
23 & 300--350 & $>$1000 & 2 & 2 & $0.0^{+2.2+0.0}_{-0.0-0.0}$ & $0.00^{+0.46+0.00}_{-0.00-0.00}$ & $0.19^{+0.02+0.07}_{-0.01-0.07}$ & $0.03^{+0.01+0.04}_{-0.01-0.02}$ & $0.2^{+2.6+0.1}_{-0.0-0.1}$ & 0 \\
24 & 350--500 & 350--500 & 2 & 2 & $11.4^{+4.5+2.5}_{-3.9-2.5}$ & $6.3^{+1.7+2.1}_{-1.6-2.1}$ & $35^{+ 0+12}_{- 0-12}$ & $1.0^{+0.5+1.2}_{-0.5-0.6}$ & $53^{+ 6+13}_{- 6-13}$ & 84 \\
25 & 350--500 & 500--1000 & 2 & 2 & $6.1^{+2.9+1.5}_{-2.4-1.5}$ & $2.9^{+1.2+0.8}_{-1.1-0.8}$ & $9.3^{+0.1+3.3}_{-0.1-3.3}$ & $0.44^{+0.05+0.52}_{-0.05-0.39}$ & $18.7^{+4.1+3.8}_{-3.5-3.7}$ & 23 \\ \hline
26 & 350--500 & $>$1000 & 2 & 2 & $0.0^{+1.1+0.0}_{-0.0-0.0}$ & $0.00^{+0.46+0.00}_{-0.00-0.00}$ & $0.35^{+0.02+0.13}_{-0.02-0.13}$ & $0.06^{+0.04+0.08}_{-0.04-0.02}$ & $0.4^{+1.5+0.1}_{-0.0-0.1}$ & 2 \\
27 & 500--750 & 500--1000 & 2 & 2 & $1.4^{+2.9+0.4}_{-1.4-0.0}$ & $2.03^{+0.84+0.61}_{-0.70-0.61}$ & $8.6^{+0.1+3.1}_{-0.1-3.1}$ & $0.03^{+0.01+0.04}_{-0.01-0.03}$ & $12.1^{+3.7+3.2}_{-2.1-3.2}$ & 16 \\
28 & 500--750 & $>$1000 & 2 & 2 & $0.0^{+2.2+0.0}_{-0.0-0.0}$ & $0.00^{+0.46+0.00}_{-0.00-0.00}$ & $0.24^{+0.02+0.09}_{-0.02-0.09}$ & $0.00^{+0.01+0.00}_{-0.00-0.00}$ & $0.2^{+2.7+0.1}_{-0.0-0.1}$ & 0 \\
29 & $>$750 & 750--1500 & 2 & 2 & $0.0^{+1.6+0.0}_{-0.0-0.0}$ & $0.07^{+0.46+0.07}_{-0.04-0.06}$ & $1.09^{+0.04+0.41}_{-0.04-0.41}$ & $0.01^{+0.01+0.01}_{-0.01-0.00}$ & $1.2^{+2.1+0.4}_{-0.1-0.4}$ & 4 \\
30 & $>$750 & $>$1500 & 2 & 2 & $0.0^{+2.0+0.0}_{-0.0-0.0}$ & $0.00^{+0.46+0.00}_{-0.00-0.00}$ & $0.05^{+0.01+0.02}_{-0.01-0.02}$ & $0.00^{+0.01+0.00}_{-0.00-0.00}$ & $0.0^{+2.5+0.0}_{-0.0-0.0}$ & 0 \\ 
\end{scotch}}
\end{table*}

\begin{table*}[htbp]
\renewcommand{\arraystretch}{1.25}
\centering
\topcaption{
Observed numbers of events and prefit background predictions
in the $3\leq\njets\leq4$ search regions.
The first uncertainty is statistical and second systematic.
}
\label{tab:pre-fit-results-nj1}
\resizebox{\textwidth}{!}{
\begin{scotch}{ ccccccccccc }
Bin & $\MHT$ [\GeVns{}] & $\HT$ [\GeVns{}] & $\njets$ & $\nbjets$ & Lost-$\Pe/\Pgm$ & $\Pgt\to\text{had}$ & $\cPZ\to\Pgn\cPagn$ & QCD & Total pred. & Obs. \\ \hline
31 & 300--350 & 300--500 & 3--4 & 0 & $2830^{+45+200}_{-45-200}$ & $2152^{+29+160}_{-29-150}$ & $8353^{+52+480}_{-52-470}$ & $273^{+68+120}_{-68-100}$ & $13608^{+110+560}_{-110-540}$ & 14520 \\
32 & 300--350 & 500--1000 & 3--4 & 0 & $1125^{+25+120}_{-25-120}$ & $909^{+18+100}_{-18-100}$ & $2487^{+29+140}_{-28-140}$ & $119^{+ 8+51}_{- 8-45}$ & $4640^{+52+220}_{-52-210}$ & 4799 \\
33 & 300--350 & $>$1000 & 3--4 & 0 & $72.7^{+7.1+6.1}_{-7.1-6.1}$ & $65.3^{+5.2+6.4}_{-5.2-6.3}$ & $176^{+ 8+14}_{- 8-12}$ & $41^{+ 2+18}_{- 2-16}$ & $356^{+15+24}_{-15-22}$ & 354 \\
34 & 350--500 & 350--500 & 3--4 & 0 & $1439^{+37+110}_{-37-110}$ & $930^{+19+120}_{-19-110}$ & $5014^{+41+280}_{-41-280}$ & $114^{+ 6+48}_{- 6-43}$ & $7496^{+70+330}_{-69-320}$ & 7973 \\
35 & 350--500 & 500--1000 & 3--4 & 0 & $1402^{+27+140}_{-27-140}$ & $1253^{+22+120}_{-22-120}$ & $4811^{+40+270}_{-40-260}$ & $80^{+ 9+34}_{- 9-31}$ & $7547^{+65+330}_{-64-320}$ & 7735 \\ \hline
36 & 350--500 & $>$1000 & 3--4 & 0 & $103^{+ 8+11}_{- 8-11}$ & $77.0^{+5.9+7.6}_{-5.9-7.5}$ & $303^{+11+24}_{-10-21}$ & $24^{+ 1+10}_{- 1- 9}$ & $506^{+18+30}_{-17-26}$ & 490 \\
37 & 500--750 & 500--1000 & 3--4 & 0 & $339^{+15+33}_{-15-33}$ & $297^{+10+26}_{-10-26}$ & $2143^{+28+150}_{-28-140}$ & $5.5^{+0.2+2.3}_{-0.2-2.1}$ & $2785^{+37+160}_{-37-150}$ & 2938 \\
38 & 500--750 & $>$1000 & 3--4 & 0 & $33.8^{+4.4+3.6}_{-4.3-3.6}$ & $30.5^{+3.4+2.9}_{-3.4-2.9}$ & $219^{+10+16}_{- 9-15}$ & $1.29^{+0.53+0.55}_{-0.53-0.49}$ & $284^{+12+17}_{-12-16}$ & 303 \\
39 & $>$750 & 750--1500 & 3--4 & 0 & $28.2^{+4.4+3.7}_{-4.3-3.7}$ & $26.0^{+2.9+3.4}_{-2.9-3.4}$ & $319^{+11+44}_{-11-40}$ & $0.32^{+0.03+0.14}_{-0.03-0.12}$ & $373^{+14+44}_{-13-41}$ & 334 \\
40 & $>$750 & $>$1500 & 3--4 & 0 & $2.9^{+2.0+0.7}_{-1.5-0.7}$ & $1.38^{+0.66+0.17}_{-0.48-0.17}$ & $27.8^{+3.9+4.1}_{-3.5-3.8}$ & $0.10^{+0.01+0.04}_{-0.01-0.04}$ & $32.2^{+4.8+4.2}_{-4.0-3.9}$ & 46 \\ \hline
41 & 300--350 & 300--500 & 3--4 & 1 & $746^{+25+55}_{-25-55}$ & $627^{+15+48}_{-15-47}$ & $1235^{+ 8+130}_{- 8-120}$ & $59^{+ 4+24}_{- 4-22}$ & $2667^{+41+150}_{-41-150}$ & 2677 \\
42 & 300--350 & 500--1000 & 3--4 & 1 & $296^{+15+25}_{-15-25}$ & $262^{+ 9+27}_{- 9-27}$ & $385^{+ 4+39}_{- 4-39}$ & $38^{+ 4+15}_{- 4-14}$ & $981^{+24+56}_{-24-56}$ & 1048 \\
43 & 300--350 & $>$1000 & 3--4 & 1 & $20.8^{+4.1+2.1}_{-4.0-2.1}$ & $19.0^{+2.6+1.8}_{-2.5-1.8}$ & $27.6^{+1.3+3.2}_{-1.2-3.0}$ & $11.4^{+0.8+4.7}_{-0.8-4.4}$ & $78.8^{+6.9+6.3}_{-6.6-6.0}$ & 92 \\
44 & 350--500 & 350--500 & 3--4 & 1 & $321^{+17+25}_{-17-25}$ & $263^{+10+22}_{-10-21}$ & $738^{+ 6+74}_{- 6-74}$ & $22.3^{+1.4+9.1}_{-1.4-8.5}$ & $1343^{+28+82}_{-28-81}$ & 1332 \\
45 & 350--500 & 500--1000 & 3--4 & 1 & $329^{+14+26}_{-14-26}$ & $324^{+11+26}_{-11-26}$ & $737^{+ 6+74}_{- 6-74}$ & $17.6^{+3.4+7.2}_{-3.4-6.7}$ & $1407^{+26+83}_{-26-83}$ & 1515 \\ \hline
46 & 350--500 & $>$1000 & 3--4 & 1 & $20.4^{+4.0+2.0}_{-3.8-2.0}$ & $19.9^{+2.9+1.8}_{-2.9-1.7}$ & $47.5^{+1.7+5.5}_{-1.6-5.1}$ & $5.7^{+0.5+2.3}_{-0.5-2.2}$ & $93.4^{+7.1+6.5}_{-6.9-6.2}$ & 113 \\
47 & 500--750 & 500--1000 & 3--4 & 1 & $69.7^{+7.4+6.6}_{-7.3-6.6}$ & $56.0^{+4.1+5.0}_{-4.1-4.9}$ & $322^{+ 4+35}_{- 4-35}$ & $1.34^{+0.10+0.55}_{-0.10-0.51}$ & $449^{+12+36}_{-12-36}$ & 472 \\
48 & 500--750 & $>$1000 & 3--4 & 1 & $15.3^{+3.4+1.9}_{-3.3-1.9}$ & $7.0^{+1.4+0.7}_{-1.4-0.7}$ & $34.4^{+1.5+3.8}_{-1.4-3.8}$ & $0.38^{+0.14+0.16}_{-0.14-0.15}$ & $57.0^{+5.1+4.4}_{-4.9-4.3}$ & 57 \\
49 & $>$750 & 750--1500 & 3--4 & 1 & $3.3^{+1.5+0.5}_{-1.3-0.5}$ & $4.8^{+1.3+0.8}_{-1.2-0.8}$ & $48.5^{+1.7+7.9}_{-1.7-7.3}$ & $0.13^{+0.01+0.05}_{-0.01-0.05}$ & $56.8^{+3.3+7.9}_{-3.0-7.4}$ & 61 \\
50 & $>$750 & $>$1500 & 3--4 & 1 & $1.0^{+1.2+0.3}_{-0.7-0.3}$ & $0.77^{+0.75+0.16}_{-0.59-0.16}$ & $4.40^{+0.62+0.75}_{-0.55-0.71}$ & $0.03^{+0.01+0.01}_{-0.01-0.01}$ & $6.2^{+2.0+0.8}_{-1.4-0.8}$ & 8 \\ \hline
51 & 300--350 & 300--500 & 3--4 & 2 & $137^{+11+11}_{-11-11}$ & $133^{+ 7+11}_{- 7-11}$ & $145^{+ 1+26}_{- 1-26}$ & $9.0^{+1.1+3.9}_{-1.1-3.4}$ & $424^{+18+31}_{-17-31}$ & 464 \\
52 & 300--350 & 500--1000 & 3--4 & 2 & $92.3^{+9.1+9.5}_{-9.0-9.5}$ & $85.6^{+5.7+7.5}_{-5.7-7.4}$ & $53.0^{+0.6+9.6}_{-0.6-9.6}$ & $3.8^{+1.2+1.6}_{-1.2-1.4}$ & $235^{+15+16}_{-15-15}$ & 227 \\
53 & 300--350 & $>$1000 & 3--4 & 2 & $3.4^{+2.2+0.8}_{-1.7-0.8}$ & $2.41^{+0.91+0.50}_{-0.78-0.50}$ & $3.95^{+0.18+0.75}_{-0.17-0.73}$ & $2.23^{+0.18+0.96}_{-0.18-0.86}$ & $12.0^{+3.1+1.6}_{-2.5-1.5}$ & 17 \\
54 & 350--500 & 350--500 & 3--4 & 2 & $39.6^{+6.1+3.8}_{-5.9-3.8}$ & $39.8^{+3.9+3.8}_{-3.8-3.8}$ & $84^{+ 1+15}_{- 1-15}$ & $2.7^{+0.6+1.1}_{-0.6-1.0}$ & $166^{+10+16}_{-10-16}$ & 208 \\
55 & 350--500 & 500--1000 & 3--4 & 2 & $83.9^{+8.2+7.8}_{-8.1-7.8}$ & $69.4^{+4.9+5.9}_{-4.9-5.8}$ & $97^{+ 1+18}_{- 1-17}$ & $3.1^{+0.2+1.3}_{-0.2-1.2}$ & $254^{+13+20}_{-13-20}$ & 286 \\ \hline
56 & 350--500 & $>$1000 & 3--4 & 2 & $6.2^{+4.0+1.0}_{-3.6-1.0}$ & $3.8^{+1.1+0.6}_{-1.0-0.6}$ & $6.8^{+0.2+1.3}_{-0.2-1.3}$ & $0.95^{+0.16+0.41}_{-0.16-0.36}$ & $17.7^{+5.2+1.8}_{-4.6-1.8}$ & 25 \\
57 & 500--750 & 500--1000 & 3--4 & 2 & $11.8^{+3.3+2.0}_{-3.1-2.0}$ & $10.5^{+1.8+1.6}_{-1.7-1.6}$ & $39.7^{+0.5+7.4}_{-0.5-7.3}$ & $0.22^{+0.04+0.09}_{-0.04-0.08}$ & $62.1^{+5.1+7.8}_{-4.8-7.7}$ & 64 \\
58 & 500--750 & $>$1000 & 3--4 & 2 & $2.6^{+2.3+0.6}_{-1.6-0.6}$ & $2.9^{+1.5+0.6}_{-1.5-0.6}$ & $4.90^{+0.21+0.92}_{-0.21-0.91}$ & $0.10^{+0.03+0.04}_{-0.03-0.04}$ & $10.5^{+3.8+1.2}_{-3.1-1.2}$ & 13 \\
59 & $>$750 & 750--1500 & 3--4 & 2 & $0.0^{+1.1+0.0}_{-0.0-0.0}$ & $0.32^{+0.48+0.09}_{-0.13-0.09}$ & $6.3^{+0.2+1.4}_{-0.2-1.3}$ & $0.03^{+0.02+0.01}_{-0.02-0.01}$ & $6.6^{+1.6+1.4}_{-0.3-1.3}$ & 4 \\
60 & $>$750 & $>$1500 & 3--4 & 2 & $0.0^{+1.1+0.0}_{-0.0-0.0}$ & $0.03^{+0.46+0.01}_{-0.02-0.01}$ & $0.65^{+0.09+0.15}_{-0.08-0.14}$ & $0.01^{+0.01+0.01}_{-0.01-0.00}$ & $0.7^{+1.6+0.1}_{-0.1-0.1}$ & 1 \\ \hline
61 & 300--350 & 300--500 & 3--4 & $\geq$3 & $6.4^{+2.8+0.7}_{-2.3-0.7}$ & $10.3^{+1.9+2.7}_{-1.9-2.7}$ & $5.0^{+0.0+2.8}_{-0.0-2.8}$ & $0.35^{+0.18+0.42}_{-0.18-0.16}$ & $22.0^{+4.7+3.9}_{-4.2-3.9}$ & 27 \\
62 & 300--350 & 500--1000 & 3--4 & $\geq$3 & $4.9^{+2.7+0.6}_{-2.2-0.6}$ & $6.2^{+1.4+1.7}_{-1.3-1.7}$ & $2.5^{+0.0+1.4}_{-0.0-1.4}$ & $0.75^{+0.52+0.90}_{-0.52-0.24}$ & $14.4^{+4.2+2.4}_{-3.6-2.2}$ & 20 \\
63 & 300--350 & $>$1000 & 3--4 & $\geq$3 & $0.0^{+1.1+0.0}_{-0.0-0.0}$ & $0.94^{+0.87+0.44}_{-0.74-0.44}$ & $0.21^{+0.01+0.12}_{-0.01-0.12}$ & $1.6^{+0.2+1.9}_{-0.2-1.4}$ & $2.7^{+2.0+2.0}_{-0.8-1.5}$ & 4 \\
64 & 350--500 & 350--500 & 3--4 & $\geq$3 & $0.6^{+1.2+0.1}_{-0.6-0.0}$ & $4.2^{+1.5+1.3}_{-1.4-1.3}$ & $2.5^{+0.0+1.4}_{-0.0-1.4}$ & $0.09^{+0.04+0.11}_{-0.04-0.05}$ & $7.4^{+2.6+1.9}_{-1.9-1.9}$ & 8 \\
65 & 350--500 & 500--1000 & 3--4 & $\geq$3 & $10.2^{+6.3+2.1}_{-5.7-2.1}$ & $7.0^{+1.5+1.9}_{-1.5-1.9}$ & $4.3^{+0.0+2.4}_{-0.0-2.4}$ & $0.78^{+0.18+0.94}_{-0.18-0.60}$ & $22.3^{+7.9+3.8}_{-7.2-3.7}$ & 26 \\ \hline
66 & 350--500 & $>$1000 & 3--4 & $\geq$3 & $0.0^{+1.1+0.0}_{-0.0-0.0}$ & $0.21^{+0.49+0.13}_{-0.16-0.13}$ & $0.36^{+0.01+0.20}_{-0.01-0.20}$ & $0.54^{+0.15+0.65}_{-0.15-0.39}$ & $1.1^{+1.6+0.7}_{-0.2-0.5}$ & 5 \\
67 & 500--750 & 500--1000 & 3--4 & $\geq$3 & $1.4^{+2.9+0.4}_{-1.4-0.0}$ & $1.13^{+0.74+0.45}_{-0.58-0.45}$ & $1.50^{+0.02+0.83}_{-0.02-0.83}$ & $0.10^{+0.10+0.13}_{-0.10-0.00}$ & $4.1^{+3.6+1.0}_{-2.0-0.9}$ & 0 \\
68 & 500--750 & $>$1000 & 3--4 & $\geq$3 & $0.00^{+0.95+0.00}_{-0.00-0.00}$ & $0.12^{+0.46+0.09}_{-0.06-0.09}$ & $0.26^{+0.01+0.15}_{-0.01-0.15}$ & $0.02^{+0.03+0.02}_{-0.02-0.00}$ & $0.4^{+1.4+0.2}_{-0.1-0.2}$ & 2 \\
69 & $>$750 & 750--1500 & 3--4 & $\geq$3 & $0.00^{+0.97+0.00}_{-0.00-0.00}$ & $0.00^{+0.46+0.00}_{-0.00-0.00}$ & $0.29^{+0.01+0.16}_{-0.01-0.16}$ & $0.01^{+0.02+0.01}_{-0.01-0.00}$ & $0.3^{+1.4+0.2}_{-0.0-0.2}$ & 1 \\
70 & $>$750 & $>$1500 & 3--4 & $\geq$3 & $0.0^{+1.4+0.0}_{-0.0-0.0}$ & $0.00^{+0.46+0.00}_{-0.00-0.00}$ & $0.04^{+0.01+0.02}_{-0.00-0.02}$ & $0.01^{+0.03+0.02}_{-0.01-0.00}$ & $0.0^{+1.8+0.0}_{-0.0-0.0}$ & 0 \\ 
\end{scotch}}
\end{table*}

\begin{table*}[htbp]
\renewcommand{\arraystretch}{1.25}
\centering
\topcaption{
Observed numbers of events and prefit background predictions
in the $5\leq\njets\leq6$ search regions.
The first uncertainty is statistical and second systematic.
}
\label{tab:pre-fit-results-nj2}
\resizebox{\textwidth}{!}{
\begin{scotch}{ ccccccccccc }
Bin & $\MHT$ [\GeVns{}] & $\HT$ [\GeVns{}] & $\njets$ & $\nbjets$ & Lost-$\Pe/\Pgm$ & $\Pgt\to\text{had}$ & $\cPZ\to\Pgn\cPagn$ & QCD & Total pred. & Obs. \\ \hline
71 & 300--350 & 300--500 & 5--6 & 0 & $217^{+11+22}_{-11-22}$ & $166^{+ 6+27}_{- 6-27}$ & $489^{+12+42}_{-12-39}$ & $49^{+ 5+21}_{- 5-19}$ & $922^{+21+58}_{-21-56}$ & 1015 \\
72 & 300--350 & 500--1000 & 5--6 & 0 & $397^{+13+37}_{-13-37}$ & $403^{+ 9+36}_{- 9-36}$ & $772^{+16+61}_{-15-57}$ & $113^{+ 4+47}_{- 4-43}$ & $1686^{+27+93}_{-27-88}$ & 1673 \\
73 & 300--350 & $>$1000 & 5--6 & 0 & $49.6^{+4.5+5.4}_{-4.5-5.4}$ & $55.1^{+3.8+8.3}_{-3.8-8.3}$ & $100.0^{+6.4+8.2}_{-6.0-7.1}$ & $49^{+ 1+21}_{- 1-19}$ & $254^{+11+24}_{-10-22}$ & 226 \\
74 & 350--500 & 350--500 & 5--6 & 0 & $71^{+ 7+11}_{- 6-11}$ & $47^{+ 3+16}_{- 3-16}$ & $242^{+ 9+20}_{- 9-19}$ & $12.7^{+2.3+5.3}_{-2.3-4.8}$ & $372^{+13+29}_{-13-28}$ & 464 \\
75 & 350--500 & 500--1000 & 5--6 & 0 & $384^{+12+33}_{-12-33}$ & $412^{+11+32}_{-11-32}$ & $1110^{+19+84}_{-19-78}$ & $65^{+ 2+27}_{- 2-25}$ & $1971^{+30+99}_{-29-93}$ & 2018 \\ \hline
76 & 350--500 & $>$1000 & 5--6 & 0 & $76.9^{+6.4+8.9}_{-6.4-8.9}$ & $72.4^{+4.8+9.3}_{-4.8-9.3}$ & $170^{+ 8+14}_{- 8-12}$ & $28^{+ 1+12}_{- 1-11}$ & $347^{+14+22}_{-14-21}$ & 320 \\
77 & 500--750 & 500--1000 & 5--6 & 0 & $66.7^{+5.1+7.3}_{-5.0-7.3}$ & $70.1^{+4.3+6.1}_{-4.2-6.0}$ & $302^{+10+23}_{-10-22}$ & $3.2^{+0.1+1.3}_{-0.1-1.2}$ & $442^{+14+25}_{-14-24}$ & 460 \\
78 & 500--750 & $>$1000 & 5--6 & 0 & $23.9^{+2.9+4.5}_{-2.9-4.5}$ & $31.2^{+3.1+4.0}_{-3.1-4.0}$ & $123.5^{+7.3+9.4}_{-6.9-8.9}$ & $2.5^{+0.1+1.1}_{-0.1-1.0}$ & $181^{+10+11}_{- 9-11}$ & 170 \\
79 & $>$750 & 750--1500 & 5--6 & 0 & $4.0^{+1.2+0.7}_{-1.1-0.7}$ & $4.90^{+0.89+0.52}_{-0.76-0.52}$ & $52.2^{+4.6+7.5}_{-4.2-6.8}$ & $0.23^{+0.04+0.10}_{-0.04-0.09}$ & $61.3^{+5.0+7.5}_{-4.6-6.9}$ & 74 \\
80 & $>$750 & $>$1500 & 5--6 & 0 & $0.90^{+0.61+0.19}_{-0.45-0.19}$ & $1.46^{+0.67+0.16}_{-0.49-0.16}$ & $16.5^{+2.9+2.7}_{-2.5-2.5}$ & $0.25^{+0.06+0.11}_{-0.06-0.10}$ & $19.1^{+3.2+2.7}_{-2.7-2.5}$ & 19 \\ \hline
81 & 300--350 & 300--500 & 5--6 & 1 & $130^{+ 8+11}_{- 8-11}$ & $131^{+ 6+17}_{- 6-17}$ & $133^{+ 3+19}_{- 3-19}$ & $12.8^{+2.8+5.2}_{-2.8-4.9}$ & $407^{+15+29}_{-15-28}$ & 450 \\
82 & 300--350 & 500--1000 & 5--6 & 1 & $290^{+11+25}_{-11-25}$ & $302^{+ 8+25}_{- 8-25}$ & $218^{+ 4+31}_{- 4-30}$ & $41^{+ 4+17}_{- 4-16}$ & $851^{+20+50}_{-20-49}$ & 781 \\
83 & 300--350 & $>$1000 & 5--6 & 1 & $25.8^{+3.4+2.5}_{-3.4-2.5}$ & $31.6^{+2.9+5.9}_{-2.9-5.9}$ & $29.0^{+1.8+4.1}_{-1.7-4.0}$ & $18.4^{+0.8+7.5}_{-0.8-7.1}$ & $105^{+ 7+11}_{- 6-10}$ & 100 \\
84 & 350--500 & 350--500 & 5--6 & 1 & $45.4^{+5.5+5.4}_{-5.4-5.4}$ & $32^{+ 3+11}_{- 3-11}$ & $65.1^{+2.4+9.3}_{-2.3-9.1}$ & $3.7^{+0.5+1.5}_{-0.5-1.4}$ & $146^{+ 9+16}_{- 8-16}$ & 160 \\
85 & 350--500 & 500--1000 & 5--6 & 1 & $228^{+10+20}_{-10-20}$ & $269^{+ 8+21}_{- 8-21}$ & $310^{+ 5+43}_{- 5-42}$ & $28^{+ 3+11}_{- 3-11}$ & $834^{+19+53}_{-19-52}$ & 801 \\ \hline
86 & 350--500 & $>$1000 & 5--6 & 1 & $40.5^{+5.5+4.2}_{-5.4-4.2}$ & $36.0^{+3.3+4.3}_{-3.3-4.2}$ & $49.4^{+2.3+7.0}_{-2.2-6.7}$ & $11.9^{+0.7+4.8}_{-0.7-4.5}$ & $138^{+ 9+10}_{- 9-10}$ & 138 \\
87 & 500--750 & 500--1000 & 5--6 & 1 & $23.4^{+3.5+2.6}_{-3.4-2.6}$ & $32.1^{+2.8+3.3}_{-2.8-3.3}$ & $84^{+ 3+12}_{- 3-12}$ & $1.45^{+0.11+0.59}_{-0.11-0.55}$ & $141^{+ 7+13}_{- 7-12}$ & 135 \\
88 & 500--750 & $>$1000 & 5--6 & 1 & $8.5^{+1.8+1.1}_{-1.7-1.1}$ & $13.0^{+1.8+1.5}_{-1.7-1.5}$ & $35.3^{+2.1+4.9}_{-2.0-4.8}$ & $1.33^{+0.17+0.54}_{-0.17-0.51}$ & $58.0^{+4.1+5.3}_{-3.9-5.2}$ & 49 \\
89 & $>$750 & 750--1500 & 5--6 & 1 & $3.7^{+1.4+0.7}_{-1.2-0.7}$ & $2.9^{+1.0+0.4}_{-0.9-0.4}$ & $14.9^{+1.3+2.8}_{-1.2-2.6}$ & $0.07^{+0.01+0.03}_{-0.01-0.03}$ & $21.6^{+2.8+2.9}_{-2.5-2.7}$ & 16 \\
90 & $>$750 & $>$1500 & 5--6 & 1 & $1.06^{+0.74+0.26}_{-0.56-0.26}$ & $1.16^{+0.73+0.18}_{-0.57-0.18}$ & $4.79^{+0.85+0.96}_{-0.73-0.92}$ & $0.16^{+0.07+0.07}_{-0.07-0.06}$ & $7.2^{+1.7+1.0}_{-1.3-1.0}$ & 6 \\ \hline
91 & 300--350 & 300--500 & 5--6 & 2 & $60.1^{+7.1+6.0}_{-7.0-6.0}$ & $50.2^{+3.3+4.9}_{-3.3-4.9}$ & $23.8^{+0.6+7.1}_{-0.6-7.1}$ & $2.9^{+0.9+1.1}_{-0.9-1.1}$ & $137^{+10+11}_{-10-11}$ & 143 \\
92 & 300--350 & 500--1000 & 5--6 & 2 & $137^{+ 9+13}_{- 9-13}$ & $160^{+ 6+14}_{- 6-14}$ & $39^{+ 1+12}_{- 1-11}$ & $11.8^{+1.8+4.6}_{-1.8-4.5}$ & $347^{+15+22}_{-15-22}$ & 332 \\
93 & 300--350 & $>$1000 & 5--6 & 2 & $16.9^{+3.8+2.0}_{-3.7-2.0}$ & $15.9^{+2.1+2.1}_{-2.1-2.1}$ & $5.1^{+0.3+1.5}_{-0.3-1.5}$ & $5.6^{+0.4+2.2}_{-0.4-2.2}$ & $43.5^{+5.9+3.9}_{-5.8-3.9}$ & 36 \\
94 & 350--500 & 350--500 & 5--6 & 2 & $13.3^{+3.1+1.9}_{-2.9-1.9}$ & $7.0^{+1.1+2.3}_{-1.0-2.3}$ & $11.7^{+0.4+3.5}_{-0.4-3.5}$ & $1.02^{+0.54+0.40}_{-0.54-0.39}$ & $32.9^{+4.3+4.6}_{-4.0-4.6}$ & 28 \\
95 & 350--500 & 500--1000 & 5--6 & 2 & $107.5^{+7.6+9.6}_{-7.6-9.6}$ & $121.2^{+5.8+9.9}_{-5.8-9.8}$ & $55^{+ 1+16}_{- 1-16}$ & $5.9^{+1.0+2.3}_{-1.0-2.2}$ & $290^{+14+22}_{-13-21}$ & 288 \\ \hline
96 & 350--500 & $>$1000 & 5--6 & 2 & $14.2^{+2.8+1.8}_{-2.7-1.8}$ & $15.7^{+2.2+2.0}_{-2.1-2.0}$ & $8.7^{+0.4+2.6}_{-0.4-2.6}$ & $3.2^{+0.1+1.2}_{-0.1-1.2}$ & $41.8^{+5.0+4.0}_{-4.8-3.9}$ & 44 \\
97 & 500--750 & 500--1000 & 5--6 & 2 & $8.4^{+2.3+1.1}_{-2.2-1.1}$ & $8.3^{+1.3+1.0}_{-1.2-1.0}$ & $15.0^{+0.5+4.4}_{-0.5-4.4}$ & $0.34^{+0.05+0.13}_{-0.05-0.13}$ & $32.1^{+3.7+4.7}_{-3.4-4.7}$ & 35 \\
98 & 500--750 & $>$1000 & 5--6 & 2 & $2.1^{+1.3+0.3}_{-1.0-0.3}$ & $4.0^{+1.1+0.6}_{-1.0-0.6}$ & $6.2^{+0.4+1.9}_{-0.3-1.8}$ & $0.16^{+0.05+0.06}_{-0.05-0.06}$ & $12.5^{+2.4+2.0}_{-2.0-2.0}$ & 18 \\
99 & $>$750 & 750--1500 & 5--6 & 2 & $0.74^{+0.87+0.22}_{-0.53-0.22}$ & $0.68^{+0.64+0.16}_{-0.45-0.16}$ & $2.64^{+0.23+0.85}_{-0.21-0.83}$ & $0.05^{+0.05+0.02}_{-0.05-0.00}$ & $4.1^{+1.5+0.9}_{-1.0-0.9}$ & 8 \\
100 & $>$750 & $>$1500 & 5--6 & 2 & $0.77^{+0.65+0.24}_{-0.45-0.24}$ & $1.07^{+0.72+0.33}_{-0.56-0.33}$ & $0.84^{+0.15+0.28}_{-0.13-0.27}$ & $0.03^{+0.03+0.01}_{-0.03-0.00}$ & $2.7^{+1.4+0.5}_{-1.0-0.5}$ & 3 \\ \hline
101 & 300--350 & 300--500 & 5--6 & $\geq$3 & $2.8^{+1.5+0.3}_{-1.2-0.3}$ & $5.1^{+1.0+0.8}_{-0.9-0.8}$ & $2.0^{+0.0+1.1}_{-0.0-1.1}$ & $0.50^{+0.37+0.57}_{-0.37-0.13}$ & $10.4^{+2.5+1.5}_{-2.1-1.4}$ & 18 \\
102 & 300--350 & 500--1000 & 5--6 & $\geq$3 & $17.0^{+3.2+1.6}_{-3.1-1.6}$ & $23.5^{+2.4+3.2}_{-2.3-3.2}$ & $4.2^{+0.1+2.3}_{-0.1-2.3}$ & $3.9^{+2.3+4.5}_{-2.3-1.6}$ & $48.7^{+6.0+6.2}_{-5.9-4.5}$ & 44 \\
103 & 300--350 & $>$1000 & 5--6 & $\geq$3 & $4.4^{+2.1+0.6}_{-1.8-0.6}$ & $2.50^{+0.86+0.47}_{-0.73-0.47}$ & $0.65^{+0.04+0.35}_{-0.04-0.35}$ & $3.3^{+0.4+3.7}_{-0.4-2.8}$ & $10.8^{+3.0+3.8}_{-2.6-3.0}$ & 6 \\
104 & 350--500 & 350--500 & 5--6 & $\geq$3 & $0.8^{+1.7+0.2}_{-0.8-0.0}$ & $1.14^{+0.75+0.33}_{-0.59-0.33}$ & $0.87^{+0.03+0.47}_{-0.03-0.47}$ & $0.18^{+0.08+0.21}_{-0.08-0.10}$ & $3.0^{+2.4+0.6}_{-1.4-0.6}$ & 4 \\
105 & 350--500 & 500--1000 & 5--6 & $\geq$3 & $15.2^{+2.6+1.5}_{-2.6-1.5}$ & $17.6^{+2.2+2.7}_{-2.1-2.7}$ & $5.7^{+0.1+3.1}_{-0.1-3.1}$ & $1.7^{+0.1+1.9}_{-0.1-1.6}$ & $40.2^{+4.8+4.8}_{-4.7-4.6}$ & 34 \\ \hline
106 & 350--500 & $>$1000 & 5--6 & $\geq$3 & $1.9^{+1.1+0.3}_{-0.8-0.3}$ & $3.8^{+1.1+0.7}_{-1.0-0.7}$ & $1.14^{+0.05+0.62}_{-0.05-0.62}$ & $2.4^{+0.3+2.7}_{-0.3-2.1}$ & $9.2^{+2.2+2.8}_{-1.9-2.3}$ & 8 \\
107 & 500--750 & 500--1000 & 5--6 & $\geq$3 & $1.8^{+1.1+0.3}_{-0.8-0.3}$ & $1.71^{+0.77+0.67}_{-0.61-0.67}$ & $1.48^{+0.05+0.81}_{-0.05-0.80}$ & $0.20^{+0.04+0.23}_{-0.04-0.17}$ & $5.2^{+1.8+1.1}_{-1.5-1.1}$ & 4 \\
108 & 500--750 & $>$1000 & 5--6 & $\geq$3 & $1.13^{+0.96+0.25}_{-0.66-0.25}$ & $0.94^{+0.67+0.27}_{-0.49-0.27}$ & $0.73^{+0.04+0.40}_{-0.04-0.40}$ & $0.11^{+0.03+0.12}_{-0.03-0.08}$ & $2.9^{+1.6+0.6}_{-1.1-0.6}$ & 2 \\
109 & $>$750 & 750--1500 & 5--6 & $\geq$3 & $0.00^{+0.72+0.00}_{-0.00-0.00}$ & $0.07^{+0.46+0.04}_{-0.06-0.04}$ & $0.31^{+0.03+0.17}_{-0.03-0.17}$ & $0.02^{+0.04+0.03}_{-0.02-0.00}$ & $0.4^{+1.2+0.2}_{-0.1-0.2}$ & 0 \\
110 & $>$750 & $>$1500 & 5--6 & $\geq$3 & $0.00^{+0.63+0.00}_{-0.00-0.00}$ & $0.03^{+0.46+0.01}_{-0.02-0.01}$ & $0.11^{+0.02+0.06}_{-0.02-0.06}$ & $0.00^{+0.02+0.01}_{-0.00-0.00}$ & $0.1^{+1.1+0.1}_{-0.0-0.1}$ & 1 \\ 
\end{scotch}}
\end{table*}

\begin{table*}[htbp]
\renewcommand{\arraystretch}{1.25}
\centering
\topcaption{
Observed numbers of events and prefit background predictions
in the $7\leq\njets\leq8$ search regions.
The first uncertainty is statistical and second systematic.
}
\label{tab:pre-fit-results-nj3}
\resizebox{\textwidth}{!}{
\begin{scotch}{ ccccccccccc }
Bin & $\MHT$ [\GeVns{}] & $\HT$ [\GeVns{}] & $\njets$ & $\nbjets$ & Lost-$\Pe/\Pgm$ & $\Pgt\to\text{had}$ & $\cPZ\to\Pgn\cPagn$ & QCD & Total pred. & Obs. \\ \hline
111 & 300--350 & 500--1000 & 7--8 & 0 & $48.0^{+3.9+5.4}_{-3.8-5.4}$ & $60.8^{+3.4+6.0}_{-3.4-6.0}$ & $76^{+ 5+11}_{- 5-10}$ & $30^{+ 2+12}_{- 2-11}$ & $215^{+ 9+18}_{- 9-17}$ & 218 \\
112 & 300--350 & $>$1000 & 7--8 & 0 & $21.2^{+2.9+2.3}_{-2.9-2.3}$ & $20.3^{+2.2+2.8}_{-2.1-2.8}$ & $23.9^{+3.3+2.8}_{-2.9-2.5}$ & $20.5^{+0.5+8.5}_{-0.5-7.8}$ & $85.9^{+6.1+9.6}_{-5.8-9.0}$ & 85 \\
113 & 350--500 & 500--1000 & 7--8 & 0 & $43.2^{+3.9+4.9}_{-3.9-4.9}$ & $54.2^{+3.6+5.7}_{-3.5-5.7}$ & $89^{+ 6+11}_{- 5-10}$ & $14.3^{+1.9+5.9}_{-1.9-5.4}$ & $201^{+10+14}_{- 9-14}$ & 215 \\
114 & 350--500 & $>$1000 & 7--8 & 0 & $22.5^{+2.8+2.7}_{-2.7-2.7}$ & $23.3^{+2.5+2.3}_{-2.4-2.3}$ & $48.3^{+4.7+5.4}_{-4.3-4.8}$ & $12.6^{+0.7+5.2}_{-0.7-4.8}$ & $106.7^{+7.1+8.3}_{-6.7-7.7}$ & 75 \\
115 & 500--750 & 500--1000 & 7--8 & 0 & $6.9^{+1.8+1.4}_{-1.7-1.4}$ & $4.96^{+0.95+0.77}_{-0.84-0.77}$ & $26.5^{+3.6+3.3}_{-3.2-3.0}$ & $0.88^{+0.10+0.36}_{-0.10-0.34}$ & $39.2^{+4.5+3.7}_{-4.1-3.5}$ & 34 \\ \hline
116 & 500--750 & $>$1000 & 7--8 & 0 & $5.4^{+1.1+0.9}_{-1.0-0.9}$ & $9.9^{+1.6+1.7}_{-1.5-1.7}$ & $27.2^{+3.7+3.1}_{-3.2-2.8}$ & $1.56^{+0.12+0.64}_{-0.12-0.59}$ & $44.1^{+4.5+3.7}_{-4.1-3.5}$ & 38 \\
117 & $>$750 & 750--1500 & 7--8 & 0 & $1.26^{+0.70+0.50}_{-0.58-0.50}$ & $1.44^{+0.74+0.24}_{-0.57-0.24}$ & $3.6^{+1.4+0.7}_{-1.0-0.6}$ & $0.07^{+0.02+0.03}_{-0.02-0.03}$ & $6.4^{+2.0+0.9}_{-1.5-0.8}$ & 5 \\
118 & $>$750 & $>$1500 & 7--8 & 0 & $0.69^{+0.47+0.16}_{-0.35-0.16}$ & $1.03^{+0.69+0.15}_{-0.51-0.15}$ & $1.5^{+1.2+0.3}_{-0.7-0.3}$ & $0.07^{+0.01+0.03}_{-0.01-0.03}$ & $3.3^{+1.7+0.4}_{-1.1-0.4}$ & 5 \\
119 & 300--350 & 500--1000 & 7--8 & 1 & $64.7^{+5.1+6.4}_{-5.1-6.4}$ & $77.0^{+3.9+7.5}_{-3.8-7.4}$ & $31.7^{+2.1+8.6}_{-1.9-8.4}$ & $11.2^{+0.5+4.7}_{-0.5-4.3}$ & $184^{+ 9+14}_{- 9-14}$ & 146 \\
120 & 300--350 & $>$1000 & 7--8 & 1 & $16.3^{+2.4+1.7}_{-2.4-1.7}$ & $19.9^{+2.2+2.1}_{-2.1-2.1}$ & $10.3^{+1.4+2.7}_{-1.2-2.6}$ & $8.3^{+0.2+3.5}_{-0.2-3.2}$ & $54.8^{+4.8+5.2}_{-4.7-5.0}$ & 68 \\ \hline
121 & 350--500 & 500--1000 & 7--8 & 1 & $46.9^{+4.4+5.0}_{-4.4-5.0}$ & $58.6^{+3.7+5.7}_{-3.7-5.7}$ & $37.0^{+2.4+9.7}_{-2.2-9.5}$ & $7.5^{+0.4+3.2}_{-0.4-2.9}$ & $150^{+ 8+13}_{- 8-12}$ & 113 \\
122 & 350--500 & $>$1000 & 7--8 & 1 & $19.5^{+2.5+2.1}_{-2.4-2.1}$ & $19.5^{+2.3+2.0}_{-2.3-2.0}$ & $21.0^{+2.0+5.4}_{-1.9-5.3}$ & $5.3^{+0.5+2.2}_{-0.5-2.0}$ & $65.3^{+5.2+6.5}_{-5.1-6.4}$ & 67 \\
123 & 500--750 & 500--1000 & 7--8 & 1 & $7.6^{+2.0+1.4}_{-1.9-1.4}$ & $5.5^{+1.1+0.8}_{-1.1-0.8}$ & $11.5^{+1.6+3.0}_{-1.4-3.0}$ & $0.36^{+0.04+0.15}_{-0.04-0.14}$ & $24.9^{+3.5+3.4}_{-3.3-3.4}$ & 19 \\
124 & 500--750 & $>$1000 & 7--8 & 1 & $9.3^{+2.1+1.3}_{-2.0-1.3}$ & $7.5^{+1.5+0.8}_{-1.4-0.8}$ & $11.4^{+1.5+3.0}_{-1.4-2.9}$ & $0.98^{+0.12+0.41}_{-0.12-0.37}$ & $29.2^{+3.9+3.3}_{-3.7-3.3}$ & 22 \\
125 & $>$750 & 750--1500 & 7--8 & 1 & $0.14^{+0.30+0.05}_{-0.14-0.00}$ & $0.44^{+0.51+0.10}_{-0.22-0.10}$ & $1.48^{+0.56+0.44}_{-0.42-0.43}$ & $0.07^{+0.03+0.03}_{-0.03-0.03}$ & $2.14^{+0.99+0.46}_{-0.56-0.45}$ & 4 \\ \hline
126 & $>$750 & $>$1500 & 7--8 & 1 & $0.00^{+0.47+0.00}_{-0.00-0.00}$ & $0.14^{+0.47+0.02}_{-0.08-0.02}$ & $0.70^{+0.55+0.22}_{-0.34-0.21}$ & $0.03^{+0.01+0.01}_{-0.01-0.01}$ & $0.9^{+1.1+0.2}_{-0.3-0.2}$ & 6 \\
127 & 300--350 & 500--1000 & 7--8 & 2 & $34.7^{+3.5+3.6}_{-3.5-3.6}$ & $47.7^{+3.0+4.4}_{-3.0-4.4}$ & $8.1^{+0.5+3.6}_{-0.5-3.5}$ & $5.3^{+0.5+2.1}_{-0.5-2.1}$ & $95.8^{+6.6+7.1}_{-6.5-7.0}$ & 95 \\
128 & 300--350 & $>$1000 & 7--8 & 2 & $9.0^{+2.1+1.2}_{-2.1-1.2}$ & $10.8^{+1.4+1.3}_{-1.4-1.3}$ & $2.4^{+0.3+1.0}_{-0.3-1.0}$ & $3.2^{+0.1+1.3}_{-0.1-1.3}$ & $25.4^{+3.6+2.4}_{-3.4-2.4}$ & 26 \\
129 & 350--500 & 500--1000 & 7--8 & 2 & $26.2^{+3.0+2.9}_{-3.0-2.9}$ & $31.0^{+2.5+3.3}_{-2.5-3.2}$ & $9.6^{+0.6+4.1}_{-0.6-4.1}$ & $2.5^{+0.2+1.0}_{-0.2-1.0}$ & $69.3^{+5.6+6.1}_{-5.5-6.1}$ & 84 \\
130 & 350--500 & $>$1000 & 7--8 & 2 & $13.3^{+2.5+1.5}_{-2.4-1.5}$ & $13.3^{+1.8+1.3}_{-1.7-1.3}$ & $4.7^{+0.5+2.0}_{-0.4-2.0}$ & $1.95^{+0.13+0.78}_{-0.13-0.75}$ & $33.3^{+4.3+3.0}_{-4.2-2.9}$ & 35 \\ \hline
131 & 500--750 & 500--1000 & 7--8 & 2 & $2.5^{+1.4+0.5}_{-1.2-0.5}$ & $0.86^{+0.50+0.21}_{-0.18-0.21}$ & $2.6^{+0.3+1.1}_{-0.3-1.1}$ & $0.10^{+0.01+0.04}_{-0.01-0.04}$ & $6.0^{+1.9+1.3}_{-1.4-1.3}$ & 7 \\
132 & 500--750 & $>$1000 & 7--8 & 2 & $6.0^{+2.3+1.0}_{-2.2-1.0}$ & $3.3^{+1.0+0.6}_{-0.9-0.6}$ & $2.9^{+0.4+1.2}_{-0.3-1.2}$ & $0.22^{+0.06+0.09}_{-0.06-0.08}$ & $12.4^{+3.4+1.7}_{-3.1-1.7}$ & 12 \\
133 & $>$750 & 750--1500 & 7--8 & 2 & $0.16^{+0.34+0.08}_{-0.16-0.00}$ & $0.44^{+0.56+0.15}_{-0.32-0.15}$ & $0.39^{+0.15+0.18}_{-0.11-0.18}$ & $0.03^{+0.01+0.01}_{-0.01-0.01}$ & $1.03^{+0.91+0.25}_{-0.49-0.23}$ & 2 \\
134 & $>$750 & $>$1500 & 7--8 & 2 & $0.53^{+0.62+0.20}_{-0.38-0.20}$ & $0.61^{+0.57+0.22}_{-0.33-0.22}$ & $0.13^{+0.10+0.06}_{-0.06-0.06}$ & $0.06^{+0.02+0.02}_{-0.02-0.02}$ & $1.3^{+1.2+0.3}_{-0.7-0.3}$ & 2 \\
135 & 300--350 & 500--1000 & 7--8 & $\geq$3 & $8.1^{+1.8+1.0}_{-1.7-1.0}$ & $9.4^{+1.4+1.3}_{-1.3-1.3}$ & $4.1^{+0.3+2.3}_{-0.2-2.3}$ & $2.9^{+0.6+3.3}_{-0.6-2.3}$ & $24.6^{+3.2+4.3}_{-3.1-3.7}$ & 12 \\ \hline
136 & 300--350 & $>$1000 & 7--8 & $\geq$3 & $4.7^{+2.0+0.7}_{-1.8-0.7}$ & $5.4^{+1.2+0.8}_{-1.1-0.8}$ & $1.51^{+0.21+0.85}_{-0.18-0.84}$ & $2.4^{+0.3+2.7}_{-0.3-2.1}$ & $13.9^{+3.2+3.0}_{-2.9-2.5}$ & 8 \\
137 & 350--500 & 500--1000 & 7--8 & $\geq$3 & $5.9^{+1.9+0.8}_{-1.7-0.8}$ & $7.4^{+1.4+1.2}_{-1.3-1.2}$ & $4.7^{+0.3+2.7}_{-0.3-2.7}$ & $1.2^{+0.1+1.3}_{-0.1-1.1}$ & $19.2^{+3.2+3.3}_{-3.1-3.2}$ & 16 \\
138 & 350--500 & $>$1000 & 7--8 & $\geq$3 & $2.6^{+1.1+0.3}_{-1.0-0.3}$ & $4.8^{+1.3+0.7}_{-1.2-0.7}$ & $3.1^{+0.3+1.8}_{-0.3-1.8}$ & $2.1^{+0.3+2.3}_{-0.3-1.8}$ & $12.6^{+2.5+3.0}_{-2.2-2.6}$ & 8 \\
139 & 500--750 & 500--1000 & 7--8 & $\geq$3 & $0.23^{+0.48+0.08}_{-0.23-0.00}$ & $0.30^{+0.48+0.10}_{-0.13-0.10}$ & $1.70^{+0.23+0.96}_{-0.20-0.96}$ & $0.11^{+0.04+0.12}_{-0.04-0.08}$ & $2.34^{+0.99+0.98}_{-0.41-0.96}$ & 3 \\
140 & 500--750 & $>$1000 & 7--8 & $\geq$3 & $3.4^{+2.4+0.7}_{-2.1-0.7}$ & $1.59^{+0.83+0.49}_{-0.69-0.49}$ & $1.51^{+0.20+0.85}_{-0.18-0.85}$ & $0.22^{+0.08+0.24}_{-0.08-0.14}$ & $6.7^{+3.2+1.2}_{-2.7-1.2}$ & 4 \\
141 & $>$750 & 750--1500 & 7--8 & $\geq$3 & $0.00^{+0.56+0.00}_{-0.00-0.00}$ & $0.05^{+0.46+0.02}_{-0.03-0.02}$ & $0.19^{+0.07+0.11}_{-0.05-0.11}$ & $0.03^{+0.04+0.03}_{-0.03-0.00}$ & $0.3^{+1.0+0.1}_{-0.1-0.1}$ & 0 \\
142 & $>$750 & $>$1500 & 7--8 & $\geq$3 & $0.00^{+0.72+0.00}_{-0.00-0.00}$ & $0.04^{+0.46+0.02}_{-0.02-0.02}$ & $0.12^{+0.10+0.07}_{-0.06-0.07}$ & $0.01^{+0.03+0.01}_{-0.01-0.00}$ & $0.2^{+1.2+0.1}_{-0.1-0.1}$ & 0 \\ 
\end{scotch}}
\end{table*}

\begin{table*}[htbp]
\renewcommand{\arraystretch}{1.25}
\centering
\topcaption{
Observed numbers of events and prefit background predictions
in the $\njets\geq9$ search regions.
The first uncertainty is statistical and second systematic.
}
\label{tab:pre-fit-results-nj4}
\resizebox{\textwidth}{!}{
\begin{scotch}{ ccccccccccc }
Bin & $\MHT$ [\GeVns{}] & $\HT$ [\GeVns{}] & $\njets$ & $\nbjets$ & Lost-$\Pe/\Pgm$ & $\Pgt\to\text{had}$ & $\cPZ\to\Pgn\cPagn$ & QCD & Total pred. & Obs. \\ \hline
143 & 300--350 & 500--1000 & $\geq$9 & 0 & $6.2^{+2.7+1.7}_{-2.6-1.7}$ & $3.46^{+0.89+0.59}_{-0.77-0.59}$ & $2.6^{+1.2+0.7}_{-0.9-0.7}$ & $2.9^{+0.3+1.3}_{-0.3-1.1}$ & $15.1^{+3.8+2.3}_{-3.5-2.2}$ & 7 \\
144 & 300--350 & $>$1000 & $\geq$9 & 0 & $3.5^{+1.2+0.6}_{-1.1-0.6}$ & $4.6^{+1.0+0.6}_{-0.9-0.6}$ & $3.0^{+1.4+0.6}_{-1.0-0.6}$ & $4.2^{+0.3+1.9}_{-0.3-1.6}$ & $15.2^{+2.7+2.1}_{-2.3-1.9}$ & 12 \\
145 & 350--500 & 500--1000 & $\geq$9 & 0 & $2.39^{+0.99+0.69}_{-0.89-0.69}$ & $2.39^{+0.86+0.48}_{-0.73-0.48}$ & $2.9^{+1.3+0.7}_{-0.9-0.6}$ & $0.97^{+0.08+0.43}_{-0.08-0.37}$ & $8.6^{+2.3+1.2}_{-1.9-1.1}$ & 6 \\
146 & 350--500 & $>$1000 & $\geq$9 & 0 & $3.7^{+1.1+0.6}_{-1.1-0.6}$ & $4.6^{+1.0+0.6}_{-0.9-0.6}$ & $5.5^{+1.9+1.0}_{-1.5-0.9}$ & $3.1^{+0.2+1.4}_{-0.2-1.2}$ & $17.0^{+2.9+1.9}_{-2.5-1.7}$ & 13 \\
147 & 500--750 & 500--1000 & $\geq$9 & 0 & $0.15^{+0.32+0.10}_{-0.15-0.00}$ & $0.35^{+0.55+0.12}_{-0.30-0.12}$ & $1.0^{+1.3+0.4}_{-0.7-0.4}$ & $0.10^{+0.05+0.04}_{-0.05-0.04}$ & $1.6^{+1.6+0.5}_{-0.8-0.4}$ & 2 \\ \hline
148 & 500--750 & $>$1000 & $\geq$9 & 0 & $0.98^{+0.50+0.26}_{-0.41-0.26}$ & $1.98^{+0.74+0.30}_{-0.58-0.30}$ & $3.5^{+1.6+0.7}_{-1.1-0.7}$ & $0.47^{+0.05+0.21}_{-0.05-0.18}$ & $6.9^{+2.0+0.8}_{-1.5-0.8}$ & 11 \\
149 & $>$750 & 750--1500 & $\geq$9 & 0 & $0.00^{+0.44+0.00}_{-0.00-0.00}$ & $0.00^{+0.46+0.00}_{-0.00-0.00}$ & $0.00^{+0.64+0.00}_{-0.00-0.00}$ & $0.01^{+0.02+0.00}_{-0.01-0.00}$ & $0.0^{+1.1+0.0}_{-0.0-0.0}$ & 0 \\
150 & $>$750 & $>$1500 & $\geq$9 & 0 & $0.23^{+0.27+0.16}_{-0.17-0.16}$ & $0.28^{+0.50+0.08}_{-0.21-0.08}$ & $0.00^{+0.82+0.00}_{-0.00-0.00}$ & $0.05^{+0.03+0.02}_{-0.03-0.02}$ & $0.6^{+1.1+0.2}_{-0.4-0.2}$ & 1 \\
151 & 300--350 & 500--1000 & $\geq$9 & 1 & $6.5^{+1.8+1.1}_{-1.7-1.1}$ & $4.57^{+0.93+0.77}_{-0.81-0.77}$ & $1.83^{+0.84+0.68}_{-0.60-0.74}$ & $1.02^{+0.06+0.42}_{-0.06-0.40}$ & $13.9^{+2.8+1.5}_{-2.6-1.6}$ & 25 \\
152 & 300--350 & $>$1000 & $\geq$9 & 1 & $5.7^{+1.6+0.7}_{-1.5-0.7}$ & $7.3^{+1.3+1.1}_{-1.2-1.1}$ & $2.08^{+0.95+0.69}_{-0.68-0.77}$ & $2.43^{+0.06+0.99}_{-0.06-0.94}$ & $17.5^{+3.0+1.8}_{-2.8-1.8}$ & 20 \\ \hline
153 & 350--500 & 500--1000 & $\geq$9 & 1 & $2.92^{+0.94+0.57}_{-0.84-0.57}$ & $2.96^{+0.77+0.60}_{-0.61-0.60}$ & $2.00^{+0.91+0.71}_{-0.65-0.78}$ & $0.53^{+0.05+0.22}_{-0.05-0.21}$ & $8.4^{+1.9+1.1}_{-1.6-1.2}$ & 8 \\
154 & 350--500 & $>$1000 & $\geq$9 & 1 & $5.4^{+1.4+0.7}_{-1.3-0.7}$ & $7.7^{+1.4+1.1}_{-1.3-1.1}$ & $3.9^{+1.3+1.3}_{-1.0-1.4}$ & $1.48^{+0.05+0.60}_{-0.05-0.57}$ & $18.4^{+3.1+1.9}_{-2.8-2.0}$ & 14 \\
155 & 500--750 & 500--1000 & $\geq$9 & 1 & $0.14^{+0.30+0.08}_{-0.14-0.00}$ & $0.24^{+0.49+0.21}_{-0.18-0.16}$ & $0.71^{+0.94+0.35}_{-0.46-0.36}$ & $0.03^{+0.03+0.01}_{-0.03-0.00}$ & $1.1^{+1.2+0.4}_{-0.6-0.4}$ & 1 \\
156 & 500--750 & $>$1000 & $\geq$9 & 1 & $0.68^{+0.58+0.12}_{-0.41-0.12}$ & $1.20^{+0.64+0.21}_{-0.44-0.21}$ & $2.4^{+1.1+0.8}_{-0.8-0.9}$ & $0.20^{+0.02+0.08}_{-0.02-0.07}$ & $4.5^{+1.6+0.8}_{-1.2-0.9}$ & 4 \\
157 & $>$750 & 750--1500 & $\geq$9 & 1 & $0.00^{+0.73+0.00}_{-0.00-0.00}$ & $0.04^{+0.46+0.02}_{-0.04-0.00}$ & $0.00^{+0.45+0.00}_{-0.00-0.00}$ & $0.01^{+0.01+0.00}_{-0.01-0.00}$ & $0.1^{+1.3+0.0}_{-0.0-0.0}$ & 0 \\ \hline
158 & $>$750 & $>$1500 & $\geq$9 & 1 & $0.13^{+0.27+0.06}_{-0.13-0.00}$ & $0.03^{+0.46+0.01}_{-0.02-0.01}$ & $0.00^{+0.57+0.00}_{-0.00-0.00}$ & $0.02^{+0.01+0.01}_{-0.01-0.01}$ & $0.18^{+0.93+0.06}_{-0.15-0.01}$ & 0 \\
159 & 300--350 & 500--1000 & $\geq$9 & 2 & $4.1^{+1.3+0.7}_{-1.2-0.7}$ & $4.68^{+0.92+0.85}_{-0.80-0.85}$ & $0.64^{+0.29+0.34}_{-0.21-0.36}$ & $0.40^{+0.06+0.24}_{-0.06-0.21}$ & $9.8^{+2.2+1.2}_{-2.0-1.2}$ & 13 \\
160 & 300--350 & $>$1000 & $\geq$9 & 2 & $5.2^{+1.6+0.7}_{-1.5-0.7}$ & $5.5^{+1.2+1.0}_{-1.1-1.0}$ & $0.73^{+0.33+0.37}_{-0.24-0.39}$ & $1.32^{+0.15+0.68}_{-0.15-0.58}$ & $12.7^{+2.8+1.4}_{-2.6-1.4}$ & 10 \\
161 & 350--500 & 500--1000 & $\geq$9 & 2 & $3.01^{+0.91+0.63}_{-0.82-0.63}$ & $4.7^{+1.1+0.9}_{-1.0-0.9}$ & $0.70^{+0.32+0.36}_{-0.23-0.39}$ & $0.30^{+0.08+0.14}_{-0.08-0.12}$ & $8.7^{+2.0+1.1}_{-1.8-1.1}$ & 4 \\
162 & 350--500 & $>$1000 & $\geq$9 & 2 & $4.4^{+1.1+0.6}_{-1.1-0.6}$ & $6.3^{+1.4+0.8}_{-1.3-0.8}$ & $1.35^{+0.47+0.67}_{-0.36-0.72}$ & $0.63^{+0.03+0.32}_{-0.03-0.27}$ & $12.7^{+2.6+1.3}_{-2.4-1.3}$ & 12 \\ \hline
163 & 500--750 & 500--1000 & $\geq$9 & 2 & $0.00^{+0.39+0.00}_{-0.00-0.00}$ & $0.35^{+0.49+0.17}_{-0.18-0.17}$ & $0.25^{+0.33+0.15}_{-0.16-0.16}$ & $0.01^{+0.01+0.01}_{-0.01-0.00}$ & $0.61^{+0.95+0.23}_{-0.24-0.23}$ & 0 \\
164 & 500--750 & $>$1000 & $\geq$9 & 2 & $2.0^{+1.1+0.4}_{-0.9-0.4}$ & $1.95^{+0.87+0.45}_{-0.73-0.45}$ & $0.84^{+0.39+0.43}_{-0.28-0.46}$ & $0.09^{+0.02+0.04}_{-0.02-0.04}$ & $4.9^{+2.0+0.7}_{-1.7-0.7}$ & 7 \\
165 & $>$750 & 750--1500 & $\geq$9 & 2 & $0.00^{+0.60+0.00}_{-0.00-0.00}$ & $0.01^{+0.46+0.01}_{-0.00-0.00}$ & $0.00^{+0.16+0.00}_{-0.00-0.00}$ & $0.00^{+0.01+0.00}_{-0.00-0.00}$ & $0.0^{+1.1+0.0}_{-0.0-0.0}$ & 0 \\
166 & $>$750 & $>$1500 & $\geq$9 & 2 & $0.00^{+0.38+0.00}_{-0.00-0.00}$ & $0.00^{+0.46+0.00}_{-0.00-0.00}$ & $0.00^{+0.20+0.00}_{-0.00-0.00}$ & $0.01^{+0.02+0.00}_{-0.01-0.00}$ & $0.01^{+0.87+0.00}_{-0.01-0.00}$ & 0 \\
167 & 300--350 & 500--1000 & $\geq$9 & $\geq$3 & $1.06^{+0.63+0.27}_{-0.50-0.27}$ & $1.06^{+0.57+0.29}_{-0.34-0.29}$ & $0.37^{+0.17+0.26}_{-0.12-0.28}$ & $0.47^{+0.13+0.56}_{-0.13-0.34}$ & $3.0^{+1.2+0.7}_{-0.9-0.6}$ & 1 \\ \hline
168 & 300--350 & $>$1000 & $\geq$9 & $\geq$3 & $3.5^{+1.7+0.5}_{-1.5-0.5}$ & $2.6^{+1.0+0.7}_{-0.9-0.7}$ & $0.42^{+0.19+0.29}_{-0.14-0.31}$ & $2.1^{+0.3+2.4}_{-0.3-1.8}$ & $8.6^{+2.7+2.6}_{-2.4-2.0}$ & 4 \\
169 & 350--500 & 500--1000 & $\geq$9 & $\geq$3 & $1.03^{+0.60+0.30}_{-0.47-0.30}$ & $1.58^{+0.71+0.43}_{-0.55-0.43}$ & $0.40^{+0.18+0.28}_{-0.13-0.31}$ & $0.10^{+0.03+0.11}_{-0.03-0.07}$ & $3.1^{+1.3+0.6}_{-1.0-0.6}$ & 3 \\
170 & 350--500 & $>$1000 & $\geq$9 & $\geq$3 & $0.81^{+0.56+0.14}_{-0.41-0.14}$ & $0.96^{+0.54+0.16}_{-0.27-0.16}$ & $0.77^{+0.27+0.53}_{-0.20-0.58}$ & $1.3^{+0.2+1.5}_{-0.2-1.1}$ & $3.8^{+1.1+1.6}_{-0.7-1.3}$ & 2 \\
171 & 500--750 & 500--1000 & $\geq$9 & $\geq$3 & $0.00^{+0.43+0.00}_{-0.00-0.00}$ & $0.03^{+0.46+0.03}_{-0.02-0.03}$ & $0.14^{+0.19+0.11}_{-0.09-0.11}$ & $0.01^{+0.02+0.01}_{-0.01-0.00}$ & $0.18^{+0.91+0.11}_{-0.09-0.11}$ & 0 \\
172 & 500--750 & $>$1000 & $\geq$9 & $\geq$3 & $0.00^{+0.48+0.00}_{-0.00-0.00}$ & $0.53^{+0.56+0.13}_{-0.31-0.13}$ & $0.48^{+0.22+0.33}_{-0.16-0.37}$ & $0.13^{+0.14+0.15}_{-0.13-0.00}$ & $1.1^{+1.1+0.4}_{-0.4-0.4}$ & 3 \\
173 & $>$750 & 750--1500 & $\geq$9 & $\geq$3 & $0.00^{+0.50+0.00}_{-0.00-0.00}$ & $0.00^{+0.46+0.00}_{-0.00-0.00}$ & $0.00^{+0.09+0.00}_{-0.00-0.00}$ & $0.01^{+0.05+0.02}_{-0.01-0.00}$ & $0.01^{+0.97+0.02}_{-0.01-0.00}$ & 0 \\
174 & $>$750 & $>$1500 & $\geq$9 & $\geq$3 & $0.00^{+0.42+0.00}_{-0.00-0.00}$ & $0.00^{+0.46+0.00}_{-0.00-0.00}$ & $0.00^{+0.11+0.00}_{-0.00-0.00}$ & $0.02^{+0.05+0.02}_{-0.02-0.00}$ & $0.02^{+0.89+0.02}_{-0.02-0.00}$ & 0 \\ 
\end{scotch}}
\end{table*}

\begin{table*}[htbp]
\renewcommand{\arraystretch}{1.25}
  \centering
  \topcaption{
  Observed numbers of events and prefit background predictions
 in the aggregate search regions.
 The first uncertainty is statistical and second systematic.
}
\label{tab:pre-fit-results-asr}
  \resizebox{\textwidth}{!}{
\begin{scotch}{ccccccccccc}
Bin & $\MHT$ [\GeVns{}] & $\HT$ [\GeVns{}] & $\njets$ & $\nbjets$ & Lost-$\Pe/\Pgm$ & $\Pgt\to\text{had}$ & $\cPZ\to\Pgn\cPagn$ & QCD & Total pred. & Obs. \\ \hline
1 & $>$500 & $>$500 & $\geq$2 & 0 & $842^{+25+48}_{-25-46}$ & $753^{+16+65}_{-16-65}$ & $5968^{+48+360}_{-47-350}$ & $21.4^{+0.6+8.5}_{-0.6-7.1}$ & $7584^{+63+370}_{-62-360}$ & 7838 \\[1mm]
2 & $>$750 & $>$1500 & $\geq$3 & 0 & $4.8^{+2.2+0.6}_{-1.6-0.6}$ & $4.2^{+1.3+0.3}_{-0.9-0.3}$ & $45.8^{+5.1+5.2}_{-4.3-4.9}$ & $0.47^{+0.06+0.18}_{-0.06-0.16}$ & $55.2^{+6.2+5.3}_{-5.0-4.9}$ & 71 \\[1mm]
3 & $>$500 & $>$500 & $\geq$5 & 0 & $111.0^{+6.4+8.3}_{-6.3-7.9}$ & $127.6^{+5.9+8.5}_{-5.7-8.6}$ & $558^{+15+36}_{-14-34}$ & $9.4^{+0.2+3.5}_{-0.2-3.1}$ & $806^{+19+38}_{-18-37}$ & 819 \\[1mm]
4 & $>$750 & $>$1500 & $\geq$5 & 0 & $1.82^{+0.82+0.26}_{-0.59-0.21}$ & $2.8^{+1.1+0.2}_{-0.7-0.2}$ & $18.1^{+3.3+2.7}_{-2.6-2.6}$ & $0.37^{+0.06+0.15}_{-0.06-0.13}$ & $23.0^{+3.8+2.7}_{-2.9-2.6}$ & 25 \\[1mm]
5 & $>$750 & $>$1500 & $\geq$9 & 0 & $0.23^{+0.27+0.14}_{-0.17-0.07}$ & $0.28^{+0.50+0.08}_{-0.21-0.07}$ & $0.00^{+0.82+0.00}_{-0.00-0.00}$ & $0.05^{+0.03+0.02}_{-0.03-0.02}$ & $0.6^{+1.1+0.2}_{-0.4-0.1}$ & 1 \\[1mm]
6 & $>$500 & $>$500 & $\geq$2 & $\geq$2 & $46.9^{+8.9+3.1}_{-5.9-3.0}$ & $44.0^{+4.4+3.2}_{-3.4-3.2}$ & $102^{+ 2+14}_{- 1-14}$ & $2.5^{+0.3+1.5}_{-0.2-1.3}$ & $196^{+13+15}_{- 9-15}$ & 216 \\[1mm]
7 & $>$750 & $>$750 & $\geq$3 & $\geq$1 & $11.5^{+4.1+1.0}_{-2.2-0.9}$ & $13.7^{+3.0+1.2}_{-2.0-1.2}$ & $87^{+ 3+10}_{- 3-10}$ & $0.87^{+0.15+0.34}_{-0.11-0.31}$ & $113^{+ 8+10}_{- 5-10}$ & 123 \\[1mm]
8 & $>$500 & $>$500 & $\geq$5 & $\geq$3 & $6.6^{+3.3+0.6}_{-2.3-0.6}$ & $5.3^{+1.9+0.9}_{-1.1-0.9}$ & $6.8^{+0.5+2.8}_{-0.3-2.8}$ & $0.87^{+0.20+0.96}_{-0.17-0.70}$ & $19.5^{+5.2+3.2}_{-3.4-3.1}$ & 17 \\[1mm]
9 & $>$750 & $>$1500 & $\geq$5 & $\geq$2 & $1.3^{+1.4+0.2}_{-0.6-0.2}$ & $1.8^{+1.3+0.4}_{-0.7-0.4}$ & $1.20^{+0.41+0.33}_{-0.19-0.33}$ & $0.13^{+0.07+0.06}_{-0.04-0.05}$ & $4.4^{+2.8+0.6}_{-1.3-0.6}$ & 6 \\[1mm]
10 & $>$750 & $>$750 & $\geq$9 & $\geq$3 & $0.00^{+0.66+0.00}_{-0.00-0.00}$ & $0.00^{+0.65+0.00}_{-0.00-0.00}$ & $0.00^{+0.15+0.00}_{-0.00-0.00}$ & $0.03^{+0.07+0.04}_{-0.02-0.01}$ & $0.0^{+1.3+0.0}_{-0.0-0.0}$ & 0 \\[1mm]
11 & $>$300 & $>$300 & $\geq$7 & $\geq$1 & $328^{+12+21}_{-12-20}$ & $380^{+10+22}_{- 9-22}$ & $193^{+ 8+38}_{- 6-38}$ & $69^{+ 1+29}_{- 1-26}$ & $969^{+23+57}_{-22-55}$ & 890 \\[1mm]
12 & $>$750 & $>$750 & $\geq$5 & $\geq$1 & $7.2^{+2.8+0.8}_{-1.6-0.7}$ & $7.7^{+2.4+0.8}_{-1.4-0.8}$ & $26.6^{+2.4+3.9}_{-1.8-3.7}$ & $0.65^{+0.14+0.26}_{-0.11-0.23}$ & $42.2^{+5.7+4.0}_{-3.5-3.9}$ & 48 \\
  \end{scotch}}
\end{table*}
\cleardoublepage \section{The CMS Collaboration \label{app:collab}}\begin{sloppypar}\hyphenpenalty=5000\widowpenalty=500\clubpenalty=5000\input{SUS-16-033-authorlist.tex}\end{sloppypar}
\end{document}

%% file: SUS-16-033-authorlist.tex
\textbf{Yerevan Physics Institute,  Yerevan,  Armenia}\\*[0pt]
A.M.~Sirunyan, A.~Tumasyan
\vskip\cmsinstskip
\textbf{Institut f\"{u}r Hochenergiephysik,  Wien,  Austria}\\*[0pt]
W.~Adam, F.~Ambrogi, E.~Asilar, T.~Bergauer, J.~Brandstetter, E.~Brondolin, M.~Dragicevic, J.~Er\"{o}, M.~Flechl, M.~Friedl, R.~Fr\"{u}hwirth\cmsAuthorMark{1}, V.M.~Ghete, J.~Grossmann, J.~Hrubec, M.~Jeitler\cmsAuthorMark{1}, A.~K\"{o}nig, N.~Krammer, I.~Kr\"{a}tschmer, D.~Liko, T.~Madlener, I.~Mikulec, E.~Pree, D.~Rabady, N.~Rad, H.~Rohringer, J.~Schieck\cmsAuthorMark{1}, R.~Sch\"{o}fbeck, M.~Spanring, D.~Spitzbart, J.~Strauss, W.~Waltenberger, J.~Wittmann, C.-E.~Wulz\cmsAuthorMark{1}, M.~Zarucki
% ...
\vskip\cmsinstskip
% ...
1:~~Also at Vienna University of Technology, Vienna, Austria\\
% ...
